\documentclass[manuscript,screen]{acmart}

\usepackage{babel}
\usepackage{multicol}
\usepackage{subfigure}
\usepackage{array}
\usepackage{textcomp}
\usepackage{multirow}
\usepackage{graphicx}
\usepackage{caption}
\usepackage{enumitem}

\newcommand{\lb}{\mathcal{LB}}
\newcommand{\cb}{\mathcal{CB}}
\newcommand{\ib}{\mathcal{IB}}
\newcommand{\mv}{\mathcal{MV}}
\newcommand{\hb}{\mathcal{HB}}
\newcommand{\skb}{\mathcal{SKB}}

\providecommand{\tabularnewline}{\\}

\setcopyright{acmcopyright}
\copyrightyear{2023}
\acmYear{2023}
\acmDOI{XXXXXXX.XXXXXXX}

\acmConference[ACM Computing Surveys]{Make sure to enter the correct
  conference title from your rights confirmation emai}

\begin{document}

\title{Envisioning the Future of Cyber Security in Post-Quantum Era: A Survey on PQ Standardization, Applications, Challenges and Opportunities}

\author{Saleh Darzi}
\email{salehdarzi@usf.edu}
\affiliation{%
  \institution{University of South Florida}
  \streetaddress{4202 E Fowler Ave.}
  \city{Tampa}
  \state{Florida}
  \country{USA}
  \postcode{33620}
}

\author{Kasra Ahmadi}
\email{Ahmadi1@usf.edu}
\affiliation{%
  \institution{University of South Florida}
  \streetaddress{4202 E Fowler Ave.}
  \city{Tampa}
  \country{USA}}

\author{Saeed Aghapour}
\email{Aghapour@usf.edu}
\affiliation{%
  \institution{University of South Florida}
   \streetaddress{4202 E Fowler Ave.}
  \city{Tamap}
  \country{USA}
}

\author{Attila Altay Yavuz}
\affiliation{%
	\email{attilaayavuz@usf.edu}
 \institution{University of South Florida}
 \streetaddress{4202 E Fowler Ave.}
 \city{Tampa}
 \state{Florida}
 \country{USA}}

\author{Mehran Mozaffari Kermani}
\affiliation{%
\email{mehran2@usf.edu}
  \institution{University of South Florida}
  \streetaddress{4202 E Fowler Ave.}
  \city{Tampa}
  \state{Florida}
  \country{USA}}

\renewcommand{\shortauthors}{Saleh et al.}

\begin{abstract}
	The rise of quantum computers exposes vulnerabilities in current public key cryptographic protocols, necessitating the development of secure post-quantum (PQ) schemes. Hence, we conduct a comprehensive study on various PQ approaches, covering the constructional design, structural vulnerabilities, and offer security assessments, implementation evaluations, and a particular focus on side-channel attacks. We analyze global standardization processes, evaluate their metrics in relation to real-world applications, and primarily focus on standardized PQ schemes, selected additional signature competition candidates, and PQ-secure cutting-edge schemes beyond standardization. Finally, we present visions and potential future directions for a seamless transition to the PQ era.
\end{abstract}

\ccsdesc[500]{Security and privacy~Cryptography}

\keywords{Post quantum cryptography, cybersecurity, standardization, quantum safe applications}


\maketitle

\section{Introduction}
\label{section1}
\vspace{-2mm}

While symmetric key cryptography (SKC) is faster and computationally lighter, the challenges of distributing keys in large-scale systems and scalability issues in online communications have driven the adoption of public-key cryptography (PKC) methods such as RSA and elliptic curve cryptography (ECC). 
PKC-based methods rely on the computational difficulty of solving specific mathematical problems. The RSA scheme, for instance, is based on the integer factorization problem (IFP), while the Diffie-Hellman scheme is based on the discrete logarithm problem (DLP). The hardness of these problems is determined by the absence of proof and the inability of probabilistic computers to solve them in polynomial time \cite{katz2020introduction}. 
But this changed in 1994 when Peter Shor \cite{shor1994algorithms} introduced two probabilistic polynomial-time quantum algorithms capable of efficiently solving IFP and DLP, achieved through the utilization of period finding using Quantum Fourier Transformation. 
This development posed a significant threat to the security of contemporary communication, network systems, and the current standardized schemes (\textit{e.g.}, RSA, DH, DSA, ECDH, ESDSA) that are widely employed in various applications globally. 
Within SKC, the vulnerabilities often revolve around the reliance on brute force attacks as the primary threat. In 1996, Grover's probabilistic algorithm \cite{grover1996fast} was proposed, primarily targeting SKC. Specifically, Grover's algorithm accelerates brute force attacks on symmetric primitives by approximately a quadratic factor, thereby reducing the security of the standardized SKC schemes (\textit{e.g.}, AES, SHA2, SHA3) to half of the key size.

Quantum computers, while a significant technological development, currently face practicality limitations associated with performance metrics, state stability, hardware size, memory constraints, and the need for fault tolerance.
Defining qubits (\textit{i.e.}, quantum bits) as a quantifying metric for the potency of quantum computers, the existing number of qubits remains insufficient for practical real-world problem-solving, necessitating a substantial increase by several orders of magnitude \cite{yang2023survey}.
Over the past decade, there has been a global endeavor involving academia and industry with leading companies such as Google, IBM, Microsoft, and Intel aimed at realizing quantum technology. Notably, Google has achieved a significant milestone by constructing a 433-qubit Osprey in 2022. They anticipate surpassing the milestone of 4000 qubits in a quantum computer by the year 2025 \cite{IBM}. However, it is worth noting that achieving the capability to break current cryptography schemes implemented on classical computers would necessitate thousands or even millions of error-free qubits.
However, the global efforts and billions of dollars investment in this technology has shifted the question from " whether " these systems can be breached to "when" they might be compromised \cite{mosca2021quantum}.

The necessity of standardization is self-evident and represents the initial formal stride into the emerging era of PQ technology.
It's crucial to acknowledge that there are pressing concerns and ongoing threats that demand immediate consideration. One notable concern is the store-now-decrypt-later (SNDL) attack, which relies on a future quantum computer capable of decrypting stored sensitive data that has been eavesdropped and stored. Thus, delaying the standardization process of PQC allows adversaries to accumulate more sensitive data, such as medical records, national data, and more. Another concern relates to existing technologies and applications that heavily rely on classical cryptography, rendering them insecure in the PQ era. This poses challenges, costs, and even impossibilities in terms of updating infrastructure for vehicles, networks like smart grids \cite{darzi2022lpm2da}, and other large-scale systems \cite{qassim2017post}.  
Given the intricacies involved in standardizing PQC compared to symmetric primitives, including the diverse array of PQC approaches and their associated implementation costs, the shift from PKC to PQC necessitates multiple rounds of assessment. 
In light of these, it became imperative to initiate a global initiative at the earliest opportunity.
Also, transitioning from current cryptosystems and widely-used internet protocols, among other networks, to PQ security schemes is a time-consuming process that may take years to achieve widespread adoption even if standardized promptly.

In addition to the general standardization of PQC, certain applications necessitate specialized treatment and potentially separate competitions tailored to their specific criteria. For instance, in the context of blockchain technology, the need arises for a signature scheme that goes beyond conventional methods \cite{behnia2021lattice}. Specifically, the signature must possess unique characteristics such as unlikability and decentralization, among others, in order to fulfill diverse security requirements across various application-specific roles \cite{behnia2017high}. Similarly, prevalent machine learning (ML) protocols demand specialized methods like fully homomorphic encryption (FHE), multi-party computation (MPC), and zero-knowledge proofs (ZKP) with PQ security. Thus, beyond signature and encryption schemes, novel approaches have been developed to address specific challenges and offer unique capabilities.  
Furthermore, apart from addressing security concerns, the standardization process also enables academia and industry to consider real-world needs and criteria such as implementation considerations, power consumption, and various attack vectors like side-channel and fault injection attacks. Notably, many of the proposed schemes aimed at achieving higher security standards require computationally intensive operations that may pose challenges when implemented in resource-constrained networks \cite{behnia2022lightweight}. Consequently, it becomes apparent, based on the aforementioned analysis, that it is imperative to prioritize the standardization process for PQC and thoroughly survey all potential PQ approaches.

\vspace{-3mm}
\subsection{Related Works}
\vspace{-2mm}
Extensive research endeavors have concentrated on PQC, leading to the creation of numerous surveys. Table \ref{tab:surveyofsurvey} offers a comprehensive evaluation of PQC surveys that covering all major approaches and candidate schemes, including our own survey.   
The criteria for our comparison, represented in the table's columns, have been developed through a comprehensive review of global standardization procedures, evaluation metrics, scrutiny of PQC candidates submitted to NIST, and schemes outside standardization, rigorous security assessments, performance evaluations, implementation considerations, potential attack scenarios, examination of exotic/advance features, and the inclusion of a visions for each PQC approach and application-specific projections.  
Due to the inherent limitations and non-comprehensive nature, our comparison excludes surveys that only discuss encryption algorithms (\textit{e.g.}, \cite{ravi2021lattice}), digital signatures (\textit{e.g.}, \cite{buser2023survey}), or exclusively focus on one PQC approach (\textit{e.g.}, lattice-based  \cite{ravi2021lattice, nejatollahi2019post}, code-based \cite{weger2022survey}, isogeny-based \cite{beullens2023proving}, hash-based \cite{li2022hash}, multivariate cryptography \cite{dey2023progress}, quantum key distribution (QKD)\cite{mehic2020quantum}).

Besides surveys tailored to a particular approach, a plethora of studies focus on applications with unique requirements and widespread utility. This includes research on constraint devices within the realm of the Internet of Everything (or IoT) \cite{9787987, doi.org/10.1002/spe.3121,doi.org/10.1002/spy2.200}, investigations into blockchain-based schemes necessitating novel methodologies \cite{buser2023survey, 8967098, 8932459}, exploration of exotic features pertinent to ML/AI applications \cite{9363165}, as well as surveys considering various networks such as the Internet, vehicular networks, and wireless sensor networks \cite{en15030714,9646494}, among others.   
Nonetheless, it's evident that these application-specific surveys are confined to particular types of PQ solutions and are ill-suited for a fair comparison with our all-encompassing survey. As an illustration, PQC surveys in the IoT domain primarily concentrate on computational cost, communication overhead, and implementation considerations tailored to constrained devices.

Table \ref{tab:surveyofsurvey} highlights that, to the best of our knowledge, we are the sole survey conducting an exhaustive examination of global PQC standardization endeavors, initiatives, and their associated metrics.  
Furthermore, only one other survey attempted to carry out a constructional analysis encompassing security, implementation, and attack assessments concurrently, though it did not delve into a scheme-level analysis.  
In addition, most surveys primarily focus on analyzing schemes within the NIST standardization process and do not extend their scope to consider schemes beyond this sphere, which may hold promise in the PQ era.  
Also, only a few surveys have explored unconventional features beyond generic use cases, advanced cryptographic constructions, and application-specific analyses, especially for major applications in the PQ era. 
\begin{table*}
	\caption{A Comprehensive Analysis and Comparison of PQC Surveys}
	\vspace{-4mm}
	\label{tab:surveyofsurvey}
	\begin{tabular}{@{}l|@{}c|@{}c|@{}c|@{}c|@{}c|@{}c|@{}c|@{}c|@{}c|@{}c}
		\toprule
		Surveys & \multicolumn{2}{c|}{Standardization} & \multicolumn{2}{c|}{Candidates} & Security & \small{Implementation}& Attack & \small{Application} & Exotic & Future \\
		\& Year & Processes &  Metrics & NIST & Beyond &Analysis & Analysis & Analysis & \small{Assessment} & Features& Visions\\
		\midrule
				\small{Our Survey} &Global& $\boldsymbol{\checkmark}$&$\boldsymbol{\checkmark}$&$\boldsymbol{\checkmark}$&$\boldsymbol{\checkmark}$ &$\boldsymbol{\checkmark}$&$\boldsymbol{\checkmark}$&$\boldsymbol{\checkmark}$& $\boldsymbol{\checkmark}$&$\boldsymbol{\checkmark}$\\
		\midrule
		\cite{10048976} 2023& NIST&$\boldsymbol{\times}$& $\boldsymbol{\checkmark}$&$\boldsymbol{\times}$& $\boldsymbol{\checkmark}$& $\boldsymbol{\checkmark}$&$\boldsymbol{\times}$ &$\boldsymbol{\times}$&$\boldsymbol{\times}$&$\boldsymbol{\times}$ \\
		\midrule
		\cite{dam2023survey} 2023& NIST&$\boldsymbol{\times}$&$\boldsymbol{\checkmark}$&$\boldsymbol{\times}$ & $\boldsymbol{\checkmark}$& $\boldsymbol{\checkmark}$&$\boldsymbol{\times}$& $\boldsymbol{\times}$& $\boldsymbol{\times}$& $\boldsymbol{\times}$\\
		\midrule		
		\cite{buser2023survey} 2023& $\boldsymbol{\times}$&$\boldsymbol{\times}$&$\boldsymbol{\checkmark}$&$\boldsymbol{\checkmark}$ &$\boldsymbol{\checkmark}$&$\boldsymbol{\times}$&$\boldsymbol{\times}$ &$\boldsymbol{\checkmark}$  &$\boldsymbol{\checkmark}$& $\boldsymbol{\times}$\\
		\midrule
		\cite{Hekkala2023} 2023& NIST&$\boldsymbol{\times}$& $\boldsymbol{\checkmark}$&$\boldsymbol{\times}$ &$\boldsymbol{\checkmark}$&$\boldsymbol{\checkmark}$& $\boldsymbol{\times}$&$\boldsymbol{\times}$& $\boldsymbol{\times}$&$\boldsymbol{\times}$ \\
		\midrule
		\cite{en15030714} 2022&$\boldsymbol{\times}$ &$\boldsymbol{\times}$ &$\boldsymbol{\times}$ &$\boldsymbol{\times}$& $\boldsymbol{\times}$&$\boldsymbol{\checkmark}$&$\boldsymbol{\checkmark}$&$\boldsymbol{\checkmark}$&$\boldsymbol{\times}$ &$\boldsymbol{\times}$ \\
		\midrule
		\cite{9835864} 2022& NIST& $\boldsymbol{\times}$&$\boldsymbol{\checkmark}$&$\boldsymbol{\times}$ &$\boldsymbol{\checkmark}$ &$\boldsymbol{\times}$&$\boldsymbol{\times}$ &$\boldsymbol{\times}$&$\boldsymbol{\times}$&$\boldsymbol{\times}$ \\
		\midrule
		\cite{9646494} 2022&$\boldsymbol{\times}$&$\boldsymbol{\times}$&$\boldsymbol{\checkmark}$  &$\boldsymbol{\checkmark}$ & $\boldsymbol{\checkmark}$&$\boldsymbol{\checkmark}$& $\boldsymbol{\checkmark}$ &$\boldsymbol{\times}$&$\boldsymbol{\times}$&$\boldsymbol{\checkmark}$\\
		\midrule		
		\cite{9787987} 2022&$\boldsymbol{\times}$ &$\boldsymbol{\times}$&$\boldsymbol{\times}$& $\boldsymbol{\times}$& $\boldsymbol{\times}$&  $\boldsymbol{\checkmark}$&$\boldsymbol{\times}$&$\boldsymbol{\checkmark}$&$\boldsymbol{\times}$&$\boldsymbol{\times}$ \\
		\midrule
		\cite{KUMAR2022100242} 2022& $\boldsymbol{\checkmark}$&$\boldsymbol{\times}$&$\boldsymbol{\checkmark}$&$\boldsymbol{\times}$ &$\boldsymbol{\checkmark}$&$\boldsymbol{\checkmark}$&$\boldsymbol{\times}$&$\boldsymbol{\times}$ &$\boldsymbol{\times}$ &$\boldsymbol{\checkmark}$ \\
		\midrule
		\cite{doi.org/10.1002/spe.3121} 2022&$\boldsymbol{\checkmark}$&$\boldsymbol{\checkmark}$ &$\boldsymbol{\checkmark}$&$\boldsymbol{\checkmark}$ &$\boldsymbol{\checkmark}$& $\boldsymbol{\checkmark}$&$\boldsymbol{\times}$ &$\boldsymbol{\checkmark}$&$\boldsymbol{\times}$& $\boldsymbol{\checkmark}$\\
		\midrule
		\cite{dichiano2021survey} 2021&$\boldsymbol{\times}$&$\boldsymbol{\times}$&$\boldsymbol{\checkmark}$&$\boldsymbol{\times}$&$\boldsymbol{\checkmark}$&$\boldsymbol{\checkmark}$&$\boldsymbol{\times}$&$\boldsymbol{\times}$&$\boldsymbol{\times}$&$\boldsymbol{\times}$ \\
		\midrule		
		\cite{9363165} 2021&$\boldsymbol{\times}$ &$\boldsymbol{\times}$ &$\boldsymbol{\checkmark}$&$\boldsymbol{\checkmark}$ &$\boldsymbol{\checkmark}$ &$\boldsymbol{\checkmark}$& $\boldsymbol{\times}$&$\boldsymbol{\checkmark}$&$\boldsymbol{\checkmark}$ &$\boldsymbol{\checkmark}$\\
		\midrule
		\cite{doi.org/10.1002/spy2.200} 2021&$\boldsymbol{\times}$&$\boldsymbol{\times}$&$\boldsymbol{\checkmark}$& $\boldsymbol{\times}$&$\boldsymbol{\checkmark}$&$\boldsymbol{\times}$ & $\boldsymbol{\checkmark}$& $\boldsymbol{\checkmark}$& $\boldsymbol{\times}$ &$\boldsymbol{\checkmark}$\\
		\midrule		
		\cite{8932459} 2020& $\boldsymbol{\checkmark}$&$\boldsymbol{\times}$&$\boldsymbol{\checkmark}$&$\boldsymbol{\checkmark}$ &$\boldsymbol{\checkmark}$ &$\boldsymbol{\checkmark}$&  $\boldsymbol{\times}$&$\boldsymbol{\checkmark}$&$\boldsymbol{\times}$ &$\boldsymbol{\checkmark}$\\
		\midrule
		\cite{9286147} 2020& NIST&$\boldsymbol{\times}$&$\boldsymbol{\checkmark}$ &$\boldsymbol{\times}$& $\boldsymbol{\checkmark}$&$\boldsymbol{\times}$& $\boldsymbol{\times}$&$\boldsymbol{\times}$&$\boldsymbol{\times}$ &$\boldsymbol{\times}$\\
		\midrule
		\cite{10.1007/978-3-030-78375-4_17} 2020& NIST&$\boldsymbol{\times}$ &$\boldsymbol{\checkmark}$&$\boldsymbol{\times}$&$\boldsymbol{\checkmark}$ &$\boldsymbol{\checkmark}$& $\boldsymbol{\times}$& $\boldsymbol{\checkmark}$&$\boldsymbol{\times}$&$\boldsymbol{\checkmark}$ \\
		\midrule		
		\cite{8967098} 2020&$\boldsymbol{\times}$ &$\boldsymbol{\times}$ &$\boldsymbol{\checkmark}$ &$\boldsymbol{\checkmark}$&$\boldsymbol{\checkmark}$&$\boldsymbol{\checkmark}$&$\boldsymbol{\times}$ &$\boldsymbol{\checkmark}$ &$\boldsymbol{\checkmark}$&$\boldsymbol{\checkmark}$ \\
		\midrule
		\cite{1910089619} 2020 &$\boldsymbol{\times}$  &$\boldsymbol{\times}$ &$\boldsymbol{\checkmark}$ &$\boldsymbol{\times}$&$\boldsymbol{\checkmark}$ &$\boldsymbol{\times}$ &$\boldsymbol{\times}$&$\boldsymbol{\times}$&$\boldsymbol{\times}$&$\boldsymbol{\times}$ \\
		\midrule
		\cite{survey24} 2019 & $\boldsymbol{\times}$&$\boldsymbol{\times}$ &$\boldsymbol{\checkmark}$ &$\boldsymbol{\times}$&$\boldsymbol{\checkmark}$ &$\boldsymbol{\times}$ &$\boldsymbol{\times}$&$\boldsymbol{\times}$&$\boldsymbol{\times}$&$\boldsymbol{\times}$ \\
		\midrule
		\cite{malina2018feasibility} 2018 &$\boldsymbol{\times}$ &$\boldsymbol{\times}$ &$\boldsymbol{\checkmark}$ &$\boldsymbol{\times}$&$\boldsymbol{\checkmark}$ &$\boldsymbol{\checkmark}$ &$\boldsymbol{\times}$&$\boldsymbol{\times}$&$\boldsymbol{\times}$&$\boldsymbol{\times}$ \\
		\midrule
		\cite{Bernstein2017} 2017 &$\boldsymbol{\times}$  &$\boldsymbol{\times}$ &$\boldsymbol{\times}$ &$\boldsymbol{\times}$&$\boldsymbol{\checkmark}$ &$\boldsymbol{\times}$ &$\boldsymbol{\times}$&$\boldsymbol{\times}$&$\boldsymbol{\times}$&$\boldsymbol{\times}$ \\
		\bottomrule
	\end{tabular}
\vspace{-5mm}
\end{table*}

\vspace{-1mm}
\subsection{Contribution}
\vspace{-2mm}

\begin{itemize}[leftmargin=*]
	\item Our survey distinguishes itself by not only conducting a comprehensive study of worldwide standardization efforts and initiatives, but also meticulously assessing their evaluation metrics concerning practical applications and demands, ultimately facilitating the path forward for PQC. Additionally, we provide a thorough overview of all projects, institutions, and PQC-related products on a worldwide scale. This serves the purpose of ensuring our alignment with the correct path, understanding our current position, identifying future directions, and pinpointing areas that require improvement. This approach facilitates a more informed comparison of the proposed scheme with real-world requirements and simplifies the transition into the PQ era.
	
	\item We provide an exhaustive analysis that covers all spectrum of PQ solutions, spanning from established PQC approaches to unconventional methods (such as PQ-secure schemes derived from MPC techniques, ZKP protocols, and symmetric key primitives), as well as quantum cryptography. 
	Our analysis encompasses construction methodologies, the underlying security assumptions, structural vulnerabilities, and highlights the PQC candidates within the NIST framework, with a specific emphasis on standardized PQ schemes and selected schemes from the additional signature competition. To ensure comprehensiveness, we also extend to evaluate cutting-edge schemes beyond standardization that hold value in the PQ era. 
	In addition to a thorough security assessment, our survey delves into performance evaluations across various implementation benchmarks and conducts assessments of potential attacks, with a specific focus on side-channel attacks.  
				
	\item Considering the existence of diverse applications such as IoT, Blockchain, ML/AI, and others that necessitate specialized treatment and involve advanced methods like FHE, MPC, ZKP, signatures with exotic features (\textit{e.g.}, blind-, group-, ring-signature, etc.), an examination of these advanced methods lead to providing a solid roadmap for the future of PQC. As such, we offer a tailored assessment of these major applications for each PQC approach, shedding light on their specialized requirements and providing insights into advanced constructions in the PQ era.

	\item In addition to the aforementioned analysis, there remain unanswered but critical questions: 
	\textit{``Have there been any omitted hard problems or solid scheme proposals in PQC standardizations that possess merits?"} 
	\textit{``Do the current standardizations and schemes possess applicability to deeply embedded systems, functionality in real-world real-time cyber-physical systems, and compatibility with parallel and batching processing?"}  
	\textit{``Do the current standardizations result in a lightweight, high-speed, high-performance PQC, or is it necessary to pursue another competition?"}  
	\textit{``Is the Isogeny-based approach dead? What factors contributed to the lack of anticipation regarding schemes like SIKE?"}  
	\textit{``And what course of action should be taken following the completion of the NIST Standardization process? What steps should be undertaken subsequently?"}  
	We aim to identify potential solutions, offer insights, and pave the way for further research into pivotal questions like these.
\end{itemize}



\vspace{-4mm}
\subsection{Outline}
\vspace{-2mm}
The structure of this survey is outlined in the following manner. In Section \ref{section2}, we present an exhaustive study of PQC standardization procedures, projects, and their evaluation metrics around the world. Section \ref{section3} forms the evaluative nucleus of this survey, wherein the diverse range of approaches, inclusive of lattice-based, hash-based, code-based, multivariate,  isogeny-based, symmetric key-based, miscellaneous methods, and QKD, are scrutinized in-depth based on their underlying hard problems, constructions, structural attacks, and potential candidates. In the ensuing Section \ref{section4}, we make deductions based on the preceding analysis and the evaluation conducted on major applications, and subsequently, articulate our visions, resultant findings, and prospective trajectory for PQC approaches. Section \ref{section5} brings the survey to a formal closure, encapsulating the discussions, and suggesting possible directions for future investigation.

\vspace{-2mm}
\section{Exhaustive Study of PQC Standardizations and Metrics}
\label{section2}
\vspace{-2mm}
Within this section, our objective is to comprehensively study worldwide initiatives focused on PQC standardization, along with the various projects and institutions that prioritize PQC.
In addition to this examination, we will delve into the standardization prerequisites and evaluation metrics that facilitate the transition from the pre-quantum to the post-quantum era, establishing a foundation for this shift.

\vspace{-3mm}
\subsection{PQC Standardizations, Procedures, and Projects}
\vspace{-2mm}
The standardization process plays a significant role in facilitating the transition to a new cybersecurity infrastructure. With the advent of quantum computers, various countries, institutions, and organizations are actively engaged in global efforts to standardize, develop, and implement PQC. 


Despite the extensive historical background of PQC research and development, it was not until 2006 that the inaugural PQCrypto conference, dedicated exclusively to PQC, was held on an international scale.
Functioning as the primary authority for standardizing PQ primitives, NIST has successfully conducted two competitions. 
The stateful hash-based signature (HBS) competition has concluded, and IETF has published RFCs pertaining to these signatures.
On the other hand, the call for PQC standardization was issued on December 2016, seeking the development of cryptographic algorithms, including PKE schemes, KEMs, and digital signatures that offer security against classical as well as quantum computers.
In each iteration of the PQC competition, NIST released a report that examined the chosen schemes and provided the rationale behind their selection (Report 1 \cite{alagic2019status}, Report 2 \cite{moody2020status}, Report 3 \cite{alagic2022status}).

NIST recently released Federal Information Processing Standards (FIPS) draft documents for three of the standardized schemes \cite{NISTsigCompetition}: 1) \textit{CRYSTALS-KYBER} as Module-Lattice-Based Key Encapsulation Mechanism Standard (FIPS 203), 2) \textit{CRYSTALS-Dilithium} as Module-Lattice-Based Digital Signature Standard (FIPS 204), 3) $\textit{SPHINCS}^+$ as Stateless Hash-Based Digital Signature Standard (FIPS 205).
In addition to the standard schemes, there were alternative PKE/KEM schemes (\textit{BIKE} \cite{aragon2017bike}, \textit{Classic-McEliece} \cite{bernstein2017classic}, \textit{HQC} \cite{melchor2018hamming}, \textit{SIKE} \cite{jao2017sike}) that underwent evaluation during the $4^{th}$ round of analysis in the PQC competition. Setting aside \textit{FALCON} \cite{fouque2019fast}, since no alternative signature schemes were retained in the PQC competition and the emphasis on the importance of promoting diversity in PQC standards, NIST initiated an additional competition dedicated to the standardization of general-purpose digital signatures.

The call for signature submission emphasized the need for non-lattice algorithms suitable for applications like certificate transparency. The predominant necessary condition is to offer a solution that demonstrates "quick verification and concise signature" properties.
For lattice-based proposals, the condition was to ensure security against EUF-CMA attacks while also offering additional features beyond what is provided by \textit{Dilithium} \cite{ducas2018crystals} and \textit{FALCON} \cite{fouque2019fast}. Non-lattice proposals, on the other hand, had to exhibit notable performance benefits over $\textit{SPHINCS}^+$ \cite{bernstein2019sphincs+}. 
Notably, NIST has recently disclosed the initial-round schemes in this competition, which will be thoroughly assessed based on each unique approach in the next section \cite{NISTsigCompetition}.  
Notably, other standardization bodies including ISO, ETSI, and IETF largely rely on NIST to conclude the competition, allowing them to proceed with their respective tasks securely. Besides the standardization entities, all different working groups worldwide involved in the PQC standardization process with their projects and products are depicted in Table \ref{tab:standardization}. 

\begin{table*}
	\caption{A Comprehensive analysis of standardization procedures and projects around the world}
	\vspace{-3mm}
	\label{tab:standardization}
	\begin{tabular}{c|c|c|l}
		\toprule
		Project & Country & Focus & Work, Actions, and Products \\
		\midrule		
		IETF \cite{IETF} & Inter & \small{PQC Standardization} & \small{Integrating PQC schemes into many protocols}\\
		& national & \small{Publishing RFCs} & \small{PQUIP works on engineering aspects of PQ}\\
		& & & \small{3GPP works on cellular networks}\\
		\midrule
		ETSI \cite{ETSI} & Europe & \small{PQC \& QKD} & \small{Working groups collaboration: IQC, ISG, QSC}\\
		&  & \small{Standardization} & \small{Publications, seminars, \& whitepapers}\\
		& & & \small{Real-world and industrial PQC development}\\
		& & & \small{Put forth migration strategies \& scenarios}\\
		\midrule
		NSA \cite{NSA} & USA & \small{Transition towards} & \small{Publishing numerous reports}\\
		& &\small{PQC \& QKD} & \small{PQC in military and intelligence operations}\\
		\midrule
		ENISA \cite{ENISA} & Europe & \small{Standardization Process} & \small{Publishing reports}\\
		& & & \small{Utilization of double encryption/signature}\\		
		\midrule
		PROMETHEUS \cite{FutureTPM} & Europe & \small{Privacy-Preserving PQ Systems} & \small{Ensuring confidentiality and privacy}\\
		& & \small{Lattice-based Cryptography} & \small{Using advanced mechanisms like MPC \& FHE}\\
		\midrule
		FutureTPM \cite{FutureTPM} & Europe & \small{Quantum-Resistant Platform} & \small{Ensuring long-term security in QC}\\
		& & & \small{Risk assessment of run-time, HW \& SW}\\
		\midrule
		ANSSI \cite{ANSSI} & France & \small{Advisory Services} & \small{PQC publications \& evaluation of}\\
		& & \small{Regulation Oversight} & \small{certificates with European guidelines}\\
		\midrule
		FLOQI \cite{FLOQI} & Germany & \small{Full-lifecycle PQC Schemes} & \small{Developing and implementing PKE for }\\
		& & & \small{automotive industry \& financial sectors}\\
		\midrule
		QuantumRISC \cite{koskesh} & Germany & \small{PQC on Embedded Devices} & \small{Practical implementation \& HW development}\\
		& & & \small{for industrial resource constraints devices} \\
		\midrule
		KBLS \cite{KBLS} & Germany & \small{Long-term Security} & \small{Enhancing cryptographic libraries}\\
		& & & \small{Facilitating and development of SW}\\
		\midrule
		Aquorypt \cite{Aquacrypt} & Germany & \small{PQC Implementation} & \small{Implementing on HW \& SW}\\
		& & & \small{Targets industrial embedded \& control systems}\\
		& & & \small{with constrained devices like smart cards}\\		
		\midrule
		PQC4MED \cite{PQC4MED} & Germany & \small{Medical Technology Services} & \small{Security approaches for embedded systems}\\
		& & & \small{By utilizing HW \& SW elements}\\
		\midrule
		QuaSiModO \cite{Qua} & Germany & \small{PQ Secure VPNs} & \small{Delivering secure networking solutions}\\		
		\midrule
		MTG-ERS \cite{MTG} & Germany & \small{PQC Implementation} & \small{Integration of PQC into customer applications}\\
		& & & \small{Introducing crypto-agility} metric\\
		\midrule 
		CRYPTREC \cite{CRYPTREC} & Japan & \small{Security of approaches} & \small{A lattice-based scheme called LOTUS}\\
		& & &\small{Introducing the versatility metric}\\
		\midrule
		CryotMathCREST \cite{CryptoMathCREST} & Japan & \small{Mathematical aspects} & \small{Designing an innovative model}\\
		&  & \small{PQC Cryptosystems' Security} & \small{in industrial mathematics.}\\		
		\midrule
		QAPP \cite{CACR} & China & \small{Designing and refining} & \small{Development of PQ libraries}\\
		& & \small{cryptosystems} & \small{Secure data transfer in client-server Apps}\\		
		\midrule
		CACR \cite{CACR} & China & \small{PQC Standardization} & \small{Design Lattice-based Cryptosystems:}\\
		& & \small{Organizing PQC Competition} & \small{LAC, LPN, Lepton, Aigi (Sig/Enc)}\\
		& &  & \small{Performing Cryptanalysis}\\		
		\bottomrule
	\end{tabular}
\end{table*}

\vspace{-4mm}
\subsection{Standardization Requirements and Evaluation Metrics}
\vspace{-2mm}
Here, we present the formal concepts and requirements established by standardization entities. Notably, NIST's standardization efforts encompass clearly defined metrics across three key aspects: security, implementation-cost, and algorithmic characteristics. On the other hand, other standardization organizations and projects predominantly establish application-specific criteria or regulations rather than metrics. 

\textbf{\textit{SECURITY:}}
The main criterion for consideration is undeniably the level of security achieved. NIST has established five security levels based on symmetric key cryptographic primitives that have already been standardized. These levels can be summarized as follows: level one corresponds to breaking AES-128, level two entails finding a collision on a 256-bit hash function (such as SHA2-256 or SHA3-256), level three involves breaking a 192-bit block cipher (AES-192), level four relates to finding a collision on a 384-bit hash function (SHA-384), and level five encompasses a key search on a 256-bit block cipher (AES-256). While the primary focus of the NIST competition is on the first three levels, the submission of parameter sets for level five is also encouraged. 
Apart from achieving these security levels, additional security properties including forward/backward security \cite{yavuz2022frog}, backward compatibility, resistance against attacks like side-channel and multi-key attacks, as well as resistance to misuse are also of importance\cite{alagic2019status}.  
It should be noted that both of the theoretical models such as the Random Oracle Model (ROM) and its extension, the Quantum Random Oracle Model (QROM), involve the capability to respond to queries and provide public accessibility. However, the QROM diverges by permitting quantum adversaries and facilitating quantum operations. 
Technically, for PKE/KEM, they need to demonstrate IND-CPA, IND-CCA, and IND-CCA2 security for general-purpose and ephemeral use cases. On the other hand, signature schemes are expected to provide EUF-CMA security \cite{alagic2022status}.

\textbf{\textit{COST, PERFORMANCE, \& IMPLEMENTATION:}}
The costs encompass various factors such as the size of keys, ciphertexts, and signatures, while also considering limitations on memory usage and code size. Furthermore, the evaluation of performance examines the efficiency of computational costs, the intensity of communication, and the rate of decryption failures. Both algorithmic and implementation characteristics are taken into account, including the chosen platform, software effectiveness, and hardware requirements. 
Besides, the performance of the key generation component is vital and should be prioritized alongside other aspects of the system. 
PQ cryptosystems often restrict the number of messages that can be signed with the same key, necessitating key generation for each message group. However, IoT devices may struggle with the computational resources required for key generation, leading to additional overhead and strain on their limited capabilities. 

\textbf{\textit{ATTACK RESISTANCE:}}
Rather than mathematically analyze a protocol, SCAs target on exploiting vulnerabilities in its implementation to acquire sensitive data. This means that even schemes with security proofs, can still be vulnerable to SCAs. In an SCA, the internal circuits are observed during cryptographic computations, and this information will be used to extract sensitive data. The process of conducting an SCA involves two crucial steps: gathering the leaked information and turning them into vectors, and searching through the different inputs which result into those leaked information to retrieve the secret value. SCAs can be carried out using various methods, such as those based on acoustics, light, timing, power analysis, and electromagnetic emissions \cite{SCA1,SCA2}, and \cite{SCA3}. The primary distinction between these methods lies in the type of data being analyzed.  
Fault injection attacks are another type of side channel attacks where errors and faults are deliberately introduced into an algorithm to disrupt its intended operation. The purpose of these attacks is to manipulate the system in such a way that it reveals sensitive data through its unintended leakages. It is crucial to mention that even a single faulty input can have severe consequences on the algorithm's output \cite{SCA4}. 

\textbf{\textit{DIVERSITY:}}
The necessity for alternative candidates and approaches in a standardization process is apparent from the perspective of common sense. Furthermore, when selecting a scheme for standardization from various existing approaches, a significant factor to consider is the diversity metric, which measures variations in mathematical complexity, security assumptions, and algebraic structure. This metric ensures the prevention of a single vulnerability that could compromise global security and undermine the efforts of standardization \cite{alagic2022status}. 

\textbf{\textit{VERSATILITY:}}
The principle behind the versatility metric is rooted in the understanding that a chain is only as strong as its weakest link. This metric aims to assess the security of an entire system when multiple cryptographic primitives are integrated together. While the individual security of each primitive holds significance, the versatility metric highlights the practical application and mathematical proof of their collective utilization. This is relevant as these cryptographic techniques often operate within a system that encompasses various cryptographic components.

\textbf{\textit{EXOTIC FEATURES:}}
While standardization entities primarily concentrate on cryptographic primitives for general-purpose use cases, it is crucial to avoid overlooking the easement of future transition to PQC, the current diverse range of application use cases, and the potential to support exotic features like blind/group/ring signatures, FHE, MPC, SS, etc., and advanced enhancements within these primitives. 

\textbf{\textit{SIMPLICITY \& UNIQUENESS:}}
Besides the aforementioned metrics, there exist certain implicit concepts that lack quantifiable measurements, such as uniqueness, simplicity, and elegance. While standardization entities have not explicitly defined these metrics, they have not been grounds for significant disqualification of schemes either. However, considering these metrics can contribute to a more well-rounded standardization process amidst the wide array of PQC schemes. Specifically, uniqueness, simplicity, and elegance pertain to having a transparent design, minimal reliance on complex mathematical techniques, fewer security assumptions and reductions, and adherence to simpler principles.

\section{Comprehensive Analysis of PQC Approaches}
\label{section3}
\vspace{-1.5mm}
In this section, we present an extensive exploration of different PQC approaches, encompassing lattice-based methods, code-based techniques, multivariate cryptography, symmetric key-based cryptography, hash-based schemes, isogeny-based approaches, non-typical methods, as well as QKD. For each approach, we commence by providing a thorough background, highlighting the underlying mathematical hard problems and the security guarantees they offer. We further discuss their construction and the range of potential cryptographic schemes, encompassing not only those proposed by NIST but also considering additional candidates from other sources. Through a detailed analysis, we evaluate these schemes based on their security levels, performance costs, implementation, and vulnerability to potential attacks.
\vspace{-3mm}
\subsection{Lattice-based Cryptography}
\label{subsection3.1}
\vspace{-2mm}
Lattice-based ($\lb$) cryptography stands as the most extensively studied approach in the realm of PQC with a dominant position considering that two of the PQC standards are selected from this approach.  Its inception can be traced back to the groundbreaking work of Ajtai in 1996 \cite{ajtai1996generating}, which revolutionized the whole direction of PQC. Since then, $\lb$ cryptography has evolved to offer provable security, implementation advantages, and a wide range of practical applications.  
Defining a lattice as the integer combination of a set of $n$ independent vectors in an $n$-dimensional space, it becomes possible to identify a set of basis vectors that encompass all lattice points within the $n$-dimensional space. This structural definition gives rise to two NP-complete problems associated with lattices: the Shortest Vector Problem (SVP) and the Closest Vector Problem (CVP). SVP involves finding the shortest non-zero vector given a basis, while CVP focuses on determining the closest vector to a specified target point within a given lattice. Since solving these problems exactly is computationally infeasible, researchers have turned to their approximate versions, which are known to be NP-hard \cite{micciancio2002complexity}.  
Generally, the foundation of security in $\lb$ cryptography primarily arises from the conjectures that, to date, no classical or quantum algorithms exist that can efficiently solve lattice problems. More strongly, it has been mathematically demonstrated that the exceptional security offered by lattice cryptosystems derives from their worst-case hardness. This means that, unlike other PQC approaches that rely on average-case complexity, the difficulty of breaking a $\lb$ cryptosystem on average is mathematically equivalent to the challenge of finding a solution to hard lattice problems like SVP, CVP, and LWE (Learning With Error) within polynomial factors \cite{peikert2016decade}.


\vspace{-3mm}
\subsubsection{\textbf{Construction:}} 

In essence, $\lb$ constructions rely on the specific mathematical hard problems and characteristics of the lattices they employ. A wide array of challenging problems have been formulated based on these lattices.

The initiation of serious involvement of $\lb$ structures in cryptography began with the proposal of the Shortest Integer Solution (SIS) by Ajtai. SIS plays a significant role in various cryptographic primitives and involves the task of finding a short integer solution to randomly generated vectors in a large algebraic group. This problem originates from fundamental hard problems such as approximate-SVP, SIVP, and LWE, which are introduced with a security reduction to these problems. In formal terms, given an integer vector $s$ belonging to $\mathbb{Z}^n$ as the secret, a specific error distribution used to sample the small error $e$, and the construction of $(a, b = <s,a>+e$ mod $q)$, where $q$ is a prime, $a$ is randomly chosen from the set $\mathbb{Z}_q^n$, and $<s,a>$ denotes the inner product, the search for LWE involves finding $s$ when given $a, b$, and the public parameters \cite{peikert2016decade}.   
Additionally, due to the inefficiency of the LWE cryptosystems, which require large matrices as keys and involve computationally intensive multiplications, alternative variants such as Generalized-LWE, Ring-LWE, Module-LWE, Compact-LWE, etc., have been proposed. These variants aim to strike a balance between security and performance by utilizing specific lattices, such as structured or ideal lattices, and incorporating structures like rings and modules.  Among the various LWE variants, the Learning With Rounding (LWR) problem introduces rounding techniques and eliminates randomized error sampling, resulting in a more compact representation. Similar to LWE, the Nth-order Truncated Polynomial Ring Unit (NTRU) is based on an assumption with highly algebraic structured lattices \cite{hoffstein1998ntru}. It introduces the NTRU one-way function. While NTRU lacks provable security, it assumes a hardness level equivalent to the SVP over general lattices.

Delving into signature constructions, essentially, $\lb$ signatures are primarily formed utilizing hash-and-sign (H\&S) or Fiat-Shamir (FS) techniques. In the H\&S approach, the scheme relies on finding a solution to the approximate CVP and is further strengthened by incorporating preimage sampling. Interestingly, this H\&S scheme can be transformed into an identity-based encryption scheme. Conversely, FS algorithm constructs $\lb$ signatures by addressing challenging problems such as SIS, LWE, and NTRU. Technically, it functions as a non-interactive zero-knowledge proof, featuring a short secret key and a one-way function serving as the public key. The FS algorithm leverages rejection sampling with abort and a diverse distribution, leading to improved implementation and performance \cite{di2021survey}.

\vspace{-3mm}
\subsubsection{\textbf{PKE/KEM Schemes:}}
Given the challenges associated with directly designing a cryptosystem that achieves CCA security, one alternative approach is to employ the conversion algorithm developed by Fujisaki and Okamoto (FO).  As an example, \textit{CRYSTALS-KYBER} \cite{bos2018crystals} is the only $\lb$ standard constructed based on Module-LWE and achieves a CCA-secure KEM using the FO-type paradigm.  By considering the reduction from worst-case Module-SIVP to average-case Module-LWE, the Module-LWE problem excels in $\lb$ cryptography by effectively achieving high performance without significant compromises to security. In particular variants, \textit{CRYSTALS-KYBER} exhibits strong security against the QROM and, in general, provides superior performance in both hardware and software \cite{Aghapour12023}, making it well-suited for a wide range of KEM applications.  
Similarly, \textit{SABER} \cite{d2018saber} presented an IND-CCA2 secure KEM based on the module learning with rounding (MLWR) problem, utilizing power-of-2 moduli to achieve improved implementation efficiency. Despite providing strong security in ROM and demonstrating efficiency, NIST selected \textit{CRYSTALS-KYBER} over \textit{SABER} for standardization. This decision was primarily influenced by the higher level of rational and security guarantees offered by the Module-LWE problem, which has been extensively studied, in comparison to the Module-LWR problem.

Subsequent to \textit{SABER} and \textit{CRYSTALS-KYBER}, in order to address the expenses associated with safeguarding against SCAs during the implementation of CCA-secure PKE schemes, \textit{POLKA} \cite{hoffmann2023polka} introduces specialized countermeasures. These countermeasures aim to minimize costs and ensure resilience against information leakage and incorporate additional properties and efficiency enhancements while simultaneously maintaining a high level of security. 

The recently standardized \textit{CRYSTAL-Dilithium} \cite{ducas2018crystals} combines Module-LWE for security, the FS with abort transform for structure, and uniform distribution for error sampling. This amalgamation results in robust unforgeability within QROM. Notably, the standardization of \textit{CRYSTAL-Dilithium} is primarily attributed to its straightforward implementation and performance optimizations \cite{Aghapour22023}.  
Originally proposed in 1996 and subjected to various cryptographic analyses and multiple variations, \textit{NTRU} \cite{hoffstein1998ntru} and all its alternative schemes have not been selected by the NIST for further development.  
Nevertheless, numerous schemes have endeavored to enhance \textit{NTRU}'s design with the goal of attaining a security level comparable to LWE-based schemes or employing innovative techniques to thwart cryptanalysis.   
For example, the \textit{NTTRU} \cite{lyubashevsky2019nttru} scheme utilizes the Number-Theoretic Transform (NTT) on the ring, while \textit{BAT} \cite{fouque2022bat} is an NTRU-type KEM employing an improved decapsulation and decryption algorithms, achieving IND-CCA security. \textit{NTRU+} \cite{kim2022ntru+} seeks to overcome known limitations by introducing enhanced transformation and encoding methods within a KEM framework. Also, a recent development aims to combine the structurelessness of LWE and the compactness of \textit{NTRU} by introducing the NTWE problem. In this context, \textit{CNTR} and \textit{CTRU} were proposed as NTRU-type KEMs with lattice codes, showing higher levels of performance compared to their counterparts and NIST finalists \cite{gartner2023ntwe}.

\vspace{-3.5mm}
\subsubsection{\textbf{Signature Schemes:}} 
Despite the intention behind NIST's additional digital signature competition to choose non-lattice schemes and increase diversity, several $\lb$ signatures have been chosen due to their superior performance.  
For instance, \textit{HAETAE} \cite{cheon2023haetae}, a signature submitted to the Korean and NIST competition, is an improved successor to \textit{CRYSTALS-Dilithium} with major differentiating factors such as different sampling and distribution. \textit{HAETAE} is designed using module lattices and aims to enhance efficiency for applications in TCP/UDP protocols or IoT devices.  
\textit{EHTv3} and \textit{EHTv4} \cite{semaevdigital}, chosen from the \textit{EHT} cryptosystem family for the additional signature competition, offer the advantage of generating private keys from exceptionally small seeds, resulting in compact signature sizes and potential compatibility with 8-bit IoT platforms. However, the notable drawback remains the substantial size of their public keys.

Constructed upon NTRU lattices and the SIS problem, the Fast Fourier Lattice-based Compact Signatures over NTRU (\textit{FALCON}) scheme derives its name from these foundations \cite{fouque2019fast}. \textit{FALCON} employs a H\&S paradigm and introduces a unique tree structure called FALCON-Tree. While the implementation of \textit{FALCON} poses certain challenges, it excels in achieving the smallest bandwidth among comparable schemes, making it particularly suitable for constrained devices. Due to these factors, coupled with its high-security assurances, \textit{FALCON} was selected by NIST as a finalist.  
The \textit{ModFALCON} \cite{chuengsatiansup2020modfalcon} signature is a proposed extension of the FALCON-type scheme, designed based on module lattices. It aims to achieve intermediate security levels while maintaining compactness and efficiency. As an alternative to \textit{FALCON}, which has undergone cryptanalysis, \textit{MITAKA} \cite{espitau2022mitaka} and its successor \textit{SOLMAE} were introduced. These schemes rely on NTRU lattices and offer simpler, parallelizable, and side-channel resistant designs, providing a significant advantage in terms of implementation. In addition to the FALCON-type signature and \textit{NTRU-sign} scheme, \textit{HAWK} \cite{HAWK} presents a H\&S signature based on the Lattice Isomorphism problem (LIP). It outperforms \textit{FALCON} in signing speed and eliminates rejection sampling, thanks to its compact pre-computed distribution. Additionally, \textit{HAWK} shares similarities with \textit{NTRUSign} but incorporates enhancements and modifications in discrete Gaussian sampling. 

The \textit{HuFu} \cite{HuFu} H\&S signature, technically, is built upon the renowned GPV framework, relying on the hardness of unstructured random lattices, specifically the LWE and SIS problems. Its design significantly differs from that of \textit{CRYSTALS-Dilithium} and \textit{FALCON}. Notably, it incorporates an online/offline structure, enhancing overall performance and enabling advanced applications such as attribute-based encryption, group signatures, and blind signatures.   
\textit{EagleSign} \cite{hounkpevieaglesign}, an innovative Elgamal-like signature built upon structured lattices, offers faster operations and smaller signatures compared to standard schemes. Nevertheless, it has implementation gaps at certain security levels and notable security concerns requiring further scrutiny for enhanced security.  
While \textit{SQUIRRELS} \cite{SQUIRRELS} offers signatures within the size range of $\lb$ standards, it relies on co-cyclic lattices, which permit a swift and straightforward verification process. However, due to the use of unstructured lattices, \textit{SQUIRRELS} exhibits drawbacks such as slow key generation, a sizable public key, limited compactness, and the inclusion of floating-point arithmetic, rendering it unsuitable for certain  environments.

\textit{Raccoon} \cite{Raccoon} introduced a signature scheme akin to \textit{CRYSTALS-Dilithium}, rooted in module-based lattice problems, primarily emphasizing resistance to SCAs through the incorporation of masking techniques as safeguards against SCA and memory overhead reduction methods. While \textit{Raccoon} exhibits suitability for thresholding and holds potential in NIST's Threshold Cryptography (MPTC) due to its structural simplicity and absence of rejection sampling, its prospects in the additional signature competition remain limited. This limitation arises from its resemblance to established lattice standards and its comparatively larger signature size when compared to those standards.

\vspace{-3mm}
\subsubsection{\textbf{Implementation:}} 
\begin{figure*}
	\includegraphics[width=.48\textwidth,height=7cm,keepaspectratio]{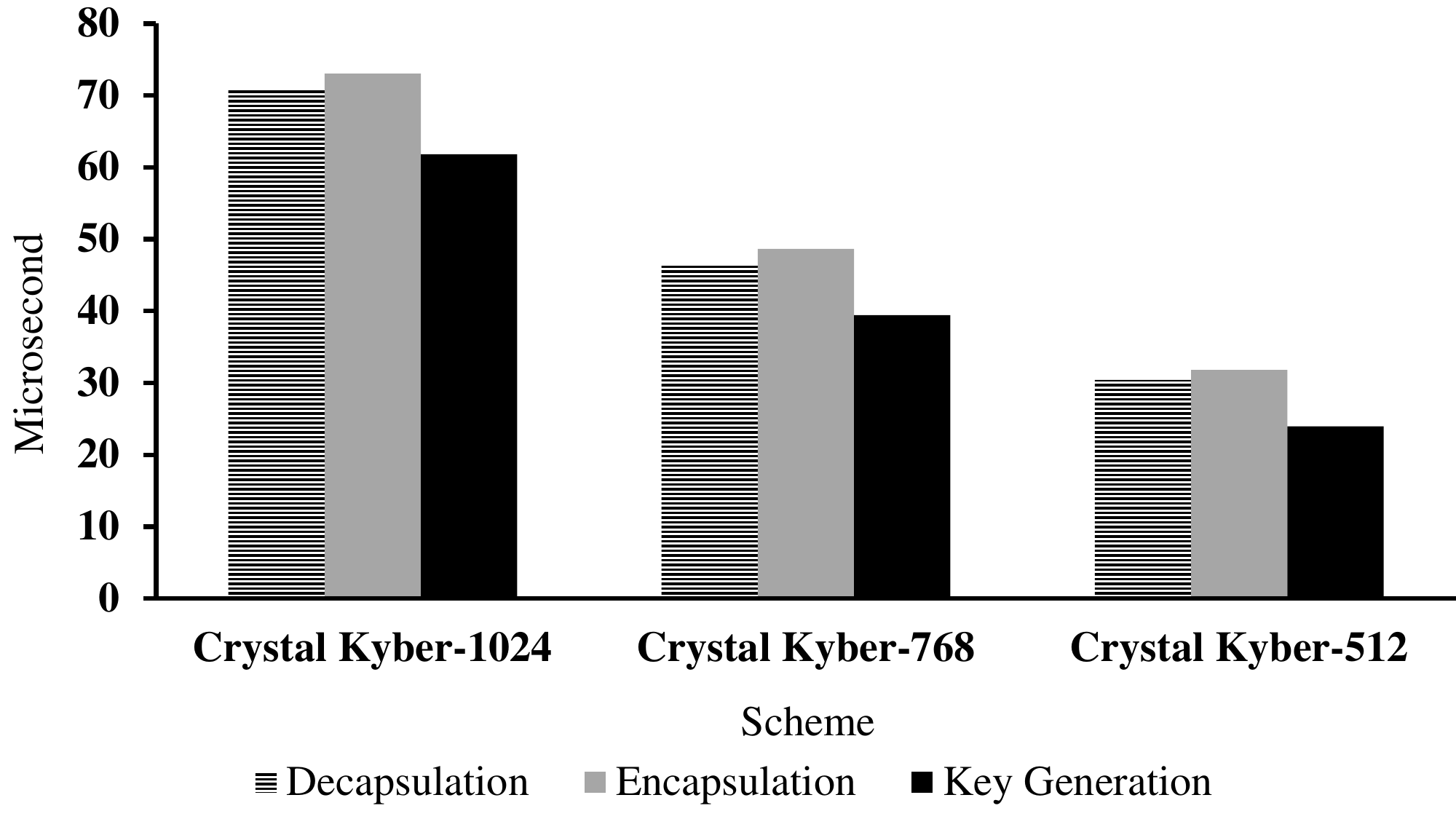}\hfill
	\includegraphics[width=.48\textwidth,height=7cm,keepaspectratio]{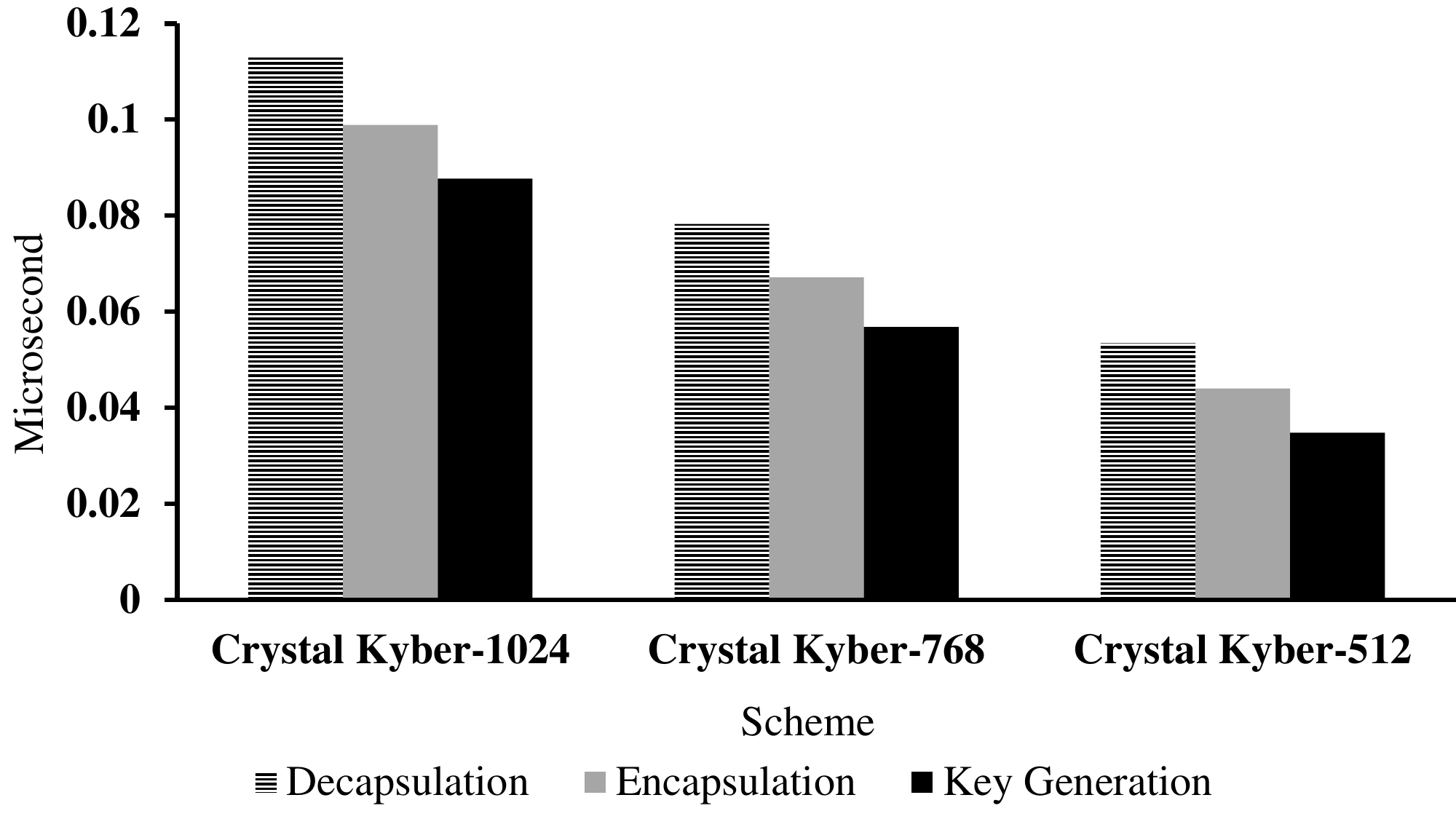}
	\vspace{-4mm}
	\caption{The time required for encapsulation, decapsulation, and key generation for Lattice based schemes on NIST's $4^{th}$round KEMs. The figure on the left corresponds to ARM platforms, while the figure on the right pertains to Intel platforms.}
	\label{fig2}
	\vspace{-2mm}
\end{figure*}

\begin{figure*}
	\includegraphics[width=.48\textwidth,height=7cm,keepaspectratio]{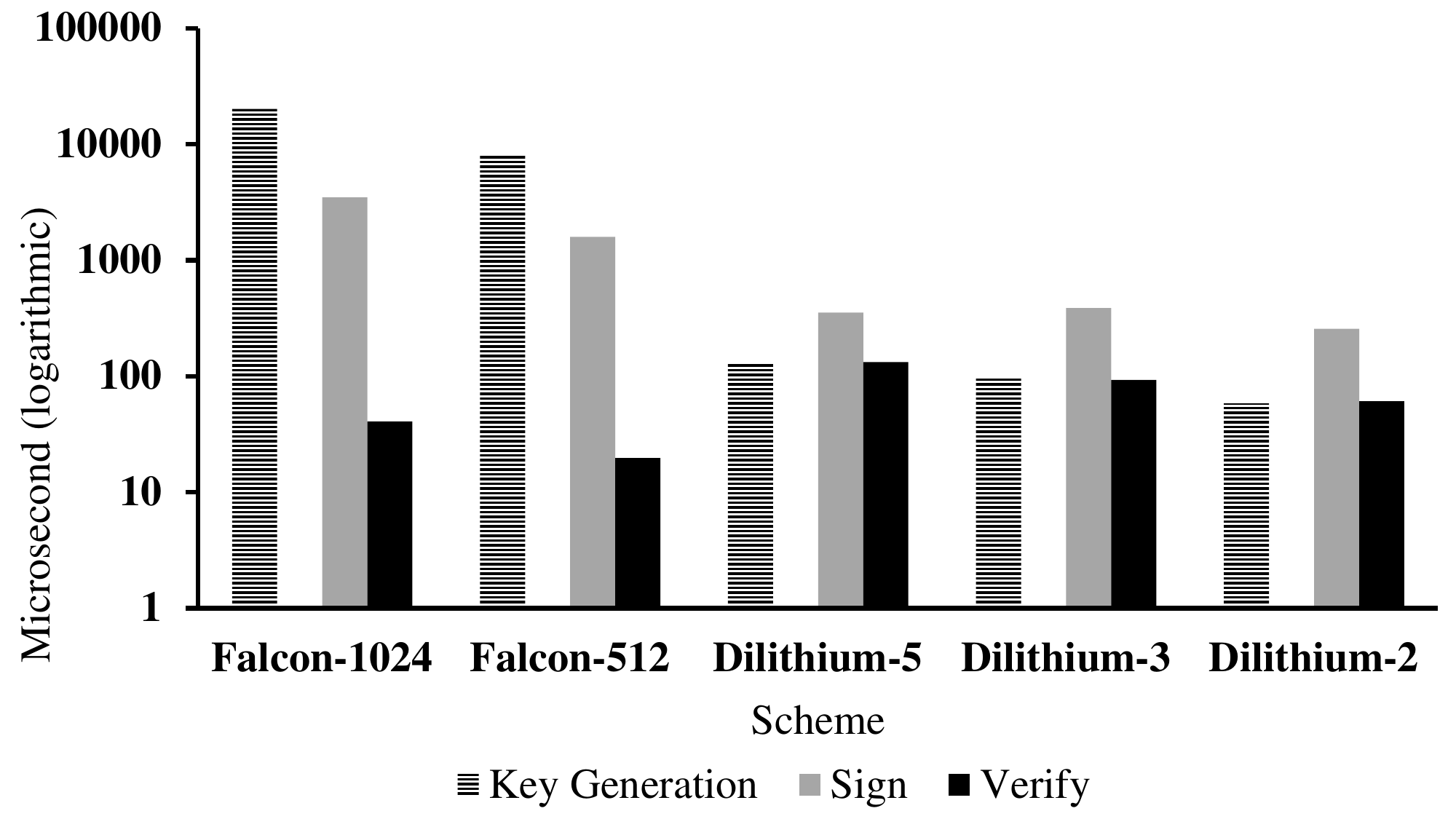}\hfill
	\includegraphics[width=.48\textwidth,height=7cm,keepaspectratio]{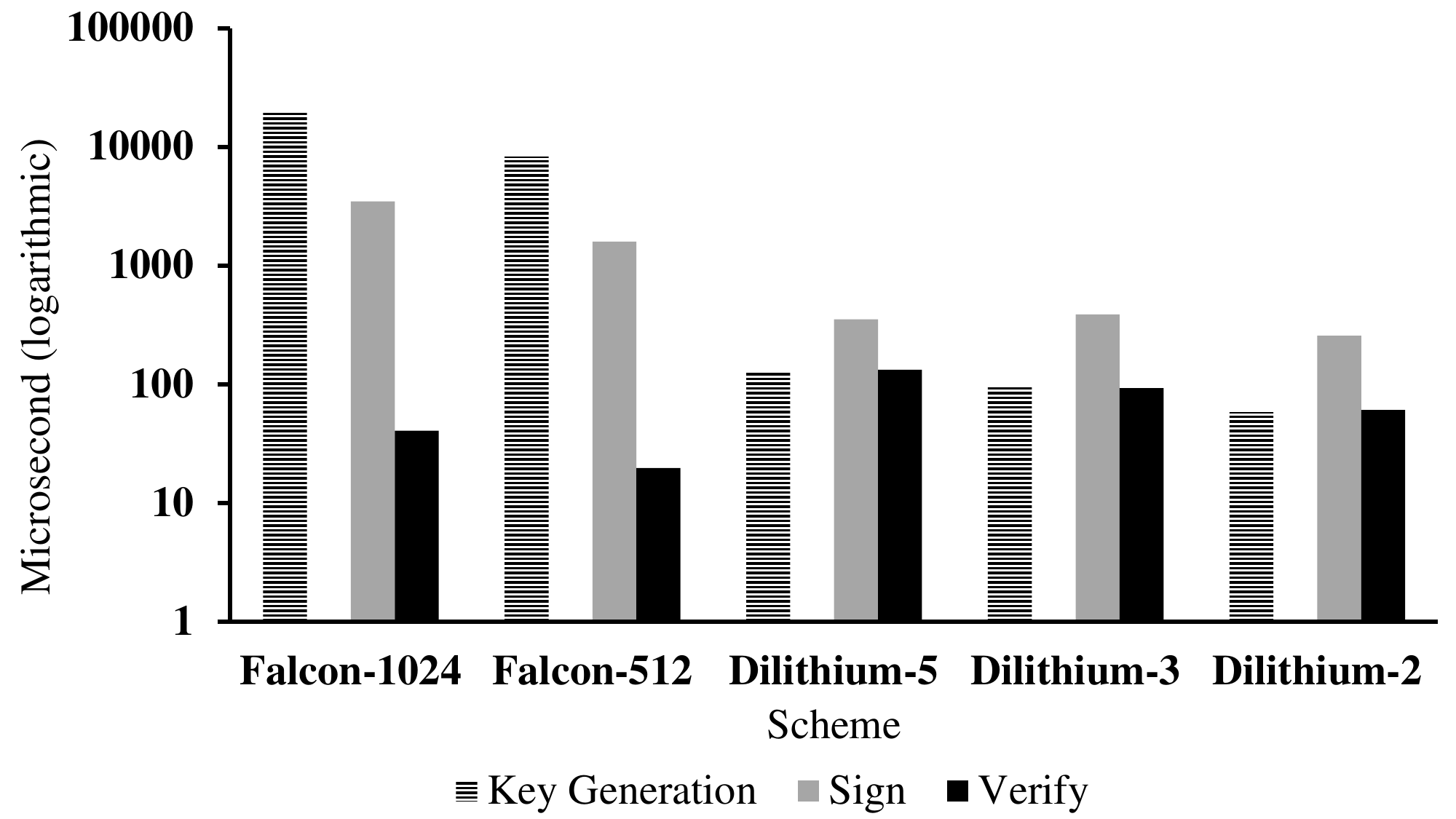}
	\vspace{-4mm}
	\caption{The time required for signing, verifying, and key generation for Lattice based schemes on NIST's $4^{th}$round signatures. The figure on the left corresponds to ARM platforms, while the figure on the right pertains to Intel platforms.}
	\label{fig5}
	\vspace{-5.5mm}
\end{figure*}

$\lb$ cryptosystems offer robust security proofs and are known for their simplicity and speed of implementation. To provide further insights, Figure \ref{fig2} and Figure \ref{fig5} illustrate the timing aspects of $\lb$ KEM and $\lb$ signatures, respectively. 
In the implementation evaluation of components within a $\lb$ cryptosystem, two critical aspects stand out. 
Firstly, the polynomial multiplication for ideal lattices and matrix multiplication for standard lattices are identified as the main bottlenecks that impact the speed of the system. Efficient algorithms and optimizations targeting these operations are crucial for achieving improved performance in $\lb$ implementations.   
Secondly, discrete Gaussian sampling plays a vital role in $\lb$ efficiency as it is utilized to generate noise that hides the secret information within the system \cite{Ahmadi12023, Ahmadi22023}.  
The effectiveness of a discrete Gaussian sampler is defined by a tuple of three parameters shaping its characteristics: $\sigma$, $\lambda$, and $\tau$.  
1) The standard deviation (\textbf{$\sigma$}) controls the dispersion of data points from the mean where a higher value results in a wider distribution with more spread-out data points.  
2) The precision parameter (\textbf{$\lambda$}) governs the statistical difference	between an ideal and implemented discrete Gaussian sampler. It ensures that the implemented sampler approximates the ideal behavior with a desired level of accuracy.  
3) The distribution tail-cut (\textbf{$\tau$}) determines the portion of the distribution that is considered as insignificant. By specifying a $\tau$ value, the sampler excludes data points beyond a certain	threshold, reducing computational complexity and enabling more efficient sampling.  
Optimizing these parameters allows for fine-tuning the discrete Gaussian sampler to meet specific requirements in terms of statistical characteristics, computational efficiency, and security considerations within the context of the overall $\lb$ cryptosystem implementation. Different Gaussian samplers are described in Table \ref{Tab4} where $\boldsymbol{\checkmark}$ and $\boldsymbol{\times}$ means being convenient or not.  
\vspace{+2mm}

\begin{minipage}[c]{\textwidth}
	\captionof{table}{Different Gaussian Samplers Referenced From \cite{nejatollahi2019post}}
	\centering
	\vspace{-4mm}
\begin{tabular}{|c|c|c|}
	\hline 
	\textbf{Sampler} & \textbf{Speed} & \textbf{Features}\tabularnewline
	\hline 
	Rejection \cite{gottert2012design}& slow & $\boldsymbol{\checkmark}$ constrained devices \tabularnewline
	\hline 
	Ziggurat \cite{marsaglia2000ziggurat}& flexible & $\boldsymbol{\checkmark}$ software implementation \tabularnewline
	\hline 
	CDT \cite{lindner2011better} & fast &  $\boldsymbol{\checkmark}$ digital signature \tabularnewline
	\hline 
	Knuth-Yao \cite{knuth1976complexity}& fastest & $\boldsymbol{\times}$ digital signature  \tabularnewline
	\hline 
	Bernoulli \cite{ducas2013lattice}& fast & $\boldsymbol{\checkmark}$ any scheme  \tabularnewline
	\hline 
	Binomial \cite{alkim2016post}& fast & $\boldsymbol{\times}$ digital signature \tabularnewline
	\hline 
\end{tabular}
	\label{Tab4}
\end{minipage}

\vspace{+2mm}


Recently, in \cite{li2023high}, a new architecture for \textit{CRYSTALS-Dilithium} is presented, which exploits parallel execution besides utilizing a three-stage pipelined modular multiplier. Overall, their design enhances related works by a factor of $3.01x$ in terms of operation frequency and $3.26x$ in terms of computational cost, respectively.   
By taking advantages of various optimization techniques, including inter-module and intra-module pipelining, the study described in \cite{ni2023hpka} introduced a new architecture for \textit{CRYSTALS-KYBER} that maximizes parallelization. After mounting their design on Artix-7 and Zynq UltraScale+ devices and comparing it with state of the art approaches, they demonstrated that their work yields greater speed improvements and area-time efficiency, all while reducing DSP consumption.  
The work in \cite{ji2023hi}, presented a GPU-based implementation of \textit{CRYSTALS-KYBER}, significantly enhancing the scheme's efficiency. Moreover, by optimizing the core operations of \textit{CRYSTALS-KYBER} such as NTT and iNTT, their work surpassed the state-of-the-art GPU implementations, achieving a performance improvement of $1.78x$.    

\vspace{-3mm}
\subsubsection{\textbf{Side-Channel Analysis:}} 
Although $\lb$ cryptography promises high efficiency, it remains susceptible to SCAs, where significant efforts have been dedicated 
to analyze the two standards and finalists chosen by NIST.


\textit{Side-Channel Analysis (KYBER):}
Dubrova et al. \cite{Kyber6} proposed a deep-learning message recovery attack that is applicable even on the masked implementation of the \textit{KYBER}. Their attack is capable of recovering a message bit on the ARM Cortex-M4 implementation of the scheme with probability of more than $99\%$.
In \cite{Kyber8}, the effect of a correlation electromagnetic analysis SCA on the FPGA implementation of the Kyber512 is investigated. By targeting the polynomial multiplication in the decapsulation algorithm, they were able to retrieve the secret key by utilizing 166,620 traces. 
In \cite{Kyber10}, by focusing on a $\lb$ PQC schemes, especially \textit{KYBER} and \textit{Dilithium}, a survey of side-channel and fault injection attacks is provided. Furthermore, they provided countermeasures to the mentioned attacks and implemented them on ARM Cortex-M4 and ARM Cortex-A53 platforms. 
Various works have proposed SCA countermeasures for \textit{CRYSTALS-KYBER}. By proposing an error detection algorithm for NTT operation, \cite{Kyber14} was able to detect both transient and permanent faults that could happen in \textit{KYBER}. Their algorithm provided high fault detection rate, while imposing low overhead to the scheme. Furthermore, the work in \cite{Kyber15}, suggested to use redundant number representation (RNR) instead of masking to prevent SCAs with a new fault detection technique based on the Chinese remainder theorem.   
Moreover, the work in \cite{yang2023stamp}, proposed a new single-trace key recovery attack, exploiting the leakages of KeyGen algorithm of \textit{KYBER} KEM scheme. The idea of their attack comes from the fact that compared to Ring-LWE problem, Module-LWE problem which is the basis of \textit{KYBER}, deploy more repeated computations on the secret key which could be then exploited to construct the attack.  
By focusing on the FPGA implementation of \textit{KYBER}, Moraitis et al. \cite{moraitis2023securing}, used duplication combined with clock randomization to secure the scheme against SCAs. Their work provided universal coverage, glitch immunity, while having zero clock cycle overhead.    

\textit{Side-Channel Analysis (Dilithium):}
By applying power analysis attacks, \cite{dilit1} were able to obtain a partial secret key which is then used to successfully forge a signature for arbitrary messages. \cite{dilit4} showed a security weakness to SCAs existing in fixed-weight polynomial sampling. This is a serious issue as fixed-weight polynomial sampling is used in different schemes such as \textit{Dilithium}, \textit{NTRU}, and \textit{NTRU Prime}. 
By exploiting leakages from bit unpacking procedure of \textit{Dilithium}, Marzougui et al. \cite{dilit6}, proposed an end-to-end full key recovery attack. To test their attack in practice, they mounted it on the ARM Cortex-M4 implementation of \textit{Dilithium} and successfully recovered the key by analyzing $756,589$ traces. In \cite{dilit7}, a differential fault attack is introduced that targets deterministic $\lb$ signatures like \textit{Dilithium}. They showed that a single random fault in signing algorithm is enough to cause key recovery through nonce-reuse. 

\textit{Side-Channel Analysis (FALCON):}
The authors in \cite{Falcon1} conducted the first electromagnetic SCA of \textit{FALCON} on the floating-point multiplications and achieved successful acquisition of the secret key employed for signing. 
Subsequently, \cite{Falcon3} 
introduced a new power analysis attack that significantly reduces the requirements and computational complexity of the attack. The first fault attack on the \textit{FALCON} was reported in \cite{Falcon4}. They employed a technique called basis extraction by aborting recursion or zeroing to carry out this attack. The results showed the susceptibility of \textit{FALCON} to fault attacks on its Gaussian sampler, ultimately leading to the disclosure of the private key. Moreover, \cite{Falcon5} proposed error detection schemes for \textit{SABER} and \textit{FALCON} that could detect transient and permanent errors with a rate of close to $100\%$, while imposing at most $22.59\%$, $19.77\%$, and $10.67\%$ overhead to area, delay, and power, respectively. The low overhead of their schemes, makes them suitable for IoT applications.

\vspace{-3mm}
\subsection{Hash-based Cryptography}
\label{subsection3.2}
\vspace{-2mm}
In the context of PQC, Hash-based ($\hb$) cryptography relies on symmetric primitives and exclusively produces signature constructions. 
NIST has not only recommended the $\hb$ approach with two Stateful hash-based signatures (\textit{XMSS} \cite{hulsing2013optimal}, \textit{LMS} \cite{mcgrew2019rfc}), but also the sole successful non-lattice-based standard to emerge from NIST's PQC competition is the $\hb$ signature, \textit{SPHINCS$^+$} \cite{bernstein2019sphincs+}. 
In essence, the combination of well-researched cryptographic hash functions and the properties of tree structures with either One-Time Signatures (OTS) or Few-Time Signatures (FTS) results in signatures that do not rely on number-theoretic assumptions, offering robust security and high performance.


\vspace{-3mm}
\subsubsection{\textbf{Construction:}}

The concept of hash-based signatures (HBSs) traces its origin back to Lamport's OTS (\textit{L-OTS}) and Merkle's Signature Scheme (\textit{MSS}) in 1979, making it the oldest approach among various signature methods. OTS allows the use of public-secret key pairs only once for signing a message. In its early stage, while HBSs having a solid construction, they were not widely employed in practical applications due to its storage requirements. Thereby, incorporating a tree structure (\textit{e.g.}, Merkle hash tree) in the construction was proposed as a solution to address the public key size issues. 	
Technically, the tree structure is responsible for managing multiple OTS keys, and we select a specific public key leaf, along with its corresponding index and authentication path, to sign a message \cite{srivastava2023overview}.

In this context, multiple OTSs have been proposed. For instance, Winternitz OTS (\textit{W-OTS}) emerged based on the concept of iterative function application and serving as the foundation for the NIST preferred signature schemes, aiming to achieve shorter signature and key sizes. 
Considering that \textit{WOTS+} serves as the foundation for standardized signatures such as \textit{SPHINCS+}, \textit{XMSS}, and \textit{LMS}, the work presented by Zhang et al. demonstrates that \textit{WOTS+} has reached an optimal version among all tree-based one-time signature (OTS) designs in terms of size \cite{srivastava2023overview}. 
Diverging from OTSs, FTSs  enable the signing and verification of multiple, albeit a limited number of plaintexts, using the same key pair. 
However, FTS offers lower security guarantees concerning the number of signatures. 
In 2001, Biba introduced the first FTS with fast verification time and a small signature size. 
Subsequently, HORS FTS was proposed as a generalization of \textit{L-OTS}, relying on the subset-resilient assumption \cite{bernstein2019sphincs+}. 

Nonetheless, the utilization limit of key pairs in OTS and FTS poses a significant concern for large-scale applications. As noted, a tree-based public key authentication approach is the optimal solution to address this issue. 
This solution encompasses three lines of work: a top-down authentication approach, a down-top authentication approach, and a hypertree structure \cite{sivasubramanian2020comparative}. 
In the down-top approach, the tree is constructed from the leaves to the root, with the root node serving as the master public key capable of authenticating all individual OTS public keys. 
On the other hand, the top-down approach involves creating all OTS keys and moving from the root to the leaves, where the parent authenticates the child nodes. 
Finally, the combination of these two approaches forms the hypertree structure, which forms the basis of the renowned \textit{SPHINCS} scheme \cite{bernstein2015sphincs} and its variants.

The introduction of \textit{SPHINCS} \cite{bernstein2015sphincs} in 2015 marked a significant milestone in the realm of stateless HBSs by offering unlimited signatures and employing a hypertree structure built with HORS FTS as a fundamental component. 	
Technically, besides the utilized component (OTS/FTS), based on the key state management methods employed, HBSs can be categorized into stateful and stateless. 
The constructions that we analyzed so far are considered as a part of stateful approach where the state of the public key must be reserved and also limited in the number of time a signature can be used.  
The stateless HBS scheme is a more recent alternative that does not maintain the signature state, eliminating the need for state management, reducing concerns related to key management, and keys are pseudo-randomly selected for each key generation process.  
Although this provides a robust solution, it results in considerably larger signatures and is less practical. 
Nonetheless, numerous research papers have focused on improving stateless HBSs for widespread adoption in various applications. 
Essentially, in large-scale scenarios and applications, there is a need to authenticate a significant number of messages using fewer key pairs.

\vspace{-3mm}
\subsubsection{\textbf{Signature Schemes:}}


Beginning with the NIST-recommended stateful hash-based signatures (HBSs), the \textit{LMS} (Leighton-Micali-Signature) scheme was developed following Merkle's signature with its own iterative function and enhanced security. 
Additionally, the \textit{XMSS} extends the \textit{MSS} with a multi-tree setup that manages \textit{W-OTS+} keys. 
$XMSS^{MT}$ (XMSS with Multi-Tree) is a stateful HBS scheme where each internal subtree represents an \textit{XMSS} tree. 
However, instead of using the Merkle tree structure, it incorporates improved variations in its algorithm. 
Ultimately, $XMSS^{MT}$ achieves an infinite number of signatures by employing a hypertree of \textit{XMSS} tree structures, utilizing a down-top approach. All variants of the \textit{XMSS} scheme can be enhanced to provide forward security \cite{hulsing2013optimal}.

\begin{figure*}
	\includegraphics[width=.48\textwidth,height=7cm,keepaspectratio]{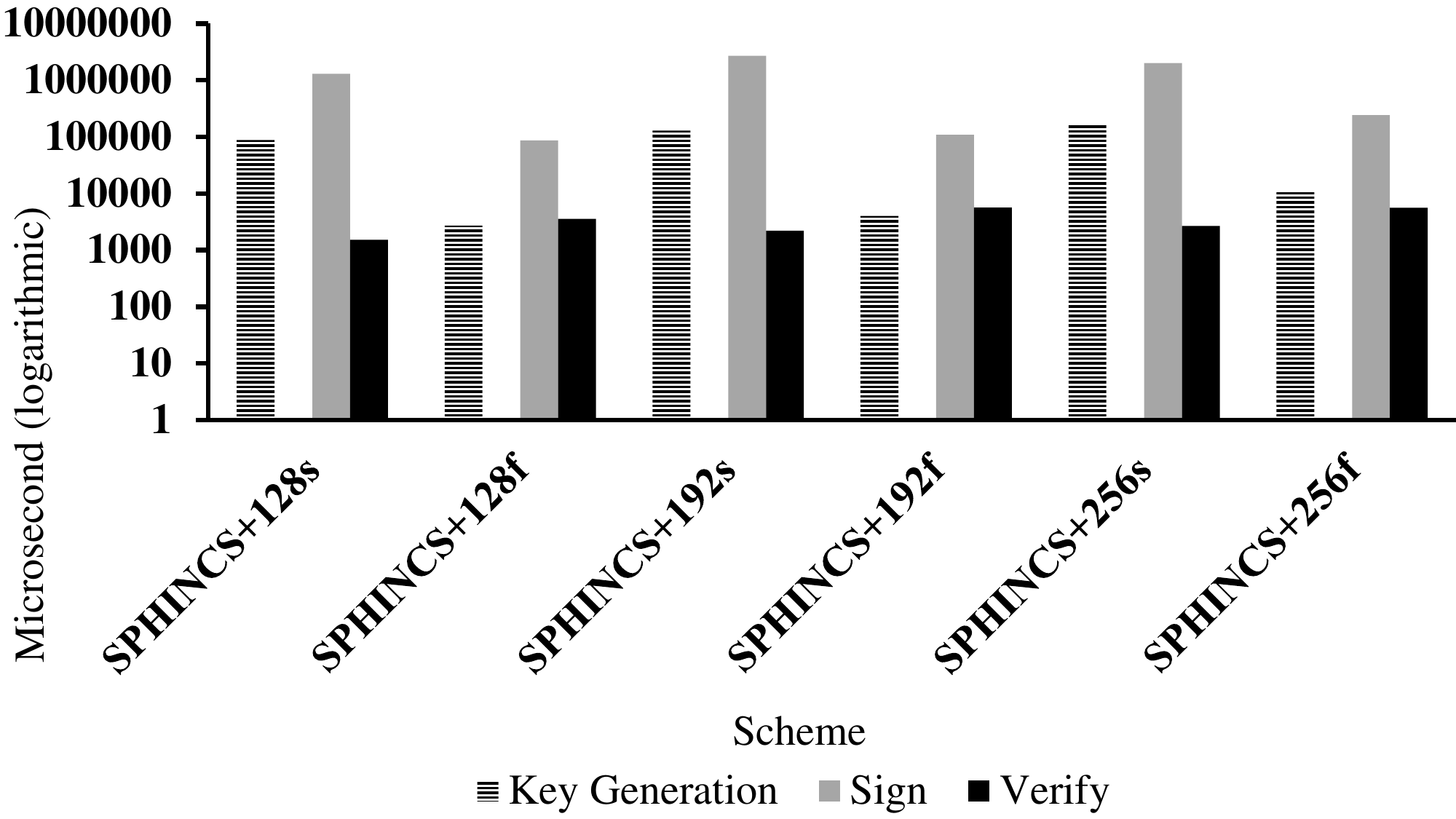}\hfill
	\includegraphics[width=.48\textwidth,height=7cm,keepaspectratio]{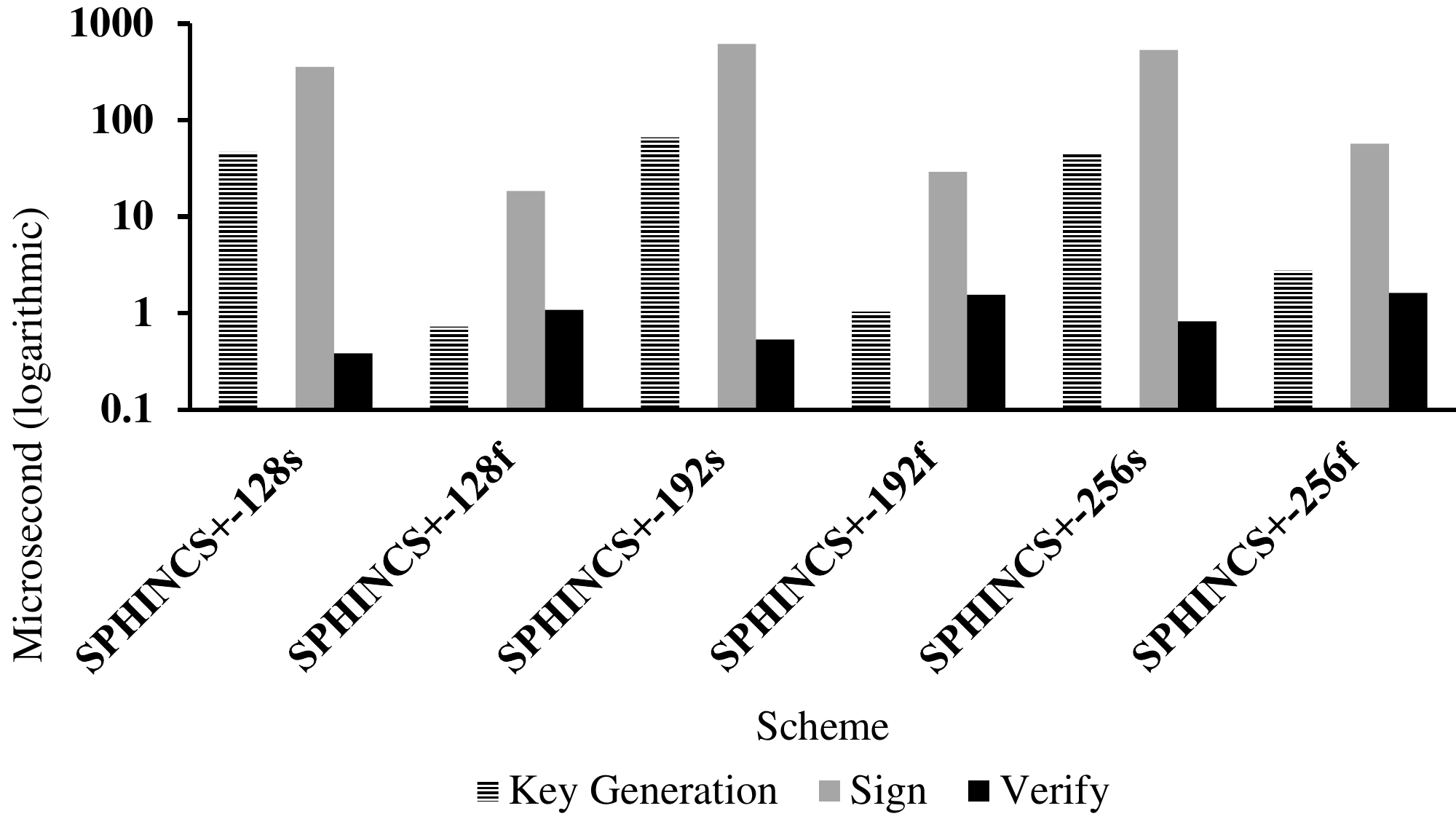}
	\vspace{-4mm}
	\caption{The time required for signing, verifying, and key generation for Hash based schemes on NIST's $4^{th}$round signatures. The figure on the left corresponds to ARM platforms, while the figure on the right pertains to Intel platforms.}
	\label{fig4}
	\vspace{-7mm}
\end{figure*}

Building upon \textit{SPHINCS}, several schemes such as 
\textit{SPHINCS-Simpira} and \textit{Gravity-SPHINCS} have been proposed with the objective of achieving higher efficiency. 
\textit{SPHINCS$^+$}, recognized as the NIST standard, represents a stateless HBS scheme constructed using a combination of various hash functions such as SHAKE-256, SHA-256, and Haraka, each with multiple parameter sets tailored to accommodate different tradeoffs. The FORS FTS is an integral component of the \textit{SPHINCS+} signature scheme. Despite the limitation on the number of derived signatures from a given public key, \textit{SPHINCS$^+$} achieves notably short public keys and demonstrates fast key generation and verification processes as shown in Figure \ref{fig4}. Its security assumptions are robust and dependent on the underlying hash function employed. While encountering some attacks and challenges in the category 5 security of their proposed parameter set, the researchers successfully addressed and resolved these issues.  
An alternative scheme utilizes the Streebog hash function, a standardized hash function from Russia, instead of SHA-256 in the \textit{SPHINCS$^+$} signature scheme to achieve improved performance. Building upon the same security model and structure as \textit{SPHINCS$^+$}, Hülsing et al. \cite{kudinov2022sphincs+} proposed a signature scheme called \textit{SPHINCS+C}, incorporating an enhanced version of FORS+C and WOTS+C. This scheme achieves greater efficiency and strikes a better balance between signature size and computation time.
In contrast, an alternative version of the \textit{SPHINCS$^+$} scheme, referred to as \textit{SPHINCS-$\alpha$}, has been introduced. It incorporates a new FORS FTS called FORC and provides various parameter sets optimized for either rapid computation or compact dimensions. By leveraging a security proof akin to \textit{SPHINCS$^+$}, \textit{SPHINCS-$\alpha$} achieves reductions in signature size, improved execution times, and enhanced computational stability. Nevertheless, it is constrained by slower speed, significantly longer verification times, and incompatibility with \textit{XMSS}, unlike \textit{SPHINCS$^+$} \cite{srivastava2023overview}. 
Additionally, the \textit{K-XMSS} and \textit{K-SPHINCS+} schemes have been proposed, leveraging the use of the Korean hash function and Korean block cipher-based hash functions, to achieve enhanced efficiency \cite{li2022hash}.

\textit{ASCON} comprises a family of cryptographic primitives, encompassing \textit{ASCON-hash}, \textit{ASCON-XOF}, \textit{ASCON-AEAD} (Authenticated Encryption with Associated Data), and \textit{ASCON-Sign}. Notably, \textit{ASCON-Sign} \cite{srivastava2023ascon} has been chosen as the designated scheme in the additional signature competition, and \textit{ASCON-AEAD} has also recently been selected as part of NIST's lightweight cryptography standardization process for constraint environments. \textit{ASCON-Sign} is built upon the foundation of \textit{SPHINCS$^+$} but distinguishes itself by replacing the inner hash function with \textit{ASCON} primitives. It offers two versions, one prioritizing robustness at the expense of slower performance and another optimized for speed. This makes \textit{ASCON-Sign} well-suited for immediate deployment in resource-constrained devices.

\vspace{-2mm}
\subsection{Code-based Cryptography}
\label{subsection3.3}
\vspace{-2mm}
Coding theory plays a pivotal role in enhancing the reliability of communication systems by incorporating additional data into the encoding process.   
Code-based ($\cb$) cryptography stems from error-correcting codes and derives its security from decoding random linear codes which are categorized in NP-Complete class of complexity.   
To design robust linear codes suitable for practical cryptographic purposes, various metrics such as Hamming, Rank, and Lee metrics are employed.  Essentially, these metrics gauge the similarity or dissimilarity between codewords, which is crucial for error correction and decryption. For example, the Hamming metric, a widely recognized metric in $\cb$ cryptography, quantifies dissimilarity between two linear codes by counting discrepancies in their strings. 



\vspace{-3mm}
\subsubsection{\textbf{Construction:}}
The origin of $\cb$ encryption construction can be traced back to 1978 when Robert McEliece introduced his cryptosystem \cite{mceliece1978public}. Essentially, $\cb$ encryption schemes take a linear code capable of correcting a certain number of errors as the private key, while the public key is derived from another version of the same code that appears (indistinguishably) random. To perform encryption, intentional errors are introduced into the code, and the decrypter can then perform a decoding procedure to recover the initial message. A wide range of codes (\textit{e.g.}, BCH, Reed-Solomon, Quasi-cyclic, Goppa, and MDPC) have been developed for use in encryption and signature construction.  

Decoding problems generally involve the task of finding a codeword or error with a specific or minimal Hamming weight. 
Specifically, given a binary matrix and a code word, the renowned computational Syndrome Decoding (SD) problem and the codeword finding NP-complete problems focus on locating a codeword with a predetermined Hamming weight. Commonly used decoding techniques in cryptography include List decoding, Minimum Distance Decoding (MDD), Maximum Likelihood Decoding (MLD), and Syndrome Decoding (SD).  

Generally, multiple renowned coding frameworks exist in most $cb$ schemes. In the original McEliece framework, the generator matrix (\textit{i.e.}, a full-rank matrix whose row forms the code) represents the secret Goppa code, whereas the Niederreiter framework, its dual variant, employs the parity-check matrix (\textit{i.e.}, a full-rank matrix whose row is orthogonal to the code) for GRS codes. Building upon these frameworks, Alekhnovich proposed another framework for random codes, which has security proof solely based on the difficulty of decoding problems. Alekhnovich's cryptosystem, with both a simple and a more generalized variant, paved the way for a new line of work based on Quasi-Cyclic (QC) schemes. The QC method offers an encryption scheme that exhibits no hidden code structure but includes errors that need to be removed before decoding-recovering the message \cite{weger2022survey}.



\vspace{-3mm}
\subsubsection{\textbf{NIST PKE/KEM Finalists:}} 
NIST has chosen all of the alternative PKE/KEM schemes from $\cb$ cryptography. 


\textit{Classic McEliece} \cite{bernstein2017classic} is not only a combination of the original \textit{McEliece} and the \textit{Niederreiter} framework but also includes the \textit{NTS-KEM} candidate. Leveraging the long-standing history and assumed one-wayness of the original \textit{McEliece}, \textit{Classic McEliece} provides strong IND-CCA2 in the QROM. Also, the utilization of the binary version of Goppa codes ensures perfect correctness. However, despite its compact ciphertext, fast encryption, and reasonably swift decryption, \textit{Classic McEliece} suffers from a significantly large key sizes (ranging from $100 kB$ to several megabytes) and inefficient key generation, limiting its applicability in certain scenarios especially for implementation on resource-constrained devices.  
To address this issue without compromising the security, various compression and decompression techniques can be explored. 
Thus, the choice of adopting this scheme is a trade-off between confidence in its security and the drawbacks related to its size and efficiency.

\textit{BIKE} \cite{aragon2017bike}, also known as Bit Flipping Key Encapsulation, is a cryptographic scheme formed upon using Moderate Density Parity-Check (MDPC) codes. It adopts the \textit{Neiderreitter} framework as the foundation for encryption, achieving IND-CPA by relying on two QC decisional variants of the Syndrome Decoding and Codeword Finding problems. Notably, \textit{BIKE} employs the well-established FO transform to construct a CCA KEM from a PKE scheme, ensuring specific correctness properties to maintain a low decryption failure rate. Despite offering competitive bandwidth and compact sizes, \textit{BIKE} is assumed to be suitable for a wide range of applications. It is worth mentioning that after the second round of NIST evaluations, \textit{BIKE} explicitly claimed to possess CCA security, positioning it as a potential candidate for the fourth round, where NIST aims to standardize one additional KEM.

The Hamming Quasi-Cyclic (\textit{HQC}) scheme \cite{melchor2018hamming} derives its name from its Quasi-cyclic construction, and it utilizes the FO transform to achieve an IND-CCA-secure KEM. \textit{HQC} is designed with the goal of providing higher security guarantees and stronger reductions to decoding problems. It employs a structure similar to the LWE-based cryptosystem, formed on top of MDPC codes, without incorporating any hidden trapdoors within the code. Consequently, \textit{HQC} exhibits a degree of independence from the decryption's decoding problem while maintaining a sufficiently low decryption failure rate. Furthermore, \textit{HQC} successfully mitigated a previous implementation's vulnerability to SCAs. The scheme shows compact sizes and achieves favorable software performance, making it suitable for various applications.

For comparing the performance and implementation of three alternative $\cb$ candidates against the $\lb$ standard, \textit{CRYSTALS-KYBER}, Table \ref{tab5:keysize} provides key sizes for different NIST security levels. Table \ref{tab6:performance} displays their performance across multiple platforms, including Intel, ARM, and FPGA. 

\vspace{-3mm}
\subsubsection{\textbf{Side-Channel Analysis:}} 

In the context of choosing alternative schemes in the $4^{th}$ round, practical implementation and performance become crucial factors, with the prerequisite of verified security. Thus, it is of paramount importance to analyze SCAs that pose a threat to their implementation and practical viability in real-world applications.

\textit{Side-Channel Analysis (Classic McEliece):} 
The work in \cite{Mc2}, presented a timing attack on the Patterson algorithm that is used for efficient decoding of Goppa codes. Since, \textit{McEliece} cryptosystem uses Goppa codes for encoding and decoding, they applied it to a FPGA implementation of the \textit{McEliece} cryptosystem directly, and showed serious vulnerabilities of the cryptosystem.  
Moreover, by applying power analysis attack on both FPGA and ARM implementation of \textit{McEliece}, Guo et al. \cite{Mc6}, presented a key recovery attack that could fully retrieve the key by gathering about $800$ traces and performing a machine-learning-based classification algorithm on them. Also, through combining machine learning with SCAs, Colombier et al. \cite{Mc7} proposed a new power analysis attack that exploits the leakage from matrix-vector multiplication operation which could eventually recover the plaintext efficiently.  
Among works dedicated to provide resiliency against SCAs, the work in \cite{Mc10} focused on the fault detection aspects of \textit{McEliece} and by targeting the finite field multiplier, the authors proposed natural and injected fault detection schemes that provide $100\%$ detection rate on the FPGA implementation of the cryptosystem. Moreover, Cintas-Canto et al. \cite{Mc11}, proposed efficient fault detection constructions that could be applied to CB cryptosystems. They applied their schemes on \textit{McEliece} cryptosystem and implemented it on the Kintex-7 FPGA device. Their implementation results indicate a $100\%$ error detection rate while adding $49\%$ overhead to area and delay, at the worst case scenario.


\begin{table*}
	\centering
	\caption{Comparing Key Size of NIST$4^{th}$ Round KEMs}
	\vspace{-3mm}
	\label{tab5:keysize}
	\begin{tabular}{@{}c|@{}c|@{}c|@{}c|@{}c|@{}c|@{}c}
		\hline 
		\textbf{Scheme} & \textbf{Type} & \textbf{Security} & \textbf{|Public Key|} & \textbf{|Private Key|} & \textbf{State} & \textbf{Ref.}\tabularnewline
		\hline 
		Crystal Kyber-512 & $\lb$ & 1 & 800 & 1,632 & Standard & \cite{Kyber_Official}\tabularnewline
		\hline 
		Crystal Kyber-512 90s & $\lb$ & 1 & 800 & 1,632 & Standard & \cite{Kyber_Official}\tabularnewline
		\hline 
		Crystal Kyber-768 & $\lb$ & 3 & 1,184 & 2,400 & Standard & \cite{Kyber_Official}\tabularnewline
		\hline 
		Crystal Kyber-768 90s & $\lb$ & 3 & 1,184 & 2,400 & Standard & \cite{Kyber_Official}\tabularnewline
		\hline 
		Crystal Kyber-1024 & $\lb$ & 5 & 1,568 & 3,168 & Standard & \cite{Kyber_Official}\tabularnewline
		\hline 
		Crystal Kyber-1024 90s & $\lb$ & 5 & 1,568 & 3,168 & Standard & \cite{Kyber_Official}\tabularnewline
		\hline 
		BIKE Level 1 & $\cb$ & 1 & 1,540 & 280 & Round 4  & \cite{Bike_Official}\tabularnewline
		\hline 
		BIKE Level 3 & $\cb$ & 3 & 3,082 & 418 & Round 4  & \cite{Bike_Official}\tabularnewline
		\hline 
		BIKE Level 5 & $\cb$ & 5 & 5,121 & 580 & Round 4  & \cite{Bike_Official}\tabularnewline
		\hline 
		McEliece 348864 & $\cb$ & 1 & 261,120 & 6,492 & Round 4  & \cite{McEliece_Official}\tabularnewline
		\hline 
		McEliece 6688128 & $\cb$ & 3 & 1,044,992 & 13,932 & Round 4  & \cite{McEliece_Official}\tabularnewline
		\hline 
		McEliece 8192128 & $\cb$ & 5 & 1,357,824 & 14,080 & Round 4  & \cite{McEliece_Official}\tabularnewline
		\hline 
		HQC Level 1 & $\cb$ & 1 & 6,170 & 252 & Round 4  & \cite{HQC_Official}\tabularnewline
		\hline 
		HQC Level 3 & $\cb$ & 3 & 10,918 & 404 & Round 4  & \cite{HQC_Official}\tabularnewline
		\hline 
		HQC Level 5 & $\cb$ & 5 & 15,898 & 532 & Round 4  & \cite{HQC_Official}\tabularnewline
		\hline 
	\end{tabular}
	\vspace{-2mm}
\end{table*}

\begin{table*}
	\centering
	\caption{Comparing The Performance of NIST$4^{th}$ Round KEMs on Intel, ARM,
		and FPGA Platforms}
	\label{tab6:performance}
	\vspace{-3mm}
	\begin{tabular}{@{}c|@{}c|@{}c|@{}c|@{}c|@{}c|@{}c|@{}c}
		\hline 
		\multirow{2}{*}{\textbf{Scheme}} & \textbf{Platform} & \textbf{Evaluation } & \multirow{2}{*}{\textbf{Frequency}} & \textbf{Key} & \multirow{2}{*}{\textbf{Encap.}} & \multirow{2}{*}{\textbf{Decap.}} & \multirow{2}{*}{\textbf{Ref.}}\tabularnewline
		& \textbf{Arch.} & \textbf{Hardware} &  & \textbf{Gen.} &  &  & \tabularnewline
		\hline 
		\multirow{3}{*}{Crystal Kyber-512} & Intel & Intel Core i7-4770K & 3.5 GHz & 122 & 154 & 187 & \cite{Kyber_Official}\tabularnewline
		\cline{2-8} \cline{3-8} \cline{4-8} \cline{5-8} \cline{6-8} \cline{7-8} \cline{8-8} 
		& ARM & Cortex-M4  & 24 MHz & 575 & 763 & 730 & \cite{botros2019memory}\tabularnewline
		\cline{2-8} \cline{3-8} \cline{4-8} \cline{5-8} \cline{6-8} \cline{7-8} \cline{8-8} 
		& FPGA & AMD Artix-7  & 161 MHz & 3.8 & 5.1 & 6.7 & \cite{xing2021compact}\tabularnewline
		\hline 
		\multirow{3}{*}{Crystal Kyber-768} & Intel & Intel Core i7-4770K & 3.5 GHz & 199 & 235 & 274 & \cite{Kyber_Official}\tabularnewline
		\cline{2-8} \cline{3-8} \cline{4-8} \cline{5-8} \cline{6-8} \cline{7-8} \cline{8-8} 
		& ARM & Cortex-M4 & 24 MHz & 946 & 1,167 & 1,117 & \cite{botros2019memory}\tabularnewline
		\cline{2-8} \cline{3-8} \cline{4-8} \cline{5-8} \cline{6-8} \cline{7-8} \cline{8-8} 
		& FPGA & AMD Artix-7  & 161 MHz & 6.3 & 7.9 & 10 & \cite{xing2021compact}\tabularnewline
		\hline 
		\multirow{3}{*}{Crystal Kyber-1024} & Intel & Intel Core i7-4770K & 3.5 GHz & 307 & 346 & 396 & \cite{Kyber_Official}\tabularnewline
		\cline{2-8} \cline{3-8} \cline{4-8} \cline{5-8} \cline{6-8} \cline{7-8} \cline{8-8} 
		& ARM & Cortex-M4 & 24 MHz & 1,483 & 1,753 & 1,698 & \cite{botros2019memory}\tabularnewline
		\cline{2-8} \cline{3-8} \cline{4-8} \cline{5-8} \cline{6-8} \cline{7-8} \cline{8-8} 
		& FPGA & AMD Artix-7  & 161 MHz & 9.4 & 11.3 & 14 & \cite{xing2021compact}\tabularnewline
		\hline 
		\multirow{3}{*}{BIKE Level 1} & Intel & Intel Xeon Platinum & 2.5 GHz & 589 & 97 & 1,135 & \cite{Bike_Official}\tabularnewline
		\cline{2-8} \cline{3-8} \cline{4-8} \cline{5-8} \cline{6-8} \cline{7-8} \cline{8-8} 
		& ARM & Cortex-M4 & 24 MHz & 24,935 & 3,253 & 49,911 & \cite{chen2021optimizing}\tabularnewline
		\cline{2-8} \cline{3-8} \cline{4-8} \cline{5-8} \cline{6-8} \cline{7-8} \cline{8-8} 
		& FPGA & AMD Artix-7  & 161 MHz & 187 & 28 & 421 & \cite{richter2021racing}\tabularnewline
		\hline 
		\multirow{3}{*}{BIKE Level 3} & Intel & Intel Xeon Platinum & 2.5 GHz & 1,823 & 223 & 3,887 & \cite{Bike_Official}\tabularnewline
		\cline{2-8} \cline{3-8} \cline{4-8} \cline{5-8} \cline{6-8} \cline{7-8} \cline{8-8} 
		& ARM & Cortex-M4 & 24 MHz & 59,820 & 8,376 & 139,234 & \cite{chen2021optimizing}\tabularnewline
		\cline{2-8} \cline{3-8} \cline{4-8} \cline{5-8} \cline{6-8} \cline{7-8} \cline{8-8} 
		& \multirow{1}{*}{FPGA} & AMD Artix-7  & 100 MHz & 693 & 80 & 1,198 & \cite{richter2021racing}\tabularnewline
		\hline 
		\multirow{2}{*}{McEliece 348864} & Intel & Intel Haswell CPU & 3.1 GHz & 56,705 & 36.4 & 127.14 & \cite{McEliece_Official}\tabularnewline
		\cline{2-8} \cline{3-8} \cline{4-8} \cline{5-8} \cline{6-8} \cline{7-8} \cline{8-8} 
		& ARM & Cortex-M4 & 24 MHz & 1,589,600 & 482.5 & 2,291 & \cite{chen2021classic}\tabularnewline
		\hline 
		\multirow{1}{*}{McEliece 6688128} & Intel & Intel Haswell CPU & 3.1 GHz & 443,746 & 171.4 & 306 & \cite{McEliece_Official}\tabularnewline
		\hline 
		\multirow{1}{*}{McEliece 8192128} & Intel & Intel Haswell CPU & 3.1 GHz & 486,195 & 156.9 & 310 & \cite{McEliece_Official}\tabularnewline
		\hline 
		HQC Level 1 & Intel & Intel Core i7-7820 & 3.6 GHz & 200.5 & 383.8 & 508.9 & \cite{HQC_Official}\tabularnewline
		\hline 
		HQC Level 3 & Intel & Intel Core i7-6700 & 3.6 GHz & 403.3 & 765.1 & 983.6 & \cite{HQC_Official}\tabularnewline
		\hline 
		HQC Level 5 & Intel & Intel Core i7-6700 & 3.6 GHz & 651.4 & 1,257.1 & 1,618.3 & \cite{HQC_Official}\tabularnewline
		\hline 
	\end{tabular}
	\vspace{-3mm}
\end{table*}

\textit{Side-Channel Analysis (BIKE):}
\textit{BIKE} has been subject of various analysis such as side-channel.  
By finding a flaw in the decryption algorithm, Guo et al. \cite{BIKE1}, proposed a key recovery attack that could retrieve the key in matter of several minutes. They took advantage of the dependency between the secret key and a failure procedure that could happen in decoding.  
Further, the work in \cite{BIKE3} found a leakage occurring in the FO transformation that is used in \textit{Kyber}, \textit{HQC} and \textit{BIKE} schemes. The leakage happens in the re-encryption step in the decapsulation algorithm because of the PRF execution. Furthermore, by analyzing the execution time differences of rejection sampling algorithm on altered and unaltered versions of the input message, Guo et al. \cite{BIKE4} proposed two new attacks that could successfully retrieve the secret key of \textit{BIKE} and \textit{HQC} schemes. The work in \cite{BIKE5}, proposed a novel attack which involved multiple trace and single track techniques. This attack proved successful in recovering secret indices, even when applied to ephemeral keys. By applying their attack on \textit{BIKE} and \textit{LEDAcrypt}, they pinpointed the vulnerabilities of these two schemes. However, these attacks can be prevented through randomization, hiding, and masking countermeasure techniques.

\textit{Side-Channel Analysis (HQC):}
Although, \textit{HQC} successfully mitigated many of the previous implementation's vulnerability to SCAs, it is still vulnerable to some attacks.  
For instance, recently Huang et al. \cite{HQC1}, proposed the first chosen-ciphertext cache-timing attack on the reference implementation of the \textit{HQC}. They exploited a vulnerability in the vector generation process of the reference implementation of \textit{HQC} and practically recovered the secret key. Moreover, the first power analysis SCA on \textit{HQC} is presented in \cite{HQC2}. Their attack requires chosen-ciphertext access model and based on the mounted decoding algorithm on the scheme, it makes the queries in a way to efficiently retrieve a large part of the secret key.  
Similarly, in \cite{HQC3} a chosen-ciphertext key recovery SCA is proposed that exploits the static secret key reuse feature of the micro controller. By taking advantage of the leakages of the diffusion property of the Reed-Muller decoding step, they were able to retrieve the whole secret key. To test their attack in practice, they mounted it on the Cortex-M4 microprocessor and obtained the whole secret key with 19,200 electromagnetic traces. 

\vspace{-4mm}
\subsubsection{\textbf{Signature Schemes:}}
When delving into $\cb$ signature schemes, it is notable that all of the initial proposals from NIST have been compromised, leaving the quest for a robust and practical signature scheme based on error-correcting codes unresolved. However, there are some promising $\cb$ signature schemes that have been selected for additional scrutiny in signature competitions.   
Of the two common signature constructions, the H\&S paradigm, which which was used in the initial $\cb$ signature scheme (\textit{CFS} \cite{courtois2001achieve}), introduces vulnerabilities. Moreover, adapting the FS transformation of a $\cb$ zero-knowledge identification scheme into a robust signature would lead to very large signature sizes. In this context, Lyubashevsky, drawing inspiration from his $\lb$ framework, designed a novel $\cb$ signature framework based on the Hamming metric. Nevertheless, this framework has encountered design issues as documented in \cite{baldi2021code}.

However, going to and beyond NIST signature competition, \textit{WAVE} \cite{WAVE} is constructed using the H\&S paradigm, generalized ternary codes, and rejection sampling techniques. It achieves provable EUF-CMA security in the ROM and boasts an efficient verification process, compact signatures, and resistance to statistical attacks. Nonetheless, potential concerns arise due to its large public keys and the need for careful implementation to accelerate its algorithms. Following the development of \textit{WAVE}, the \textit{WAVELET} scheme \cite{banegas2021wavelet} was introduced with the specific objective of being well-suited for constrained devices. While it inherits its security foundations from \textit{WAVE}, \textit{WAVELET} distinguishes itself by offering a simplified and faster verification algorithm. Additionally, \textit{WAVELET} achieves signature sizes that are shorter than those of nearly all other $\cb$ signatures, comparable to the signature sizes of PQC standards.

While the submission of \textit{pqsigRM} \cite{cho2022enhanced}, which was founded on RM codes, was initially presented but ultimately rejected in NIST's PQC competition, an enhanced iteration of it has been chosen for the additional signature competition. This enhanced version outperforms \textit{Dilithium} in terms of performance, yet it continues to contend with the challenge of maintaining large public keys. 
\textit{FuLeeca} \cite{FuLeeca} is structured around the Lee metric, utilizing a H\&S methodology with QC codes. One limitation of this scheme is its reliance on its relatively novel Lee metric, which falls between Hamming metric coding and Euclidean metric in the realm of lattices. However,  \textit{FuLeeca} boasts superior performance and smaller sizes compared to other standards, trailing only behind \textit{FALCON} in this regard. Its reliance on integer arithmetic and compact dimensions makes it highly suitable for straightforward implementation in both hardware and software.  

\textit{CROSS} \cite{CROSS} introduced a signature scheme that tackles a novel challenging problem called "restricted syndrome decoding", stemming from the NP-Complete SD problem. This scheme achieves compressed key sizes and constant-time execution without relying on any algebraic structure. While it is well-suited for IoT applications, notable drawbacks include larger signature sizes compared to other $\lb$ standards and a lack of extensive research on its underlying problem, potentially limiting its practicality.  
Formed on Linear Equivalence Problem (LEP), \textit{LESS} signature \cite{LESS} provides robust security. Notably, it is the sole $\cb$ scheme that does not primarily depend on coding hardness; instead, it employs a group action structure. Nevertheless, akin to counterparts in the $\cb$ category, it faces challenges related to large public key sizes and computational bottlenecks, particularly in scenarios where speed is a top priority. 
The Matrix Equivalence Digital Signature (\textit{MEDS}) \cite{meds} is built upon Rank metric codes and the Matrix Code Equivalence (MCE) problem. Nonetheless, it exhibits scalability limitations and faces trade-off issues between key size and signature sizes.

\vspace{-3mm}
\subsection{Multivariate Cryptography}
\label{subsection3.4}
\vspace{-2mm}
The rationale behind Multivariate Public Key Cryptography (MPKC) stems from the hardness of solving multivariate polynomials over a finite field which has been proven to have the complexity class of an NP-hard problem. Although Multivariate ($\mv$) cryptography inherently lacks formal security proofs, no classical or quantum algorithm has successfully solved this set of problems. 
While the majority of MPKC schemes are based on the Multivariate Quadratic (MQ) problem within the standard model, there are additional challenging problems such as variations of the Isomorphism Polynomials (IP) problem (including IP with one secret (IP1S), IP with two secrets (IP2S), and extended IP problems (EIP)). 
Although some of the $\mv$ problems have not been formally proven to be hard in complexity classes, they serve as the foundation for multiple signature and identification schemes.  
Among numerous schemes based on $\mv$ polynomials, we only present their most promising candidates.


\vspace{-3mm}
\subsubsection{\textbf{Construction:}}

Multiple approaches exist for constructing a PQ secure MPKC scheme, considering the hard mathematical problems associated with $\mv$ polynomials. Essentially, these constructions involve three mapping systems, denoted as $\mathcal{F}$, $\mathcal{S}$, and $\mathcal{T}$. The first mapping system, denoted as $\mathcal{F}$: $\mathbb{F}^n \rightarrow \mathbb{F}^m$, consists of $m$ multivariate quadratic polynomials with $n$ variables. This mapping system can be easily inverted. The last two mapping systems, $\mathcal{S}$ and $\mathcal{T}$ are either affine or linear invertible maps. The MQ problem states that finding $x_0 \in \mathbb{F}^n_q$ over a finite filed $\mathbb{F}_q$ where $\mathcal{F}$(\textbf{$x_0$}) = \textbf{$0$} is mathematically difficult. To construct a public key, the mapping construction $\mathcal{P}$ $=$ $\mathcal{S}$ $\circ$ $\mathcal{F}$ $\circ$ $\mathcal{T}$: $\mathbb{F}^n \rightarrow \mathbb{F}^m$ is utilized, where the two affine maps $\mathcal{S}$ and $\mathcal{T}$ conceal the underlying structure of the central map $\mathcal{F}$. Consequently, this construction treats $\mathcal{F}$, $\mathcal{S}$, and $\mathcal{T}$ as the private key because the resulting system can be considered mathematically akin to a random mapping system \cite{hashimoto2021recent}. 
Despite being based on an NP-hard problem, $mv$ cryptography exhibits certain weaknesses in its construction aiming at discovering the secret key or recovering the plaintext from a given ciphertext. 
A commonly used approach for attacking MPKC schemes is the MinRank problem, which falls under the NP-complete class and involves finding the minimum rank element in a given matrix space. In contrast to the powerful MinRank attack, which is a key-recovery protocol, differential cryptanalysis represents a message-recovery attack that leverages differentials of quadratic functions in MPKC \cite{dey2023progress}. 

Additionally, in MPKC construction, modifiers are modifications applied to the algebraic structure of multivariate maps. Their purpose is to enhance resilience against certain constructional cryptanalysis methods while improving efficiency. These modifiers encompass the addition of random equations, removal of public equations, alteration of input space, the addition of variables, inclusion of random summands, and incorporation of a non-linear component in the construction. They are referred to as Plus-modifier, Minus-modifier, projection (P-modifier), vinegar (v-modifier), internal perturbation (ip-modifier), Q-modifier, etc. \cite{smith2021new}. 
Thereby, each $\mv$ scheme has the potential to integrate various modifiers, contributing to the advancement of cryptosystems and the creation of more efficient structures.

\vspace{-3mm}
\subsubsection{\textbf{Schemes:}}

Generally, various trapdoors and algebraic fields are employed in the MPKC schemes, classified into distinct families: Big Field, Medium Field, Single Field, Mixed Field, Hidden Field Equation (HFE), Stepwise Triangular, and Oil-and-Vinegar. In this context, we exclusively introduce those that exhibit robust security potential and are under consideration for standardization. 
The \textit{C*} cryptosystem is the beginning of MPKC schemes proposed by Matsumoto and Imai (\textit{MI}) \cite{matsumoto1988public} in 1988 relying on the utilization of Big Field families with the central map defined over an extension map. Despite undergoing various cryptanalysis efforts, the \textit{MI} structure has been enhanced through the incorporation of different modifiers, leading to the development of numerous efficient cryptosystems that offer enhanced security.

The emergence  of $\mv$ signature schemes began in 1997 with the introduction of the Oil and Vinegar (\textit{OV}) scheme \cite{patarin1997oil}. This scheme was developed as a direct transformation of the attack on the \textit{MI} cryptosystem. Technically, $\mv$ polynomials over a small finite field play the role of oil and vinegar variables as the subset of them that metaphorically do not easily mix.  
Subsequently, various schemes were presented to in NIST's signature contest address constructional attacks, cryptanalysis, and efficiency concerns associated with the \textit{OV} scheme.
The Unbalanced OV (\textit{UOV}) is the scheme that was not chosen to be part of the standardization in the PQC competition, is enhanced, and managed to be selected in the additional digital signature competition. 
Subsequently, \textit{Rainbow} as the sole $\mv$ finalist in the third round of NIST's PQC competition, relies on the H\&S paradigm and incorporates multiple layers of \textit{UOV} construction along with the MQ and MinRank problems.  
However, subsequent cryptanalysis efforts, including a novel rectangular MinRank attack that outperformed previous attacks, revealed security vulnerabilities in \textit{Rainbow}. As a result, significant re-engineering is required to address these weaknesses. Other constructions have been developed based on \textit{Rainbow}'s structure, such as \textit{TriRainbow} \cite{gangulynew}, which combines \textit{Rainbow} with a triangular scheme. 
Building upon \textit{UOV} signature, there exist various signatures selected in the additional digital signature competition: 1) \textit{MAYO} \cite{MAYO} is a variant of \textit{OV} signature with new parameter choices and enhanced encoding. 2) \textit{PROV} \cite{PROV}, as the provable secure variant of \textit{UOV}. 3) \textit{QR-UOV} \cite{QR-UOV} and \textit{VOX} \cite{VOX} are a \textit{UOV} scheme with quotient ring structure to solve the large public key size problem of $\mv$ signatures. 4) \textit{SNOVA} \cite{SNOVA} is a simplified version of the \textit{UOV} signature with a non-commutative ring structure and key randomness alignment. 5) \textit{TUOV} \cite{TUOV} is a triangular variant of the \textit{UOV} scheme.

Following the linearization equation attack on the \textit{C*} cryptosystem, the concept of Hidden Field Equations (\textit{HFE}) was introduced as a countermeasure against this and other constructional attacks \cite{patarin1996hidden}. While the \textit{HFE} cryptosystem itself is not considered secure, numerous robust and efficient schemes have been developed based on this concept, incorporating various modifiers \cite{dey2023progress}.  
%
Great Multivariate Short Signature (\textit{GeMSS} \cite{casanova2017gemss}) employs the H\&S paradigm along with the \textit{HFE} construction featuring Vinegar variables and the minus modifier. However, \textit{GeMSS} suffered a severe attack that compromised key recovery, demonstrating that alternative versions or modifications would not efficiently address its inherent constructional flaw. Thus, \textit{GeMSS} was excluded by NIST during the $3^{rd}$ round. 
Additionally, \textit{HPPC}, a recently proposed HFE-type scheme, shares similarities with \textit{GeMSS} and was selected by NIST, offering defenses against known attacks targeting this class of schemes \cite{rodriguez2023hppc}. 

From a technical and preliminary perspective, one might assume that utilizing $\mv$ polynomials with degrees higher than two would pose an equally challenging or potentially even more difficult problem than solving their quadratic counterparts, given that both are classified as NP-complete. In this context, the Multivariate Cubic (MC) problem with a degree of three is employed to construct the \textit{OHV} and \textit{3WISE} signature schemes, adopting a similar construction approach as the \textit{Rainbow} signature scheme. The \textit{OHV} scheme utilizes the same central map as the original \textit{HFE} scheme, aiming to achieve enhanced security. However, a drawback of the \textit{OHV} scheme is the requirement for larger public keys, despite the intended security improvements \cite{kundu2020post}. The \textit{3WISE} signature scheme is based on cubic degree, which may offer superior security guarantees compared to MQ schemes, although it hasn't been thoroughly examined. However, it excels in terms of performance and storage efficiency, featuring faster operations.   
Additionally, the DME-Sign represents an improved iteration of the originally rejected DME proposal from the PQC competition featuring a compact signature size. Consequently, its security may necessitate further diverse cryptanalysis before advancing to the next stage, given that its large key sizes pose challenges for certain applications. 
As an innovative approach and a distinct variant of the MQ problem, \cite{yasuda2018multivariate} proposed a new hard problem known as the Constrained MQ problem. This problem serves as the foundation for the development of the pq-method and subsequent encryption schemes. Technically, the pq-method enables the conversion of an MPKC scheme to another secure MPKC scheme over a field with larger prime cardinality even if the initial scheme was not considered secure. In line with this, \textit{PERN}  \cite{yasuda2020multivariate} introduces a flexible encryption scheme accompanied by a novel decryption algorithm that involves solving equations in the domain of real numbers. 

\vspace{-4mm}
\subsection{Isogeny-based Cryptography}
\label{subsection3.6}
\vspace{-2mm}
Isogeny-Based ($\ib$) cryptography, although relatively new and less extensively studied compared to other approaches, holds potential for practical PQC solutions based on super-singular isogenies.
Essentially, $\ib$ cryptography derives its PQ security from utilizing the widely prevalent ECC in conjunction with the Diffie-Hellman theory. 
However, this approach inherently lacks certain cryptographic primitives and advance constructions, necessitates a deeper mathematical foundation for cryptanalysis, and requires further exploration in the field of PQC as it represents only one robust scheme within the global PQC standardizations.
The initial development of an $\ib$ scheme occurred in 1996, while a practical algorithm for this approach was later proposed by Jao and De Feo in 2011 \cite{jao2011towards}. In addition to the well-known SIDH isogeny problem, $\ib$ cryptography includes various challenging problems such as the general isogeny problem, maximal order representation problem, non-trivial endomorphism problem, and more \cite{de2017mathematics}.

\vspace{-4mm}
\subsubsection{\textbf{Construction:}} 
In the ECDH protocol, which employs elliptic curves, the entities engage in communication to derive a shared secret by utilizing their respective secret and public keys where its security is based on the hardness of Computational DH Problem (CDHP).
At a conceptual level, $\ib$ key exchange schemes can be regarded as PQ secure counterparts to the DH protocol.
Essentially, in the realm of $\ib$ cryptography, the isogenies of elliptic curves are rational functions that replace the role of points on curves, along with their relationships encompassing homomorphism, isomorphism, and endomorphism with the entire curve and the mappings between them \cite{de2017mathematics}.
Thereby, by defining an isogeny as a mapping or group homomorphism between two elliptic curves, we can formulate its mathematically hard problems: given two isogenous elliptic curves over a finite field, it is computationally difficult to find the isogeny connecting them.
Analogously, solving the $\ib$ problem involves finding solutions within isogeny graphs resembles solving DLP over other algebraic groups. 
Notably, quantum algorithms exist for solving $\ib$ problems over ordinary elliptic curves in exponential time. 
However, solving these problems over super-singular elliptic curves necessitates exponential time even for a quantum computer. 
Formally, consider two super-singular isogenies $E_1/\mathbb{F}_{q^2}$ and $E_2/\mathbb{F}_{q^2}$, a prime number $q$, and a fixed, smooth, and public degree isogeny $\Phi: E_1 \rightarrow E_2$. 
The Super Singular Isogeny (SSI) problem can be defined as follows \cite{galbraith2018computational}: Given points $P, Q \in E_1$ and points $\Phi(P_1), \Phi(P_2) \in E_2$, the task is to compute the isogeny, $\Phi$. The challenges in SIDH-based protocols revolve around the efficient computation of isogenies and the development of commute operations in the key agreement protocol.

\vspace{-4mm}
\subsubsection{\textbf{Schemes:}} 
Serving as the optimized instantiation of the SIDH protocol, integrated with a transformation protocol, \textit{SIKE} \cite{jao2017sike} functions as both a PKE and KEM scheme, standing out as an outlier in the NIST PQC competition.
While it demonstrated numerous advantages, including minimal communication overhead, compact key sizes, compatibility with embedded devices, exhibiting flawless correction with no decryption failure, \textit{SIKE} ultimately faced vulnerabilities and was officially deemed insecure during the 4th round of the competition.
Technically, it was revealed that the SIDH-based key exchange protocols, upon which \textit{SIKE} relied, was found to deviate from a pure isogeny problem and lacked the necessary cryptographic structure.
Consequently, enabling the development of a series of attacks targeting all parameter sets of SIDH \cite{castryck2023efficient}.
Among these attacks, a particular key recovery attack stood out. This attack leveraged the reducibility criterion and, running on a single core, successfully compromised the Microsoft \textit{SIKE} challenges in remarkably short times of 55 and 85 seconds. 
This attack significantly outperformed other attempts and redirected the focus of $\ib$ cryptography towards new challenges and research. While these attacks specifically affected schemes based on torsion points, such as \textit{SETA}, they did not impact other $\ib$ protocols like \textit{CSIDH} \cite{beullens2023proving} and \textit{SQISign} \cite{de2020sqisign}.
Technically, \textit{SQISign} employs the FS paradigm on a ZKP, with verification involving knowledge of Elliptic curve endomorphism. It stand as the sole $\ib$ scheme chosen in the additional signature contest, featuring signature and key sizes significantly smaller than other PQ standardized signatures. However, implementation challenges and a complex signing process result in slower performance compared to alternative schemes.

Efforts have been made to counter these attacks, however, these countermeasures have proven to be relatively ineffective, resulting in significantly slower protocol execution and increased communication overhead. 
\textit{FESTA} \cite{basso2023festa}, introduced as the successor to \textit{SETA}, emerged following the proposed attack on SIDH that rendered the \textit{SIKE} scheme ineligible for the competition. \textit{FESTA} aims to achieve CCA security within the QROM and has the potential to attain CCA security within the standard model. It is worth noting that \textit{FESTA} shares similarities with \textit{SIKE} in terms of key generation and encryption computations. While \textit{FESTA} presents a proof of concept, it has yet to be practically implemented, and its suitability for real-world applications needs further scrutiny.
In the ongoing research endeavors focus on developing countermeasures against SIDH attacks and introducing efficient and practical $\ib$ schemes, Basso et al. \cite{basso2023new} introduced the concept of artificially oriented curves and devised a key exchange protocol similar to \textit{SIKE}. They formulated new problems such as binary SIDH, ternary SIDH, and hybrid variants while analyzing their computational hardness and providing parameter sets to demonstrate algorithmic efficiency. Another avenue in $\ib$ cryptography involves leveraging isogenies to construct ZKPs of knowledge, which can be transformed into PQ secure signatures.

\vspace{-3mm}
\subsection{SKC-based PQ-Secure Schemes}
\label{subsection3.5}
\vspace{-2mm}
While Shor's algorithm has negligible impact on Symmetric Key-based ($\skb$) schemes and Grover's algorithm merely necessitates larger key sizes, there are novel PQ protocols that utilize symmetric key primitives such as block ciphers, stream ciphers, and hash functions. Apart from $\hb$ signatures, \textit{Picnic} \cite{chase2017post}  
was the only $\skb$ candidate in the NIST PQC competition. 
This approach traces its roots to Ishai et al. \cite{ishai2007zero}, who pioneered the development of a ZKP using an MPC protocol in a black-box manner, introducing the MPC-in-the-head (\textit{MPCitH}) paradigm.


\vspace{-3mm}
\subsubsection{\textbf{Construction:}}

The rationale behind \textit{MPCitH} is as follows. Let's assume $y = F(x)$ represents a one-way function, where $x$ is the secret key and $y$ is the public key. By employing a non-interactive ZKP of knowledge (NIZKPoK), we can construct a signature scheme where the proof of knowledge of the secret serves as the signature. In this paradigm, the security relies on two factors: the difficulty of the symmetric primitive one-way function $F(.)$ used for generating key pairs, and the security of the MPC protocol in the presence of a limited number of parties that may be honest but curious.  
For instance, \textit{Picnic} employed a modular approach where the LowMC block cipher is instantiated to generate private-public key pairs based on the plaintext and only the holder of the secret key could provide the proof with security relying on the hash function and the LowMC cipher in ROM.

Essentially, there are two lines of work regarding the symmetric primitive in these signature schemes. On one hand, employing a newly established symmetric primitive can enhance efficiency but may introduce the risk of future cryptanalysis. On the other hand, utilizing standardized and extensively studied primitives can provide robust security but often results in larger signatures and longer run times. 
For instance, the LowMC block cipher employed in \textit{Picnic}, was designed to reduce the size and depth of its circuit, catering to FHE and MPC applications. Several schemes are focused on this trend since a smaller number of multiplicative gates and circuit depth lead to more efficient signing algorithms and smaller signature sizes, as exemplified by \textit{Picnic}.  
Note that, in \textit{Picnic}-like designs, the employed LowMC block cipher is responsible for generating the secret and public keys. If an adversary manages to perform key recovery on the symmetric primitive, they can forge a signature by mimicking the behavior of an honest prover and utilizing the recovered key for signing. Consequently, various studies have attempted cryptanalysis on these symmetric primitives (like LowMC) using techniques such as linearization, man-in-the-middle attacks, and key-recovery attacks \cite{liu2022new}.

\vspace{-3mm}
\subsubsection{\textbf{Signature Schemes:}}
Following the potential side-channel vulnerabilities in a straightforward implementation of \textit{Picnic}, several schemes have been developed that either prioritize stringent security or offer improved cost and efficiency. Two such schemes, \textit{Picnic2} and \textit{Picnic3}, aim to optimize parameter sets for achieving shorter signatures, reduced computation time, and faster signing/verification. These schemes also demonstrate improvements compared to PQC standards like \textit{SPHINCS$^+$} \cite{kales2020improving}.  
\textit{Banquet} \cite{baum2021banquet} is another PQ secure signature that relies on standardized symmetric key primitives such as SHA3 and AES achieving half the signature size.   
\textit{AIMer} \cite{AIMer} employs an advanced proof system featuring its unique symmetric primitive known as \textit{AIM}. While it attains the shortest signature among $\skb$ schemes, its signature size remains larger and the signing/verification process lags behind $\lb$ standards. Nevertheless, it offers a nuanced balance between signature execution and size, while also boasting rapid key generation.






Certain schemes chosen in NIST's additional signature competition, blend mathematical problems from other PQC methods with the \textit{MPCitH} paradigm to create a robust and efficient signature.   
\textit{MIRA} \cite{aragon2023mira} originates from merging the MinRank problem with the \textit{MPCitH} paradigm. It utilizes additive secret sharing and a hypercube instance to offer decent performance for \textit{MPCitH}, though inferior to $\lb$ standards. \textit{MIRA}'s signature size increases quadratically with security levels due to the FS paradigm and MinRank structure. While it's highly parallelizable, its complexity renders it unsuitable for constrained devices.   
\textit{MiRth} \cite{MiRth} is formed on top of \textit{MPCitH} and incorporates the FS transformation into a 5-pass identification scheme, where verification involves demonstrating knowledge of a MinRank problem solution. While it doesn't achieve outstanding performance or sizes, its construction relies solely on linear algebra, facilitating straightforward implementation and the possibility of employing compression techniques for improved sizes.

The Syndrome-Decoding-in-the-Head signature (\textit{SDitH}) builds upon SD hardness using the \textit{MPCitH} framework, offering two variants: one leveraging a hypercube structure for a smaller signature size, and another using a threshold structure for quicker verification. It supports online/offline computations, outperforms \textit{SPHINCS$^+$} in terms of performance, and boasts notably shorter key sizes than $\lb$ standards. However, its reliance on symmetric key primitives means its performance could suffer without hardware support.  
By fusing the Rank SD problem with the \textit{MPCitH} paradigm, \textit{RYDE} \cite{bidoux2023ryde} attains a public key of comparable size derived from a compact seed. It encompasses two variations grounded in distinct party numbers within its additive MPC foundation, enabling either faster operations or reduced signature sizes. Despite these features, \textit{RYDE} remains slower than $\lb$ standards and its intricate nature makes it unsuitable for IoT applications.   
\textit{Biscuit} \cite{Biscuit}, evolving from \textit{MQDSS} and \textit{Picnic}, blends $\mv$ cryptography with the \textit{MPCitH} paradigm. It attains signature and public key sizes akin to \textit{Dilithium}, smaller than \textit{SPHINCS$^+$}. While grounded in the MQ problem and dependent on a non-standard hard problem, it manages to secure EUF-CMA, an unusual feat for $\mv$ schemes.

The \textit{PERK} scheme is established on the FS protocol without structured assumptions, hinging on the complexity of the Permuted Kernel problem. Like other \textit{MPCitH}-based schemes, it doesn't offer performance advantages over $\lb$ standards. Its signature size improvement is constrained, though it can be expedited through its symmetric primitive.  
The MQ-on-my-mind scheme (\textit{MQOM}) is built upon the MQ problem using the \textit{MPCitH} paradigm. Despite being highly parallelizable, it exhibits slow signing/verification times, quadratic growth with security levels, and larger sizes when compared to $\lb$ standards.   
In contrast to \textit{MPCitH} constructions, the \textit{FAEST} signature \cite{FAEST} is built upon the VOLE-in-the-Head (\textit{VOLeitH}) paradigm, which entails secure 2-party computation. Notably, its information-theoretic security relies on the deployed AES cipher, while its components are modular for potential performance enhancements. As a $\skb$ scheme, it outpaces \textit{SPHINCS$^+$} with compact key sizes and parameter set flexibility. However, its verification speed trails \textit{SPHINCS$^+$} and it does not yet exhibit an advantage over $\lb$ standards in terms of size.

\vspace{-3mm}
\subsection{Miscellaneous Methods}
\label{subsection3.7}
\vspace{-2mm}
In addition to the combinatorial schemes that leverage a blend of mathematical hard problems from diverse PQC approaches, there exist various schemes based on non-conventional methods that defy categorization within the mentioned approaches. Within this category, we emphasize the robust schemes chosen as part of NIST's additional digital signature competition.   
For instance, \textit{ALTEQ} \cite{blaseralteq} is a signature scheme created by applying the FS transform to a ZKP scheme rooted in the Alternating Trilinear Form Equivalence (ATFE) problem.  \textit{ALTEQ} incorporates two techniques, including the \textit{MPCitH}, which enables flexibility in parameter sets at the cost of a trade-off between signature size and signing/verification time. Furthermore, while it can be extended to support linkable ring signature functionality, its substantial public key and signature sizes may render it less suitable for certain scenarios.

Built upon the novel Embedded Multi-Layer Equations (eMLE) problem, \textit{eMLE-Sig2.0} \cite{liu2023emle} is an introduced signature that offers compact size, a straightforward design suitable for parallelization, and improved speed compared to existing PQC standards. Technically, the flattened version of eMLE is constructed around an equation resembling an SIS $\lb$ problem, with the distinguishing feature of eMLE being its use of large integers. Although it has been strengthened to resist lattice reduction-based attacks compared to its earlier versions, it may still be vulnerable to other SCAs, and its reliance on 32-bit integer multiplication could limit its usability on hardware lacking corresponding instructions.

\textit{KAZ-SIGN} \cite{KAZ-SIGN} is formed upon the second-order Discrete Logarithm Problem (2-DLP) to attain PQ security, featuring key and signature sizes within the range of traditional non-PQC standards such as ECC and RSA. Despite its advantages of no decryption failures and exceptionally compact sizes, the mathematical challenge it relies on, 2-DLP, lacks comprehensive study and is vulnerable to future cryptanalysis. This makes it less likely to be chosen for standardization.  
Derived from the general-purpose ZKP system Aurora, and akin to the \textit{MPCitH} paradigm, the \textit{Preon} signature  \cite{chen2023preon} is achieved by applying the FS algorithm to Zero-Knowledge Scalable Transparent Argument of Knowledge (zk-STARK). Thanks to its inherent structure, it can be extended to support exotic features such as group, attribute-based, and functional signature functionalities.   
The \textit{Xifrat1} family of cryptosystems is built upon three layers of randomly generated quasi-groups, making it mathematically infeasible to recover the input operand without full knowledge of the operands. While \textit{Xifrat1-Sign.I} boasts the significant advantage of compact signatures and constant-time key generation, its drawback lies in its high verification time. Even the authors of this cryptographic family acknowledge that the mathematical hard problem they rely on is relatively unexplored and needs further research. Thus, the likelihood of \textit{Xifrat1} being selected for standardization is minimal, but it could potentially pave the way for new directions in PQC.

\vspace{-4mm}
\subsection{Quantum Key Distribution}
\label{section7}
\vspace{-2mm}
Despite not being the main focus of our paper, we include a discussion on QKD and its significance in the PQ era for the sake of completeness. Unlike PQC, QKD serves as a quantum cryptographic primitive that offers an alternative approach to ensuring security guarantees in the PQ era. Essentially, QKD leverages the laws of physics to enable unconditional security in the generation and sharing of secret keys. Through the utilization of qubits for information encoding and secure communication via classical and quantum channels, QKD algorithms can be categorized into two groups based on the utilization of principles in quantum entanglement or Heisenberg's uncertainty. Researchers interested in a more comprehensive analysis regarding Quantum Cryptography and QKD can refer to survey papers such as \cite{mehic2020quantum, xu2023overview, giroti2022quantum}.

Note that if QKD successfully overcomes challenges and is implemented, it can achieve unconditional security regardless of the strength of quantum adversaries. Moreover, adhering to the fundamental scientific principle that any measurements of a quantum system or photons in quantum channels will inevitably alter them, the utilization of QKD enables the detection of any intrusion or unauthorized access. However, it should be noted that quantum cryptography and QKD alone serve as general-purpose tools and key-sharing mechanisms. To create a practical solution for real-world applications, they need to be combined with symmetric cryptography or standardized PQC \cite{xu2023overview}.

The QKD network comprises a layered structure consisting of six layers, each of which presents its own set of implementation limitations and algorithmic challenges. 
The practical implementation of a QKD network in real-world scenarios is accompanied by numerous concerns, such as performance costs, architectural challenges, and design complexities. Furthermore, it is important to note that QKD lacks support for certain functionalities required by the world's needs and is limited to the application layer, rendering it unsuitable for special-purpose devices. 
Therefore, despite providing unconditional security guarantees, it is crucial to recognize that the advancement of QKD should be considered and developed in conjunction with PQC. As they serve distinct roles and purposes, QKD cannot replace PQC in real-world scenarios and applications, regardless of the strength of its security guarantees.

\section{Vision, Insights, and Future Direction: Key Takeaways}
\label{section4}
\vspace{-2mm}

In this section, we offer an exposition of our observations, conclusions, and insights regarding the future of PQC. To begin, we delve into the analysis of prominent real-world applications, including IoT, blockchains, ML/AI, cloud services, and Networks, as well as specialized features such as FHE, MPC, and advanced signatures. After establishing the specialized requirements in the PQ era, we proceed to assess each PQ solution, present our key takeaways, and outline potential directions for future development.

IoT network represents a significant industrial revolution with implications for various applications. However, the IoT network consists of resource-constrained devices, characterized by limited computational power, memory size, and similar constraints. Consequently, it is crucial to employ lightweight cryptographic systems to ensure secure communication and successful implementation in such environments. Within evaluating metrics, the most vital factor within the IoT network is cost, performance, and implementation that even take precedence over security considerations.  
Drawing upon our analysis of the implementation aspect discussed in the previous section, we can confidently deduce that our trajectory for PQC, particularly in terms of performance assessment and applicability in various real-world scenarios, compares favorably to established pre-quantum cryptographic methods. However, when we turn our attention to IoT devices, which typically operate on 8-bit microcontrollers, and consider the substantial key sizes of PQ schemes in contrast to classical cryptographic systems, it becomes evident that there is still a significant journey ahead to meet the requirements of deeply-embedded systems. For example, the chosen PQC signature standards, such as "\textit{Dilithium}" and NIST-recommended stateful signature scheme, "$XMSS^{MT}$", remain impractical and unsuitable for lightweight IoT networks due to the resource-intensive computations and large data sizes they entail.

In spite of the dominant position of $\lb$ approach in the PQ era, along with their standardized schemes, it is essential to acknowledge that even when employing $\lb$ cryptosystems in IoT applications, it becomes necessary to apply compression techniques and optimization strategies to reduce the overhead associated with key sizes.  Consequently, the development of lightweight Post-Quantum Cryptography, such as PQ-secure lightweight signatures, remains an open issue. Furthermore, while certain finalist candidates prove suitable for a wide range of applications, it is crucial to recognize that current global standards and PQC protocols primarily concentrate on general-purpose scenarios. To address the specific needs of lightweight PQC, High-speed-High-performance PQC, and integration with existing real-world, real-time cyber-physical systems, additional collaborative efforts from both the industrial and academic sectors are necessary. Additionally, even though approaches such as $\lb$ or $\cb$ techniques allow for batch processing and parallelization due to their structural characteristics, further endeavors are required to ensure the fulfillment of these additional implementation and performance requirements.

Apart from the fact that all PQC approaches propose digital signatures and NIST’s additional competition for PQ secure signatures, the emphasis still remains on creating general-purpose signatures that fall short for some use cases.  However, certain applications such as blockchain, smart grids, e-voting, and particularly, emerging distributed and privacy-enhancing technologies,  necessitate the use of signatures designed to support additional features inherent to these specific domains. These applications require properties like anonymity, privacy, or traceability, which can be achieved via techniques such as ring signatures, group signatures, blind signatures, and similar approaches  \cite{csahin2023survey}.   
In the realm of advanced signatures, $\hb$ and $cb$ approaches have minimal or no significance, as the $\lb$ approach takes the lead and offers robust schemes that satisfy all desired properties and mentioned techniques.  Multivariate and $\ib$ methods have also been utilized to construct advanced signatures, providing comparable performance, although security guarantees remain a concern in these approaches.  Furthermore, in the context of cloud services, large-scale networks, and blockchain technology, which serve as widely adopted decentralized platforms across various domains, privacy preservation becomes crucial. In this regard, ZKP algorithms are the most applicable technique. Notably, ZKP is an advanced cryptographic technique that can be built upon all PQC approaches, and it can be transformed into an efficient signature scheme. However, the $\lb$ method has been extensively studied in the literature and exhibits significant performance in this regard \cite{buser2023survey}. Hence, a substantial research gap  awaits exploration in the realm of advanced cryptographic constructions, particularly in the domain of signatures endowed with unconventional features.

ML algorithms have versatile applications, serving as tools, techniques, or even used as a service in various domains, including network security for attack detection. However, the concept of Privacy-Preserving Machine Learning (PPML) has risen to address major privacy concerns. In the PQ era, ensuring security and privacy in ML typically involves three methods: FHE, Secure MPC, and differential privacy (DP). Similarly, cloud services offer scalability, cost-efficiency, and accessibility benefits but also present specific security challenges. These challenges can be addressed using cryptographic techniques such as MPC, FHE, secret sharing, and secure key management. Lattice-based cryptography heavily underpins FHE, MPC, and advanced secret-sharing methods, although there are a few schemes based on other approaches like code-based or multivariate cryptography. Based on these analyses, it is evident that standardizing a lattice-based encryption scheme would significantly advance FHE, MPC, ZKP, and other applications. While other real-world applications may not be the primary focus of PQC standardization efforts, network protocols such as TLS, SSH, certificates, IPsec, etc., are among the first applicable criteria considered by NIST.

\textbf{\textit{Lattice-based Cryptography: }}
With two PQC standards, $\lb$ cryptography has emerged as a prominent paradigm within the landscape of PQC and KEM algorithms, despite its large communication sizes. It not only exhibits superiority over other PQC approaches but also offers the best balance between security guarantees and performance while providing exceptional features and functionalities such as homomorphic properties, attribute-based constructions, and privacy-preserving applications. Moreover, $\lb$ cryptography encompasses a wide range of solutions, ranging from unstructured $\lb$ constructions with high-security assurances to structured lattices with excellent performance. The module version of the LWE construction, specifically, achieves high efficiency while maintaining security guarantees. Considering $\lb$ approaches for KEM standardization can have additional advantages, such as using the same framework as $\lb$ signatures. This leads to fewer integration requirements in real-world systems, providing versatility, cohesiveness, and compatibility. However, a potential drawback is the presence of a single point of failure, wherein if $\lb$ were heavily cryptanalyzed and lost its functionality, the security of the entire system would be compromised.

\textbf{\textit{Hash-based Cryptography: }}
%
Generally, HBSs offer numerous valuable and distinctive features in the context of PQC, as they provide security without relying on number-theoretic or hard problem assumptions. Unlike other approaches, the security of $\hb$ signatures is solely dependent on the properties of the utilized hash function, including one-wayness, collision resistance, pre-image resistance, and second pre-image resistance. Since the best-known quantum collision search algorithms for hash functions are slower than their classical counterparts, HBSs outperform other approaches in the PQ era.  
Another advantage of HBSs lies in their flexibility regarding the choice of the hash function. In the event that the underlying hash function becomes vulnerable to attacks, a straightforward and efficient solution is to replace it with another secure hash function. Furthermore, HBS schemes can provide forward security, ensuring that the possession of the current private key does not grant any advantage to an adversary attempting to compromise previous keys. Finally, the primary advantage of the $\hb$ approach stems from over four decades of study, which has established its solidity, efficient implementation, and ability to tailor parameter sets to accommodate various applications and scenarios.

\textbf{\textit{Code-based Cryptography: }} 
The $\cb$ PKE/KEMs offer robust security assurances, albeit with reduced efficiency and large key sizes. The $\cb$ schemes are considered a potential alternative candidate for standardization, alongside structured $\lb$ KEMs. Nevertheless, ongoing research in $\cb$ cryptography aims to enhance the feasibility of implementation on resource-constrained devices without compromising security guarantees by exploring compression techniques and alternative code selections. While $\cb$ signatures have long faced challenges in achieving practicality and resilience to cryptanalysis, and considering the lack of a robust $\cb$ signature scheme in the PQC competition, a significant portion of the schemes introduced in the NIST additional signature competition are either entirely based on $\cb$ techniques or utilize a combination of $\cb$ approaches.

\textbf{\textit{Multivariate Cryptography: }}
Despite the inherent lack of security proofs, active research is being conducted in the mathematical aspects of $\mv$ cryptography constructions, focusing on new hard problems that provide cryptographic functionalities and solid security assurances. It is important to acknowledge that although the MQ problem is classified as NP-hard, the level of hardness can vary for different instances of the problem. Nonetheless, numerous MPKC schemes have been proposed, aiming to achieve PQ security while striving for high efficiency and offering unique features such as blind signatures, ring signatures, and homomorphic properties. For example, the \textit{Sidon} scheme proposes a cryptosystem with homomorphic properties, based on the MQ and MinRank problems \cite{briaud2021polynomial}. While this approach is currently a subject of debate regarding standardization and may have limitations in certain aspects for general-purpose signature use cases in PQ era, it holds promise for the development of advanced signatures with unique and unconventional features in the future.  Additionally, $\mv$ schemes can be effectively combined with other approaches such as linear codes and $\lb$ structures, enhancing their versatility and applicability. Finally, note that the flexibility of their construction allows for modifications and various parameter sets, which suggests that new constructional designs will likely offer advanced features and comparable performance especially on low cost devices.

\textbf{\textit{SKC-based PQ-Secure Schemes: }}
For the $\skb$ PQ approach, it is worth mentioning that a recent line of research explores the use of other well-established computationally hard problems (\textit{e.g.}, syndrome decoding, MinRank, subset sum, MQ, and $\lb$ problems) within the MPCitH paradigm.  
However, further theoretical and practical investigations are necessary to ascertain their competitiveness compared to schemes based on symmetric primitives.  
The primary advantage of constructing signatures from symmetric primitives using the \textit{MPCitH} paradigm is lies in the absence of structured assumptions found in number theoretic cryptographic schemes, which have been extensively studied for many years, as well as established symmetric primitives. 
Secure communication, confidentiality, and privacy often require the use of $\skb$ cryptosystems after employing a standardized KEM scheme. 
Thus, the significance of symmetric primitives in the PQ era has increased twofold, and its importance remains even after standardization is complete.

\textbf{\textit{Isogeny-based Cryptography: }}
As discussed in subsection \ref{subsection3.6}, it should be noted that the attack on the SIDH problem does not encompass all isogeny problems. While the $\ib$ approach has limitations in terms of cryptographic functionalities and lacks a viable candidate for PQ-secure KEM standardization, it possesses unique characteristics, elegance, and perfect correctness. Encouragingly, researchers continue to propose new mathematical countermeasures to address the structural weaknesses of isogenies while considering tighter security considerations. It is important to recognize that overcoming the design issues associated with $\ib$ schemes, which are based on Elliptic Curve Cryptography, could potentially alter the trajectory of this approach and facilitate the transition to PQC.

\textbf{\textit{Miscellaneous Methods: }}
Many of the schemes falling within this category lack a well-established historical track record and necessitate additional cryptanalysis and evaluation, even when they boast adequate security assurances. Furthermore, a majority of these schemes encounter implementation challenges, and there is a shortage of parameter sets for certain security levels, or they are already subjected to intensive cryptanalysis, as evident from the formal comments submitted to NIST's website regarding additional digital signatures \cite{NISTsigCompetition}. Consequently, these methods are relatively novel, and their prospects for the future warrant further investigation.


\textbf{\textit{Side-Channel Analysis: }}
As previously discussed, all standardized PQC schemes still suffer from SCAs. Since side-channel resistance plays a crucial task in many applications, various countermeasure techniques must be applied to their implementation to secure these schemes against such attacks. However, note that additional computational overhead, increased design complexity, and potential impact on performance are some of the trade-offs associated with implementing these countermeasures. For instance, masking, especially useful in lattice-based cryptography, while effective in preventing power and timing side-channel analysis, can impose a significant computational overhead on the implementation. Thus, careful consideration is required to strike a balance \cite{SCA5}. To address this issue, instead of securing the implementation after proposing the scheme, some new approaches prioritize incorporating side channel resiliency right from the beginning of the scheme design process. As a result, they can propose side-channel resistant schemes which are more efficient than the schemes that their implementations have become side channel proved afterward. An example of such work is Raccoon \cite{Raccoon}, a novel $\lb$ signature inspired by \textit{Dilithium}. 

\textbf{\textit{Transition to PQC: }}
Crypto-agility, which refers to the ability to smoothly transition to PQC, should be our top priority. Given the vulnerability posed by the SNDL attack, achieving confidentiality is of utmost importance in high-risk applications. Thus, the initial focus should be the completion of a general-purpose PKE/KEM development, followed by the development of a general-purpose digital signature enabling the achievement of the CIA triad in major applications. 
After conducting a comprehensive analysis of various evaluation metrics considered in standardization processes and beyond, along with the exploration of new hard problems and constructional countermeasures in each approach, it is evident that standardizing a single scheme for general use cases would not be practical for all real-world applications. The diversity of metrics, the disparities between them, and the existing PQ secure approaches  contribute to this conclusion. Different real-world applications, such as IoT, Blockchain, ML/AI, etc., have distinct primary and even secondary requirements that often vary or even contradict one another. For instance, IoT prioritizes cost and implementation considerations, while blockchain emphasizes sizes and advanced features. Meanwhile, ML focuses on achieving higher security levels to ensure privacy. These varying objectives and requirements make it challenging to propose a standardized scheme that satisfies all application domains uniformly.

Considering the nature of PQC and the fact that the realization of a quantum computer is anticipated to be several years away, the viewpoint on IoT platforms and implementation can be categorized into three perspectives: short-term, mid-term, and long-term. The long-term perspective refers to a future decade where quantum computers possess sufficient computational power to break public key cryptosystems. Given this perspective, while the transition to PQC is of utmost importance and should be expedited, determining the ideal scheme or approach for IoT devices in the PQ era is not the immediate priority. It requires substantial efforts in various technologies, considering the potential implications of future quantum computing advancements. Therefore, the focus should be on addressing the transition to PQC promptly, while understanding that the specific scheme or approach for IoT devices in the PQ era necessitates additional attention and efforts in the long run.
Furthermore, the clarity of exotic features and versatility metrics is expected to improve once standardization is completed. Consequently, while the development of application-specific $\lb$ schemes appears to be a straightforward solution in the PQ era, it is important to acknowledge that achieving all evaluation metrics simultaneously in a single PQ cryptosystem is not feasible. In light of this, for instance, although the competition for stateful HBSs has reached its conclusion, it is conceivable that a specialized signature competition emerging after the finalized standardization of PQ secure signatures is not unexpected.

\vspace{-3mm}
\section{Conclusion}
\label{section5}
\vspace{-2mm}
Extensive academic and industrial efforts have been devoted to the study of PQC on a global scale. This paper presents a comprehensive survey encompassing all potential cryptographic approaches that offer PQ security solutions. Our analysis delves into worldwide standardization efforts, initiatives,  and projects related to PQC. By studying their evaluation metrics, their alignment with real-world requirements, and the specific features necessary for major applications, we provide a thorough examination of various PQC approaches, including lattice-based, hash-based, code-based, isogeny-based, symmetric key-based , and multivariate cryptography along with non-conventional methods. Our assessment adopts an approach-specific methodology, covering mathematical hard problems, security analysis, implementation considerations, and vulnerability to attacks. Notably, we explore side-channel and fault injection attacks for the standards and finalists of the NIST competition, accompanied by comprehensive demonstrations of their implementation and performance costs. Finally, we critically scrutinized each specific approach and offered our insights, vision, and projections to guide future research and considerations. 
\vspace{-3mm}


	\bibliographystyle{ACM-Reference-Format}
	\bibliography{main}


\begin{thebibliography}{202}


\ifx \showCODEN    \undefined \def \showCODEN     #1{\unskip}     \fi
\ifx \showDOI      \undefined \def \showDOI       #1{#1}\fi
\ifx \showISBNx    \undefined \def \showISBNx     #1{\unskip}     \fi
\ifx \showISBNxiii \undefined \def \showISBNxiii  #1{\unskip}     \fi
\ifx \showISSN     \undefined \def \showISSN      #1{\unskip}     \fi
\ifx \showLCCN     \undefined \def \showLCCN      #1{\unskip}     \fi
\ifx \shownote     \undefined \def \shownote      #1{#1}          \fi
\ifx \showarticletitle \undefined \def \showarticletitle #1{#1}   \fi
\ifx \showURL      \undefined \def \showURL       {\relax}        \fi
\providecommand\bibfield[2]{#2}
\providecommand\bibinfo[2]{#2}
\providecommand\natexlab[1]{#1}
\providecommand\showeprint[2][]{arXiv:#2}

\bibitem[Qua({[n.\,d.]})]%
        {Qua}
 \bibinfo{year}{[n.\,d.]}\natexlab{}.
\newblock \bibinfo{title}{QuaSiModo}.
\newblock
  \bibinfo{howpublished}{\url{https://www.forschung-it-sicherheit-kommunikationssysteme.de/projekte/quasimodo}}.
\newblock
\newblock
\shownote{Accessed: 2023}.


\bibitem[AIM(2023)]%
        {AIMer}
 \bibinfo{year}{2023}\natexlab{}.
\newblock \bibinfo{title}{AIMer}.
\newblock \bibinfo{howpublished}{\url{https://aimer-signature.org/}}.
\newblock
\newblock
\shownote{Accessed: Oct., 2023}.


\bibitem[ANS(2023)]%
        {ANSSI}
 \bibinfo{year}{2023}\natexlab{}.
\newblock \bibinfo{title}{ANSSI}.
\newblock
  \bibinfo{howpublished}{\url{https://www.ssi.gouv.fr/en/publication/anssi-views-on-the-post-quantum-cryptography-transition/}}.
\newblock
\newblock
\shownote{Accessed: Oct., 2023}.


\bibitem[Aqu(2023)]%
        {Aquacrypt}
 \bibinfo{year}{2023}\natexlab{}.
\newblock \bibinfo{title}{Aquacrypt}.
\newblock
  \bibinfo{howpublished}{\url{https://www.forschung-it-sicherheit-kommunikationssysteme.de/projekte/aquorypt}}.
\newblock
\newblock
\shownote{Accessed: Oct., 2023}.


\bibitem[Bik(2023)]%
        {Bike_Official}
 \bibinfo{year}{2023}\natexlab{}.
\newblock \bibinfo{title}{Bike}.
\newblock \bibinfo{howpublished}{\url{https://bikesuite.org}}.
\newblock
\newblock
\shownote{Accessed: Oct., 2023}.


\bibitem[Bis(2023)]%
        {Biscuit}
 \bibinfo{year}{2023}\natexlab{}.
\newblock \bibinfo{title}{Biscuit}.
\newblock \bibinfo{howpublished}{\url{https://www.biscuit-pqc.org/}}.
\newblock
\newblock
\shownote{Accessed: Oct., 2023}.


\bibitem[CAC(2023)]%
        {CACR}
 \bibinfo{year}{2023}\natexlab{}.
\newblock \bibinfo{title}{CACR}.
\newblock \bibinfo{howpublished}{\url{https://en.qapp.tech/help/cacr}}.
\newblock
\newblock
\shownote{Accessed: Oct., 2023}.


\bibitem[McE(2023)]%
        {McEliece_Official}
 \bibinfo{year}{2023}\natexlab{}.
\newblock \bibinfo{title}{Classical McEliece}.
\newblock \bibinfo{howpublished}{\url{https://classic.mceliece.org}}.
\newblock
\newblock
\shownote{Accessed: Oct., 2023}.


\bibitem[CRO(2023)]%
        {CROSS}
 \bibinfo{year}{2023}\natexlab{}.
\newblock \bibinfo{title}{CROSS}.
\newblock \bibinfo{howpublished}{\url{http://cross-crypto.com/}}.
\newblock
\newblock
\shownote{Accessed: Oct., 2023}.


\bibitem[Cry(2023)]%
        {CryptoMathCREST}
 \bibinfo{year}{2023}\natexlab{}.
\newblock \bibinfo{title}{CryptoMathic}.
\newblock \bibinfo{howpublished}{\url{https://www.cryptomathic.com}}.
\newblock
\newblock
\shownote{Accessed: Oct., 2023}.


\bibitem[CRY(2023)]%
        {CRYPTREC}
 \bibinfo{year}{2023}\natexlab{}.
\newblock \bibinfo{title}{CRYPTREC}.
\newblock
  \bibinfo{howpublished}{\url{https://www.cryptrec.go.jp/en/ex_reports.html}}.
\newblock
\newblock
\shownote{Accessed: Oct., 2023}.


\bibitem[ENI(2023)]%
        {ENISA}
 \bibinfo{year}{2023}\natexlab{}.
\newblock \bibinfo{title}{ENISA}.
\newblock
  \bibinfo{howpublished}{\url{https://www.enisa.europa.eu/publications/post-quantum-cryptography-current-state-and-quantum-mitigation}}.
\newblock
\newblock
\shownote{Accessed: 2023}.


\bibitem[ETS(2023)]%
        {ETSI}
 \bibinfo{year}{2023}\natexlab{}.
\newblock \bibinfo{title}{ETSI, QSC Working Group}.
\newblock
  \bibinfo{howpublished}{\url{https://www.etsi.org/technologies/quantum-safe-cryptography}}.
\newblock
\newblock
\shownote{Accessed: Oct., 2023}.


\bibitem[FAE(2023)]%
        {FAEST}
 \bibinfo{year}{2023}\natexlab{}.
\newblock \bibinfo{title}{FAEST}.
\newblock \bibinfo{howpublished}{\url{https://faest.info/}}.
\newblock
\newblock
\shownote{Accessed: Oct., 2023}.


\bibitem[FLO(2023)]%
        {FLOQI}
 \bibinfo{year}{2023}\natexlab{}.
\newblock \bibinfo{title}{FLOQI}.
\newblock \bibinfo{howpublished}{\url{https://floqi.org/en/}}.
\newblock
\newblock
\shownote{Accessed: Oct., 2023}.


\bibitem[FuL(2023)]%
        {FuLeeca}
 \bibinfo{year}{2023}\natexlab{}.
\newblock \bibinfo{title}{FuLeeca}.
\newblock
  \bibinfo{howpublished}{\url{https://www.ce.cit.tum.de/en/lnt/research/associate-professorship-of-coding-and-cryptography/fuleeca/}}.
\newblock
\newblock
\shownote{Accessed: Oct., 2023}.


\bibitem[Fut(2023)]%
        {FutureTPM}
 \bibinfo{year}{2023}\natexlab{}.
\newblock \bibinfo{title}{FutureTPM}.
\newblock
  \bibinfo{howpublished}{\url{https://futuretpm.eu/events/workshops/1-st-futuretpm-workshop}}.
\newblock
\newblock
\shownote{Accessed: Oct., 2023}.


\bibitem[HAW(2023)]%
        {HAWK}
 \bibinfo{year}{2023}\natexlab{}.
\newblock \bibinfo{title}{HAWK}.
\newblock \bibinfo{howpublished}{\url{https://hawk-sign.info/}}.
\newblock
\newblock
\shownote{Accessed: Oct., 2023}.


\bibitem[HuF(2023)]%
        {HuFu}
 \bibinfo{year}{2023}\natexlab{}.
\newblock \bibinfo{title}{HuFu}.
\newblock \bibinfo{howpublished}{\url{http://123.56.244.4/}}.
\newblock
\newblock
\shownote{Accessed: Oct., 2023}.


\bibitem[IBM(2023)]%
        {IBM}
 \bibinfo{year}{2023}\natexlab{}.
\newblock \bibinfo{title}{IBM}.
\newblock \bibinfo{howpublished}{\url{https://www.ibm.com/quantum/roadmap}}.
\newblock
\newblock
\shownote{Accessed: Oct., 2023}.


\bibitem[IET(2023)]%
        {IETF}
 \bibinfo{year}{2023}\natexlab{}.
\newblock \bibinfo{title}{IETF, PQUIP Working Group}.
\newblock \bibinfo{howpublished}{\url{https://www.ietf.org/blog/pquip/}}.
\newblock
\newblock
\shownote{Accessed: Oct., 2023}.


\bibitem[KAZ(2023)]%
        {KAZ-SIGN}
 \bibinfo{year}{2023}\natexlab{}.
\newblock \bibinfo{title}{KAZ-SIGN}.
\newblock \bibinfo{howpublished}{\url{https://www.antrapol.com/KAZ-SIGN}}.
\newblock
\newblock
\shownote{Accessed: Oct., 2023}.


\bibitem[KBL(2023)]%
        {KBLS}
 \bibinfo{year}{2023}\natexlab{}.
\newblock \bibinfo{title}{KBLS}.
\newblock
  \bibinfo{howpublished}{\url{https://www.forschung-it-sicherheit-kommunikationssysteme.de/projekte/kbls}}.
\newblock
\newblock
\shownote{Accessed: Oct., 2023}.


\bibitem[Kyb(2023)]%
        {Kyber_Official}
 \bibinfo{year}{2023}\natexlab{}.
\newblock \bibinfo{title}{Kyber}.
\newblock
  \bibinfo{howpublished}{\url{https://pq-crystals.org/kyber/index.shtml}}.
\newblock
\newblock
\shownote{Accessed: Oct., 2023}.


\bibitem[LES(2023)]%
        {LESS}
 \bibinfo{year}{2023}\natexlab{}.
\newblock \bibinfo{title}{LESS}.
\newblock \bibinfo{howpublished}{\url{https://www.less-project.com/}}.
\newblock
\newblock
\shownote{Accessed: Oct., 2023}.


\bibitem[MAY(2023)]%
        {MAYO}
 \bibinfo{year}{2023}\natexlab{}.
\newblock \bibinfo{title}{MAYO}.
\newblock \bibinfo{howpublished}{\url{https://pqmayo.org/}}.
\newblock
\newblock
\shownote{Accessed: Oct., 2023}.


\bibitem[med(2023)]%
        {meds}
 \bibinfo{year}{2023}\natexlab{}.
\newblock \bibinfo{title}{MEDS}.
\newblock \bibinfo{howpublished}{\url{https://www.meds-pqc.org/}}.
\newblock
\newblock
\shownote{Accessed: Oct., 2023}.


\bibitem[MiR(2023)]%
        {MiRth}
 \bibinfo{year}{2023}\natexlab{}.
\newblock \bibinfo{title}{MiRth}.
\newblock \bibinfo{howpublished}{\url{https://pqc-mirith.org/}}.
\newblock
\newblock
\shownote{Accessed: Oct., 2023}.


\bibitem[MTG(2023)]%
        {MTG}
 \bibinfo{year}{2023}\natexlab{}.
\newblock \bibinfo{title}{MTG}.
\newblock
  \bibinfo{howpublished}{\url{https://www.mtg.de/en/post-quantum-cryptography/what-is-pqc}}.
\newblock
\newblock
\shownote{Accessed: Oct., 2023}.


\bibitem[NSA(2023)]%
        {NSA}
 \bibinfo{year}{2023}\natexlab{}.
\newblock \bibinfo{title}{NSA}.
\newblock
  \bibinfo{howpublished}{\url{https://www.nsa.gov/Cybersecurity/Post-Quantum-Cybersecurity-Resources/}}.
\newblock
\newblock
\shownote{Accessed: Oct., 2023}.


\bibitem[HQC(2023)]%
        {HQC_Official}
 \bibinfo{year}{2023}\natexlab{}.
\newblock \bibinfo{title}{Official Web Page of HQC}.
\newblock \bibinfo{howpublished}{\url{https://pqc-hqc.org}}.
\newblock
\newblock
\shownote{Accessed: Oct., 2023}.


\bibitem[PQC(2023)]%
        {PQC4MED}
 \bibinfo{year}{2023}\natexlab{}.
\newblock \bibinfo{title}{PQC4MED}.
\newblock
  \bibinfo{howpublished}{\url{https://qubitreport.com/quantum-computing-science-and-research/2019/11/27/german-project-pqc4med-to-develop-post-quantum-cryptographic-update-method-for-medical-devices/}}.
\newblock
\newblock
\shownote{Accessed: Oct., 2023}.


\bibitem[PRO(2023)]%
        {PROV}
 \bibinfo{year}{2023}\natexlab{}.
\newblock \bibinfo{title}{PROV}.
\newblock \bibinfo{howpublished}{\url{https://prov-sign.github.io/}}.
\newblock
\newblock
\shownote{Accessed: Oct., 2023}.


\bibitem[QR-(2023)]%
        {QR-UOV}
 \bibinfo{year}{2023}\natexlab{}.
\newblock \bibinfo{title}{QR-UOV}.
\newblock \bibinfo{howpublished}{\url{http://info.isl.ntt.co.jp/crypt/qruov/}}.
\newblock
\newblock
\shownote{Accessed: Oct., 2023}.


\bibitem[kos(2023)]%
        {koskesh}
 \bibinfo{year}{2023}\natexlab{}.
\newblock \bibinfo{title}{QuantumRISC}.
\newblock \bibinfo{howpublished}{\url{https://www.quantumrisc.org/index.htm}}.
\newblock
\newblock
\shownote{Accessed: Oct., 2023}.


\bibitem[Rac(2023)]%
        {Raccoon}
 \bibinfo{year}{2023}\natexlab{}.
\newblock \bibinfo{title}{Raccoon}.
\newblock \bibinfo{howpublished}{\url{https://github.com/masksign/raccoon}}.
\newblock
\newblock
\shownote{Accessed: Oct., 2023}.


\bibitem[NIS(2023)]%
        {NISTsigCompetition}
 \bibinfo{year}{2023}\natexlab{}.
\newblock \bibinfo{title}{Round 1 Additional Signatures}.
\newblock
  \bibinfo{howpublished}{\url{https://csrc.nist.gov/Projects/pqc-dig-sig/round-1-additional-signatures}}.
\newblock
\newblock
\shownote{Accessed: Oct., 2023}.


\bibitem[SNO(2023)]%
        {SNOVA}
 \bibinfo{year}{2023}\natexlab{}.
\newblock \bibinfo{title}{SNOVA}.
\newblock \bibinfo{howpublished}{\url{https://snova.pqclab.org/}}.
\newblock
\newblock
\shownote{Accessed: Oct., 2023}.


\bibitem[SQU(2023)]%
        {SQUIRRELS}
 \bibinfo{year}{2023}\natexlab{}.
\newblock \bibinfo{title}{SQUIRRELS}.
\newblock \bibinfo{howpublished}{\url{https://www.squirrels-pqc.org/}}.
\newblock
\newblock
\shownote{Accessed: Oct., 2023}.


\bibitem[TUO(2023)]%
        {TUOV}
 \bibinfo{year}{2023}\natexlab{}.
\newblock \bibinfo{title}{TUOV}.
\newblock \bibinfo{howpublished}{\url{https://www.tuovsig.org/}}.
\newblock
\newblock
\shownote{Accessed: Oct., 2023}.


\bibitem[VOX(2023)]%
        {VOX}
 \bibinfo{year}{2023}\natexlab{}.
\newblock \bibinfo{title}{VOX}.
\newblock \bibinfo{howpublished}{\url{https://vox-sign.com/}}.
\newblock
\newblock
\shownote{Accessed: Oct., 2023}.


\bibitem[WAV(2023)]%
        {WAVE}
 \bibinfo{year}{2023}\natexlab{}.
\newblock \bibinfo{title}{WAVE}.
\newblock \bibinfo{howpublished}{\url{https://wave-sign.org/}}.
\newblock
\newblock
\shownote{Accessed: Oct., 2023}.


\bibitem[Aghapour and Ahmadi(2023a)]%
        {Aghapour22023}
\bibfield{author}{\bibinfo{person}{S Aghapour} {and} \bibinfo{person}{K
  Ahmadi}.} \bibinfo{year}{2023}\natexlab{a}.
\newblock \showarticletitle{{PUF-Dilithium: Design of a PUF-Based Dilithium
  Architecture Benchmarked on ARM Processors}}.
\newblock  (\bibinfo{year}{2023}).
\newblock


\bibitem[Aghapour and Ahmadi(2023b)]%
        {Aghapour12023}
\bibfield{author}{\bibinfo{person}{S Aghapour} {and} \bibinfo{person}{K
  Ahmadi}.} \bibinfo{year}{2023}\natexlab{b}.
\newblock \showarticletitle{{PUF-Kyber: Design of a PUF-Based Kyber
  Architecture Benchmarked on Diverse ARM Processors}}.
\newblock  (\bibinfo{year}{2023}).
\newblock


\bibitem[Ahmadi et~al\mbox{.}(2023a)]%
        {Ahmadi22023}
\bibfield{author}{\bibinfo{person}{Kasra Ahmadi}, \bibinfo{person}{Saeed
  Aghapour}, \bibinfo{person}{Mehran~Mozaffari Kermani}, {and}
  \bibinfo{person}{Reza Azarderakhsh}.} \bibinfo{year}{2023}\natexlab{a}.
\newblock \showarticletitle{{Efficient Error Detection Schemes for ECSM Window
  Method Benchmarked on FPGAs}}.
\newblock  (\bibinfo{date}{9} \bibinfo{year}{2023}).
\newblock
\urldef\tempurl%
\url{https://doi.org/10.36227/techrxiv.24168579.v1}
\showDOI{\tempurl}


\bibitem[Ahmadi et~al\mbox{.}(2023b)]%
        {Ahmadi12023}
\bibfield{author}{\bibinfo{person}{Kasra Ahmadi}, \bibinfo{person}{Saeed
  Aghapour}, \bibinfo{person}{Mehran~Mozaffari Kermani}, {and}
  \bibinfo{person}{Reza Azarderakhsh}.} \bibinfo{year}{2023}\natexlab{b}.
\newblock \showarticletitle{{Error Detection Schemes for $\tau$ NAF Conversion
  within Koblitz Curves Benchmarked on Various ARM Processors}}.
\newblock  (\bibinfo{date}{9} \bibinfo{year}{2023}).
\newblock
\urldef\tempurl%
\url{https://doi.org/10.36227/techrxiv.24168654.v1}
\showDOI{\tempurl}


\bibitem[Ahn et~al\mbox{.}(2022)]%
        {en15030714}
\bibfield{author}{\bibinfo{person}{Jongmin Ahn}, \bibinfo{person}{Hee-Yong
  Kwon}, \bibinfo{person}{Bohyun Ahn}, \bibinfo{person}{Kyuchan Park},
  \bibinfo{person}{Taesic Kim}, \bibinfo{person}{Mun-Kyu Lee}, {and}
  \bibinfo{person}{Kim}.} \bibinfo{year}{2022}\natexlab{}.
\newblock \showarticletitle{Toward Quantum Secured Distributed Energy
  Resources: Adoption of Post-Quantum Cryptography (PQC) and Quantum Key
  Distribution (QKD)}.
\newblock \bibinfo{journal}{\emph{Energies}} \bibinfo{volume}{15},
  \bibinfo{number}{3} (\bibinfo{year}{2022}).
\newblock
\showISSN{1996-1073}


\bibitem[Ajtai(1996)]%
        {ajtai1996generating}
\bibfield{author}{\bibinfo{person}{Mikl{\'o}s Ajtai}.}
  \bibinfo{year}{1996}\natexlab{}.
\newblock \showarticletitle{Generating hard instances of lattice problems}. In
  \bibinfo{booktitle}{\emph{28th annual ACM symposium on Theory of computing}}.
  \bibinfo{pages}{99--108}.
\newblock


\bibitem[Alagic et~al\mbox{.}(2019)]%
        {alagic2019status}
\bibfield{author}{\bibinfo{person}{Gorjan Alagic}, \bibinfo{person}{Gorjan
  Alagic}, \bibinfo{person}{Jacob Alperin-Sheriff}, \bibinfo{person}{Daniel
  Apon}, \bibinfo{person}{David Cooper}, \bibinfo{person}{Quynh Dang},
  \bibinfo{person}{Yi-Kai Liu}, \bibinfo{person}{Carl Miller},
  \bibinfo{person}{Dustin Moody}, \bibinfo{person}{Rene Peralta},
  {et~al\mbox{.}}} \bibinfo{year}{2019}\natexlab{}.
\newblock \showarticletitle{Status report on the first round of the NIST
  post-quantum cryptography standardization process}.
\newblock  (\bibinfo{year}{2019}).
\newblock


\bibitem[Alagic et~al\mbox{.}(2022)]%
        {alagic2022status}
\bibfield{author}{\bibinfo{person}{Gorjan Alagic}, \bibinfo{person}{Daniel
  Apon}, \bibinfo{person}{David Cooper}, \bibinfo{person}{Quynh Dang},
  \bibinfo{person}{Thinh Dang}, \bibinfo{person}{John Kelsey},
  \bibinfo{person}{Jacob Lichtinger}, \bibinfo{person}{Carl Miller},
  \bibinfo{person}{Dustin Moody}, \bibinfo{person}{Rene Peralta},
  {et~al\mbox{.}}} \bibinfo{year}{2022}\natexlab{}.
\newblock \showarticletitle{Status report on the third round of the NIST
  post-quantum cryptography standardization process}.
\newblock \bibinfo{journal}{\emph{US Department of Commerce, NIST}}
  (\bibinfo{year}{2022}).
\newblock


\bibitem[Alkim et~al\mbox{.}(2016)]%
        {alkim2016post}
\bibfield{author}{\bibinfo{person}{Erdem Alkim}, \bibinfo{person}{L{\'e}o
  Ducas}, \bibinfo{person}{Thomas P{\"o}ppelmann}, {and} \bibinfo{person}{Peter
  Schwabe}.} \bibinfo{year}{2016}\natexlab{}.
\newblock \showarticletitle{Post-quantum key $\{$Exchange—A$\}$ new hope}. In
  \bibinfo{booktitle}{\emph{25th USENIX Security Symposium (USENIX Security
  16)}}. \bibinfo{pages}{327--343}.
\newblock


\bibitem[Aragon et~al\mbox{.}(2017)]%
        {aragon2017bike}
\bibfield{author}{\bibinfo{person}{Nicolas Aragon}, \bibinfo{person}{Paulo~SLM
  Barreto}, \bibinfo{person}{Slim Bettaieb}, \bibinfo{person}{Loic Bidoux},
  \bibinfo{person}{Olivier Blazy}, \bibinfo{person}{Jean-Christophe
  Deneuville}, \bibinfo{person}{Philippe Gaborit}, \bibinfo{person}{Shay
  Gueron}, \bibinfo{person}{Tim Guneysu}, \bibinfo{person}{Carlos~Aguilar
  Melchor}, {et~al\mbox{.}}} \bibinfo{year}{2017}\natexlab{}.
\newblock \showarticletitle{BIKE: bit flipping key encapsulation}.
\newblock  (\bibinfo{year}{2017}).
\newblock


\bibitem[Aragon et~al\mbox{.}(2023)]%
        {aragon2023mira}
\bibfield{author}{\bibinfo{person}{Nicolas Aragon}, \bibinfo{person}{Lo{\"\i}c
  Bidoux}, \bibinfo{person}{Jes{\'u}s-Javier Chi-Dom{\'\i}nguez},
  \bibinfo{person}{Thibauld Feneuil}, \bibinfo{person}{Philippe Gaborit},
  \bibinfo{person}{Romaric Neveu}, {and} \bibinfo{person}{Matthieu Rivain}.}
  \bibinfo{year}{2023}\natexlab{}.
\newblock \showarticletitle{MIRA: a Digital Signature Scheme based on the
  MinRank problem and the MPC-in-the-Head paradigm}.
\newblock \bibinfo{journal}{\emph{arXiv preprint arXiv:2307.08575}}
  (\bibinfo{year}{2023}).
\newblock


\bibitem[Augustine Chidiebere~Onuora(2020)]%
        {1910089619}
\bibfield{author}{\bibinfo{person}{Anthony O Otiko J. N~Nworie Augustine
  Chidiebere~Onuora, Chibuike~Madubuike}.} \bibinfo{year}{2020}\natexlab{}.
\newblock \showarticletitle{Post-Quantum Cryptographic Algorithm: A systematic
  review of round-2 candidates}. In \bibinfo{booktitle}{\emph{Academia in
  Information Technology Profession AITP}}.
\newblock


\bibitem[Baldi et~al\mbox{.}(2021)]%
        {baldi2021code}
\bibfield{author}{\bibinfo{person}{Marco Baldi}, \bibinfo{person}{Franco
  Chiaraluce}, {and} \bibinfo{person}{Santini}.}
  \bibinfo{year}{2021}\natexlab{}.
\newblock \showarticletitle{Code-based signatures without trapdoors through
  restricted vectors}.
\newblock \bibinfo{journal}{\emph{Cryptol Archive}} (\bibinfo{year}{2021}).
\newblock


\bibitem[Banegas et~al\mbox{.}(2021)]%
        {banegas2021wavelet}
\bibfield{author}{\bibinfo{person}{Gustavo Banegas}, \bibinfo{person}{Thomas
  Debris-Alazard}, \bibinfo{person}{Milena Nedeljkovi{\'c}}, {and}
  \bibinfo{person}{Benjamin Smith}.} \bibinfo{year}{2021}\natexlab{}.
\newblock \showarticletitle{Wavelet: Code-based postquantum signatures with
  fast verification on microcontrollers}.
\newblock \bibinfo{journal}{\emph{arXiv preprint arXiv:2110.13488}}
  (\bibinfo{year}{2021}).
\newblock


\bibitem[Basso and Fouotsa(2023)]%
        {basso2023new}
\bibfield{author}{\bibinfo{person}{Andrea Basso} {and}
  \bibinfo{person}{Tako~Boris Fouotsa}.} \bibinfo{year}{2023}\natexlab{}.
\newblock \showarticletitle{New SIDH Countermeasures for a More Efficient Key
  Exchange}.
\newblock \bibinfo{journal}{\emph{Cryptology ePrint Archive}}
  (\bibinfo{year}{2023}).
\newblock


\bibitem[Basso et~al\mbox{.}(2023)]%
        {basso2023festa}
\bibfield{author}{\bibinfo{person}{Andrea Basso}, \bibinfo{person}{Luciano
  Maino}, {and} \bibinfo{person}{Giacomo Pope}.}
  \bibinfo{year}{2023}\natexlab{}.
\newblock \showarticletitle{FESTA: Fast Encryption from Supersingular Torsion
  Attacks}.
\newblock \bibinfo{journal}{\emph{Cryptology Archive}} (\bibinfo{year}{2023}).
\newblock


\bibitem[Baum et~al\mbox{.}(2021)]%
        {baum2021banquet}
\bibfield{author}{\bibinfo{person}{Carsten Baum},
  \bibinfo{person}{Cyprien~Delpech de Saint~Guilhem}, \bibinfo{person}{Daniel
  Kales}, \bibinfo{person}{Emmanuela Orsini}, \bibinfo{person}{Peter Scholl},
  {and} \bibinfo{person}{Greg Zaverucha}.} \bibinfo{year}{2021}\natexlab{}.
\newblock \showarticletitle{Banquet: short and fast signatures from AES}. In
  \bibinfo{booktitle}{\emph{IACR International Conference on Public-Key
  Cryptography}}. Springer, \bibinfo{pages}{266--297}.
\newblock


\bibitem[Bavdekar et~al\mbox{.}(2023)]%
        {10048976}
\bibfield{author}{\bibinfo{person}{Ritik Bavdekar}, \bibinfo{person}{Eashan
  Jayant~Chopde}, \bibinfo{person}{Ankit Agrawal}, \bibinfo{person}{Ashutosh
  Bhatia}, {and} \bibinfo{person}{Kamlesh Tiwari}.}
  \bibinfo{year}{2023}\natexlab{}.
\newblock \showarticletitle{Post Quantum Cryptography: A Review of Techniques,
  Challenges and Standardizations}. In \bibinfo{booktitle}{\emph{2023
  International Conference on Information Networking (ICOIN)}}.
  \bibinfo{pages}{146--151}.
\newblock


\bibitem[Behnia et~al\mbox{.}(2021)]%
        {behnia2021lattice}
\bibfield{author}{\bibinfo{person}{Rouzbeh Behnia}, \bibinfo{person}{Eamonn~W
  Postlethwaite}, \bibinfo{person}{Muslum~Ozgur Ozmen}, {and}
  \bibinfo{person}{Attila~Altay Yavuz}.} \bibinfo{year}{2021}\natexlab{}.
\newblock \showarticletitle{Lattice-based proof-of-work for post-quantum
  blockchains}. In \bibinfo{booktitle}{\emph{International Workshop on Data
  Privacy Management}}. Springer, \bibinfo{pages}{310--318}.
\newblock


\bibitem[Behnia and Yavuz(2022)]%
        {behnia2022lightweight}
\bibfield{author}{\bibinfo{person}{Rouzbeh Behnia} {and}
  \bibinfo{person}{Attila~Altay Yavuz}.} \bibinfo{year}{2022}\natexlab{}.
\newblock \bibinfo{title}{Lightweight post-quantum authentication}.
\newblock
\newblock
\newblock
\shownote{US Patent App. 17/739,036}.


\bibitem[Behnia et~al\mbox{.}(2017)]%
        {behnia2017high}
\bibfield{author}{\bibinfo{person}{Rouzbeh Behnia},
  \bibinfo{person}{Attila~Altay Yavuz}, {and} \bibinfo{person}{Muslum~Ozgur
  Ozmen}.} \bibinfo{year}{2017}\natexlab{}.
\newblock \showarticletitle{High-speed high-security public key encryption with
  keyword search}. In \bibinfo{booktitle}{\emph{IFIP annual conference on data
  and applications security and privacy}}. Springer, \bibinfo{pages}{365--385}.
\newblock


\bibitem[Bernstein et~al\mbox{.}(2017)]%
        {bernstein2017classic}
\bibfield{author}{\bibinfo{person}{Daniel~J Bernstein}, \bibinfo{person}{Tung
  Chou}, \bibinfo{person}{Tanja Lange}, \bibinfo{person}{Ingo von Maurich},
  \bibinfo{person}{Rafael Misoczki}, \bibinfo{person}{Ruben Niederhagen},
  \bibinfo{person}{Edoardo Persichetti}, \bibinfo{person}{Christiane Peters},
  \bibinfo{person}{Peter Schwabe}, \bibinfo{person}{Nicolas Sendrier},
  {et~al\mbox{.}}} \bibinfo{year}{2017}\natexlab{}.
\newblock \showarticletitle{Classic McEliece: conservative code-based
  cryptography}.
\newblock \bibinfo{journal}{\emph{NIST submissions}} \bibinfo{volume}{1},
  \bibinfo{number}{1} (\bibinfo{year}{2017}), \bibinfo{pages}{1--25}.
\newblock


\bibitem[Bernstein et~al\mbox{.}(2015)]%
        {bernstein2015sphincs}
\bibfield{author}{\bibinfo{person}{Daniel~J Bernstein}, \bibinfo{person}{Daira
  Hopwood}, \bibinfo{person}{Andreas H{\"u}lsing}, \bibinfo{person}{Tanja
  Lange}, \bibinfo{person}{Ruben Niederhagen}, \bibinfo{person}{Louiza
  Papachristodoulou}, \bibinfo{person}{Michael Schneider},
  \bibinfo{person}{Peter Schwabe}, {and} \bibinfo{person}{Zooko
  Wilcox-O’Hearn}.} \bibinfo{year}{2015}\natexlab{}.
\newblock \showarticletitle{SPHINCS: practical stateless hash-based
  signatures}. In \bibinfo{booktitle}{\emph{Annual international conference on
  the theory and applications of cryptographic techniques}}. Springer,
  \bibinfo{pages}{368--397}.
\newblock


\bibitem[Bernstein et~al\mbox{.}(2019)]%
        {bernstein2019sphincs+}
\bibfield{author}{\bibinfo{person}{Daniel~J Bernstein},
  \bibinfo{person}{Andreas H{\"u}lsing}, \bibinfo{person}{Stefan K{\"o}lbl},
  \bibinfo{person}{Ruben Niederhagen}, \bibinfo{person}{Joost Rijneveld}, {and}
  \bibinfo{person}{Peter Schwabe}.} \bibinfo{year}{2019}\natexlab{}.
\newblock \showarticletitle{The SPHINCS+ signature framework}. In
  \bibinfo{booktitle}{\emph{Proceedings of the 2019 ACM SIGSAC conference on
  computer and communications security}}. \bibinfo{pages}{2129--2146}.
\newblock


\bibitem[Bernstein(2017)]%
        {Bernstein2017}
\bibfield{author}{\bibinfo{person}{Lange~Tanja Bernstein, Daniel~J.}}
  \bibinfo{year}{2017}\natexlab{}.
\newblock \showarticletitle{Post-quantum cryptography}.
\newblock \bibinfo{journal}{\emph{Nature}}  \bibinfo{volume}{549}
  (\bibinfo{year}{2017}).
\newblock
Issue 7671.
\showISSN{1476-4687}
\urldef\tempurl%
\url{https://doi.org/10.1038/nature23461}
\showDOI{\tempurl}


\bibitem[Beullens and Feo(2023)]%
        {beullens2023proving}
\bibfield{author}{\bibinfo{person}{Ward Beullens} {and} \bibinfo{person}{De
  Feo}.} \bibinfo{year}{2023}\natexlab{}.
\newblock \showarticletitle{Proving knowledge of isogenies: a survey}.
\newblock \bibinfo{journal}{\emph{Designs, Codes and Cryptography}}
  (\bibinfo{year}{2023}), \bibinfo{pages}{1--32}.
\newblock


\bibitem[Bidoux et~al\mbox{.}(2023)]%
        {bidoux2023ryde}
\bibfield{author}{\bibinfo{person}{Lo{\"\i}c Bidoux},
  \bibinfo{person}{Jes{\'u}s-Javier Chi-Dom{\'\i}nguez},
  \bibinfo{person}{Thibauld Feneuil}, \bibinfo{person}{Philippe Gaborit},
  \bibinfo{person}{Antoine Joux}, \bibinfo{person}{Matthieu Rivain}, {and}
  \bibinfo{person}{Adrien Vin{\c{c}}otte}.} \bibinfo{year}{2023}\natexlab{}.
\newblock \showarticletitle{RYDE: A Digital Signature Scheme based on
  Rank-Syndrome-Decoding Problem with MPCitH Paradigm}.
\newblock \bibinfo{journal}{\emph{arXiv preprint arXiv:2307.08726}}
  (\bibinfo{year}{2023}).
\newblock


\bibitem[Bl{\"a}ser et~al\mbox{.}({[n.\,d.]})]%
        {blaseralteq}
\bibfield{author}{\bibinfo{person}{Markus Bl{\"a}ser},
  \bibinfo{person}{Dung~Hoang Duong}, \bibinfo{person}{Anand~Kumar Narayanan},
  \bibinfo{person}{Thomas Plantard}, \bibinfo{person}{Youming Qiao},
  \bibinfo{person}{Arnaud Sipasseuth}, {and} \bibinfo{person}{Gang Tang}.}
  \bibinfo{year}{[n.\,d.]}\natexlab{}.
\newblock \showarticletitle{The ALTEQ Signature Scheme: Algorithm
  Specifications and Supporting Documentation}.
\newblock  (\bibinfo{year}{[n.\,d.]}).
\newblock


\bibitem[Bos et~al\mbox{.}(2018)]%
        {bos2018crystals}
\bibfield{author}{\bibinfo{person}{Joppe Bos}, \bibinfo{person}{L{\'e}o Ducas},
  \bibinfo{person}{Eike Kiltz}, \bibinfo{person}{Tancr{\`e}de Lepoint},
  \bibinfo{person}{Vadim Lyubashevsky}, \bibinfo{person}{John~M Schanck},
  \bibinfo{person}{Peter Schwabe}, \bibinfo{person}{Gregor Seiler}, {and}
  \bibinfo{person}{Damien Stehl{\'e}}.} \bibinfo{year}{2018}\natexlab{}.
\newblock \showarticletitle{CRYSTALS-Kyber: a CCA-secure module-lattice-based
  KEM}. In \bibinfo{booktitle}{\emph{2018 IEEE European Symposium on Security
  and Privacy (EuroS\&P)}}. IEEE, \bibinfo{pages}{353--367}.
\newblock


\bibitem[Botros et~al\mbox{.}(2019)]%
        {botros2019memory}
\bibfield{author}{\bibinfo{person}{Leon Botros}, \bibinfo{person}{Matthias~J
  Kannwischer}, {and} \bibinfo{person}{Peter Schwabe}.}
  \bibinfo{year}{2019}\natexlab{}.
\newblock \showarticletitle{Memory-efficient high-speed implementation of Kyber
  on Cortex-M4}. In \bibinfo{booktitle}{\emph{Progress in
  Cryptology--AFRICACRYPT 2019: 11th International Conference on Cryptology in
  Africa, Rabat, Morocco, July 9--11, 2019, Proceedings 11}}. Springer.
\newblock


\bibitem[Briaud et~al\mbox{.}(2021)]%
        {briaud2021polynomial}
\bibfield{author}{\bibinfo{person}{Pierre Briaud}, \bibinfo{person}{Jean-Pierre
  Tillich}, {and} \bibinfo{person}{Javier Verbel}.}
  \bibinfo{year}{2021}\natexlab{}.
\newblock \showarticletitle{A polynomial time key-recovery attack on the Sidon
  cryptosystem}. In \bibinfo{booktitle}{\emph{International Conference on
  Selected Areas in Cryptography}}. Springer, \bibinfo{pages}{419--438}.
\newblock


\bibitem[Bruinderink and Pessl(2018)]%
        {dilit7}
\bibfield{author}{\bibinfo{person}{Leon~Groot Bruinderink} {and}
  \bibinfo{person}{Peter Pessl}.} \bibinfo{year}{2018}\natexlab{}.
\newblock \showarticletitle{Differential fault attacks on deterministic lattice
  signatures}.
\newblock \bibinfo{journal}{\emph{IACR Transactions on Cryptographic Hardware
  and Embedded Systems}} (\bibinfo{year}{2018}), \bibinfo{pages}{21--43}.
\newblock


\bibitem[Buser et~al\mbox{.}(2023)]%
        {buser2023survey}
\bibfield{author}{\bibinfo{person}{Maxime Buser}, \bibinfo{person}{Rafael
  Dowsley}, \bibinfo{person}{Muhammed Esgin}, \bibinfo{person}{Cl{\'e}mentine
  Gritti}, \bibinfo{person}{Shabnam Kasra~Kermanshahi},
  \bibinfo{person}{Veronika Kuchta}, \bibinfo{person}{Jason Legrow},
  \bibinfo{person}{Joseph Liu}, {and} \bibinfo{person}{Phan}.}
  \bibinfo{year}{2023}\natexlab{}.
\newblock \showarticletitle{A Survey on Exotic Signatures for Post-quantum
  Blockchain: Challenges and Research Directions}.
\newblock \bibinfo{journal}{\emph{Comput. Surveys}} \bibinfo{volume}{55},
  \bibinfo{number}{12} (\bibinfo{year}{2023}).
\newblock


\bibitem[Canto et~al\mbox{.}(2022)]%
        {Mc11}
\bibfield{author}{\bibinfo{person}{Alvaro~Cintas Canto},
  \bibinfo{person}{Mehran~Mozaffari Kermani}, {and} \bibinfo{person}{Reza
  Azarderakhsh}.} \bibinfo{year}{2022}\natexlab{}.
\newblock \showarticletitle{Reliable Constructions for the Key Generator of
  Code-Based Post-Quantum Cryptosystems on FPGA}.
\newblock \bibinfo{journal}{\emph{J. Emerg. Technol. Comput. Syst.}}
  \bibinfo{volume}{19}, \bibinfo{number}{1}, Article \bibinfo{articleno}{5}
  (\bibinfo{date}{dec} \bibinfo{year}{2022}), \bibinfo{numpages}{20}~pages.
\newblock
\showISSN{1550-4832}
\urldef\tempurl%
\url{https://doi.org/10.1145/3544921}
\showDOI{\tempurl}


\bibitem[Casanova et~al\mbox{.}(2017)]%
        {casanova2017gemss}
\bibfield{author}{\bibinfo{person}{Antoine Casanova},
  \bibinfo{person}{Jean-Charles Faugere}, \bibinfo{person}{Gilles Macario-Rat},
  \bibinfo{person}{Jacques Patarin}, \bibinfo{person}{Ludovic Perret}, {and}
  \bibinfo{person}{Jocelyn Ryckeghem}.} \bibinfo{year}{2017}\natexlab{}.
\newblock \emph{\bibinfo{title}{GeMSS: a great multivariate short signature}}.
\newblock \bibinfo{thesistype}{Ph.\,D. Dissertation}.
  \bibinfo{school}{UPMC-Paris 6 Sorbonne Universit{\'e}s; INRIA Paris Research
  Centre, MAMBA Team~…}.
\newblock


\bibitem[Castryck and Decru(2023)]%
        {castryck2023efficient}
\bibfield{author}{\bibinfo{person}{Wouter Castryck} {and}
  \bibinfo{person}{Thomas Decru}.} \bibinfo{year}{2023}\natexlab{}.
\newblock \showarticletitle{An efficient key recovery attack on SIDH}. In
  \bibinfo{booktitle}{\emph{Annual International Conference on the Theory and
  Applications of Cryptographic Techniques}}. Springer,
  \bibinfo{pages}{423--447}.
\newblock


\bibitem[Chase et~al\mbox{.}(2017)]%
        {chase2017post}
\bibfield{author}{\bibinfo{person}{Melissa Chase}, \bibinfo{person}{David
  Derler}, \bibinfo{person}{Steven Goldfeder}, \bibinfo{person}{Claudio
  Orlandi}, \bibinfo{person}{Sebastian Ramacher}, \bibinfo{person}{Christian
  Rechberger}, {and} \bibinfo{person}{Slamanig}.}
  \bibinfo{year}{2017}\natexlab{}.
\newblock \showarticletitle{Post-quantum zero-knowledge and signatures from
  symmetric-key primitives}. In \bibinfo{booktitle}{\emph{Proceedings of the
  acm sigsac conf. on comp. and comm. security}}.
\newblock


\bibitem[Chen et~al\mbox{.}(2023)]%
        {chen2023preon}
\bibfield{author}{\bibinfo{person}{Ming-Shing Chen}, \bibinfo{person}{Yu-Shian
  Chen}, \bibinfo{person}{Chen-Mou Cheng}, \bibinfo{person}{Shiuan Fu},
  \bibinfo{person}{Wei-Chih Hong}, \bibinfo{person}{Jen-Hsuan Hsiang},
  \bibinfo{person}{Sheng-Te Hu}, \bibinfo{person}{Po-Chun Kuo},
  \bibinfo{person}{Wei-Bin Lee}, \bibinfo{person}{Feng-Hao Liu},
  {et~al\mbox{.}}} \bibinfo{year}{2023}\natexlab{}.
\newblock \showarticletitle{Preon: zk-SNARK based Signature Scheme}.
\newblock  (\bibinfo{year}{2023}).
\newblock


\bibitem[Chen and Chou(2021)]%
        {chen2021classic}
\bibfield{author}{\bibinfo{person}{Ming-Shing Chen} {and} \bibinfo{person}{Tung
  Chou}.} \bibinfo{year}{2021}\natexlab{}.
\newblock \showarticletitle{Classic McEliece on the ARM cortex-M4}.
\newblock \bibinfo{journal}{\emph{IACR Trans. on Crypto. HW and Embed. Sys.}}
  (\bibinfo{year}{2021}).
\newblock


\bibitem[Chen et~al\mbox{.}(2021)]%
        {chen2021optimizing}
\bibfield{author}{\bibinfo{person}{Ming-Shing Chen}, \bibinfo{person}{Tung
  Chou}, {and} \bibinfo{person}{Markus Krausz}.}
  \bibinfo{year}{2021}\natexlab{}.
\newblock \showarticletitle{Optimizing bike for the intel haswell and arm
  cortex-m4}.
\newblock \bibinfo{journal}{\emph{Cryptology Archive}} (\bibinfo{year}{2021}).
\newblock


\bibitem[Cheon et~al\mbox{.}(2023)]%
        {cheon2023haetae}
\bibfield{author}{\bibinfo{person}{Jung~Hee Cheon}, \bibinfo{person}{Hyeongmin
  Choe}, \bibinfo{person}{Julien Devevey}, \bibinfo{person}{Tim G{\"u}neysu},
  \bibinfo{person}{Dongyeon Hong}, \bibinfo{person}{Markus Krausz},
  \bibinfo{person}{Georg Land}, \bibinfo{person}{Marc M{\"o}ller},
  \bibinfo{person}{Damien Stehl{\'e}}, {and} \bibinfo{person}{MinJune Yi}.}
  \bibinfo{year}{2023}\natexlab{}.
\newblock \showarticletitle{HAETAE: Shorter Lattice-Based Fiat-Shamir
  Signatures}.
\newblock \bibinfo{journal}{\emph{Cryptology ePrint Archive}}
  (\bibinfo{year}{2023}).
\newblock


\bibitem[Chiano et~al\mbox{.}(2021)]%
        {dichiano2021survey}
\bibfield{author}{\bibinfo{person}{Nicola~Di Chiano}, \bibinfo{person}{Riccardo
  Longo}, \bibinfo{person}{Alessio Meneghetti}, {and} \bibinfo{person}{Giordano
  Santilli}.} \bibinfo{year}{2021}\natexlab{}.
\newblock \bibinfo{title}{A survey on NIST PQ signatures}.
\newblock
\newblock
\showeprint[arxiv]{2107.11082}~[cs.CR]


\bibitem[Cho et~al\mbox{.}(2022)]%
        {cho2022enhanced}
\bibfield{author}{\bibinfo{person}{Jinkyu Cho}, \bibinfo{person}{Jong-Seon No},
  \bibinfo{person}{Yongwoo Lee}, \bibinfo{person}{Zahyun Koo}, {and}
  \bibinfo{person}{Young-Sik Kim}.} \bibinfo{year}{2022}\natexlab{}.
\newblock \showarticletitle{Enhanced pqsigRM: Code-Based Digital Signature
  Scheme with Short Signature and Fast Verification for Post-Quantum
  Cryptography}.
\newblock \bibinfo{journal}{\emph{Cryptology ePrint Archive}}
  (\bibinfo{year}{2022}).
\newblock


\bibitem[Chowdhury et~al\mbox{.}(2021)]%
        {SCA5}
\bibfield{author}{\bibinfo{person}{Sreeja Chowdhury}, \bibinfo{person}{Ana
  Covic}, \bibinfo{person}{Rabin~Yu Acharya}, \bibinfo{person}{Spencer Dupee},
  \bibinfo{person}{Fatemeh Ganji}, {and} \bibinfo{person}{Forte}.}
  \bibinfo{year}{2021}\natexlab{}.
\newblock \showarticletitle{Physical security in the post-quantum era: A survey
  on side-channel analysis, random number generators, and physically unclonable
  functions}.
\newblock \bibinfo{journal}{\emph{Journal of Cryptographic Engineering}}
  (\bibinfo{year}{2021}).
\newblock


\bibitem[Chuengsatiansup et~al\mbox{.}(2020)]%
        {chuengsatiansup2020modfalcon}
\bibfield{author}{\bibinfo{person}{Chitchanok Chuengsatiansup},
  \bibinfo{person}{Thomas Prest}, \bibinfo{person}{Damien Stehl{\'e}},
  \bibinfo{person}{Alexandre Wallet}, {and} \bibinfo{person}{Keita Xagawa}.}
  \bibinfo{year}{2020}\natexlab{}.
\newblock \showarticletitle{ModFalcon: Compact signatures based on module-NTRU
  lattices}. In \bibinfo{booktitle}{\emph{Proceedings of the 15th ACM Asia
  Conference on Computer and Communications Security}}.
  \bibinfo{pages}{853--866}.
\newblock


\bibitem[Cintas-Canto et~al\mbox{.}(2023)]%
        {Mc10}
\bibfield{author}{\bibinfo{person}{Alvaro Cintas-Canto},
  \bibinfo{person}{Mehran~Mozaffari Kermani}, {and} \bibinfo{person}{Reza
  Azarderakhsh}.} \bibinfo{year}{2023}\natexlab{}.
\newblock \showarticletitle{Reliable Architectures for Finite Field Multipliers
  Using Cyclic Codes on FPGA Utilized in Classic and Post-Quantum
  Cryptography}.
\newblock \bibinfo{journal}{\emph{IEEE Transactions on Very Large Scale
  Integration (VLSI) Systems}} \bibinfo{volume}{31}, \bibinfo{number}{1}
  (\bibinfo{year}{2023}).
\newblock


\bibitem[Colombier et~al\mbox{.}(2022)]%
        {Mc7}
\bibfield{author}{\bibinfo{person}{Brice Colombier},
  \bibinfo{person}{Vlad-Florin Drăgoi}, \bibinfo{person}{Pierre-Louis Cayrel},
  {and} \bibinfo{person}{Vincent Grosso}.} \bibinfo{year}{2022}\natexlab{}.
\newblock \showarticletitle{Profiled Side-Channel Attack on Cryptosystems Based
  on the Binary Syndrome Decoding Problem}.
\newblock \bibinfo{journal}{\emph{IEEE Transactions on Information Forensics
  and Security}}  \bibinfo{volume}{17} (\bibinfo{year}{2022}),
  \bibinfo{pages}{3407--3420}.
\newblock


\bibitem[Courtois et~al\mbox{.}(2001)]%
        {courtois2001achieve}
\bibfield{author}{\bibinfo{person}{Nicolas~T Courtois},
  \bibinfo{person}{Matthieu Finiasz}, {and} \bibinfo{person}{Nicolas
  Sendrier}.} \bibinfo{year}{2001}\natexlab{}.
\newblock \showarticletitle{How to achieve a McEliece-based digital signature
  scheme}. In \bibinfo{booktitle}{\emph{Advances in Cryptology—ASIACRYPT
  2001}}. Springer.
\newblock


\bibitem[Dam et~al\mbox{.}(2023)]%
        {dam2023survey}
\bibfield{author}{\bibinfo{person}{Duc-Thuan Dam}, \bibinfo{person}{Thai-Ha
  Tran}, \bibinfo{person}{Van-Phuc Hoang}, \bibinfo{person}{Cong-Kha Pham},
  {and} \bibinfo{person}{Trong-Thuc Hoang}.} \bibinfo{year}{2023}\natexlab{}.
\newblock \showarticletitle{A Survey of Post-Quantum Cryptography: Start of a
  New Race}.
\newblock \bibinfo{journal}{\emph{Cryptography}} \bibinfo{volume}{7},
  \bibinfo{number}{3} (\bibinfo{year}{2023}), \bibinfo{pages}{40}.
\newblock


\bibitem[Darzi et~al\mbox{.}(2022)]%
        {darzi2022lpm2da}
\bibfield{author}{\bibinfo{person}{Saleh Darzi}, \bibinfo{person}{Bahareh
  Akhbari}, {and} \bibinfo{person}{Hassan Khodaiemehr}.}
  \bibinfo{year}{2022}\natexlab{}.
\newblock \showarticletitle{LPM2DA: a lattice-based privacy-preserving
  multi-functional and multi-dimensional data aggregation scheme for smart
  grid}.
\newblock \bibinfo{journal}{\emph{Cluster Computing}} \bibinfo{volume}{25},
  \bibinfo{number}{1} (\bibinfo{year}{2022}), \bibinfo{pages}{263--278}.
\newblock


\bibitem[De~Feo(2017)]%
        {de2017mathematics}
\bibfield{author}{\bibinfo{person}{Luca De~Feo}.}
  \bibinfo{year}{2017}\natexlab{}.
\newblock \showarticletitle{Mathematics of isogeny based cryptography}.
\newblock \bibinfo{journal}{\emph{arXiv preprint arXiv:1711.04062}}
  (\bibinfo{year}{2017}).
\newblock


\bibitem[De~Feo et~al\mbox{.}(2020)]%
        {de2020sqisign}
\bibfield{author}{\bibinfo{person}{Luca De~Feo}, \bibinfo{person}{David Kohel},
  \bibinfo{person}{Antonin Leroux}, \bibinfo{person}{Christophe Petit}, {and}
  \bibinfo{person}{Benjamin Wesolowski}.} \bibinfo{year}{2020}\natexlab{}.
\newblock \showarticletitle{SQISign: compact post-quantum signatures from
  quaternions and isogenies}. In \bibinfo{booktitle}{\emph{Advances in
  Cryptology--ASIACRYPT 2020: 26th International Conference on the Theory and
  Application of Cryptology and Information Security, Daejeon, South Korea,
  December 7--11, 2020, Proceedings, Part I 26}}. Springer,
  \bibinfo{pages}{64--93}.
\newblock


\bibitem[Dey and Dutta(2023)]%
        {dey2023progress}
\bibfield{author}{\bibinfo{person}{Jayashree Dey} {and} \bibinfo{person}{Ratna
  Dutta}.} \bibinfo{year}{2023}\natexlab{}.
\newblock \showarticletitle{Progress in Multivariate Cryptography: Systematic
  Review, Challenges, and Research Directions}.
\newblock \bibinfo{journal}{\emph{Comput. Surveys}} \bibinfo{volume}{55},
  \bibinfo{number}{12} (\bibinfo{year}{2023}), \bibinfo{pages}{1--34}.
\newblock


\bibitem[Di~Chiano et~al\mbox{.}(2021)]%
        {di2021survey}
\bibfield{author}{\bibinfo{person}{Nicola Di~Chiano}, \bibinfo{person}{Riccardo
  Longo}, \bibinfo{person}{Alessio Meneghetti}, {and} \bibinfo{person}{Giordano
  Santilli}.} \bibinfo{year}{2021}\natexlab{}.
\newblock \showarticletitle{A survey on NIST PQ signatures}.
\newblock \bibinfo{journal}{\emph{arXiv preprint}} (\bibinfo{year}{2021}).
\newblock


\bibitem[Dubrova and Ngo(2022)]%
        {Kyber6}
\bibfield{author}{\bibinfo{person}{Elena Dubrova} {and} \bibinfo{person}{Kalle
  Ngo}.} \bibinfo{year}{2022}\natexlab{}.
\newblock \showarticletitle{Breaking a 5th-order masked implementation of
  crystals-kyber by copy-paste}.
\newblock \bibinfo{journal}{\emph{Cryptol. Archive}} (\bibinfo{year}{2022}).
\newblock


\bibitem[Ducas et~al\mbox{.}(2013)]%
        {ducas2013lattice}
\bibfield{author}{\bibinfo{person}{L{\'e}o Ducas}, \bibinfo{person}{Alain
  Durmus}, {and} \bibinfo{person}{Tancr{\`e}de Lepoint}.}
  \bibinfo{year}{2013}\natexlab{}.
\newblock \showarticletitle{Lattice signatures and bimodal Gaussians}. In
  \bibinfo{booktitle}{\emph{Annual Cryptology Conference}}. Springer.
\newblock


\bibitem[Ducas et~al\mbox{.}(2018)]%
        {ducas2018crystals}
\bibfield{author}{\bibinfo{person}{L{\'e}o Ducas}, \bibinfo{person}{Eike
  Kiltz}, \bibinfo{person}{Tancrede Lepoint}, \bibinfo{person}{Vadim
  Lyubashevsky}, \bibinfo{person}{Peter Schwabe}, \bibinfo{person}{Gregor
  Seiler}, {and} \bibinfo{person}{Damien Stehl{\'e}}.}
  \bibinfo{year}{2018}\natexlab{}.
\newblock \showarticletitle{Crystals-dilithium: A lattice-based digital
  signature scheme}.
\newblock \bibinfo{journal}{\emph{IACR Transactions on Cryptographic Hardware
  and Embedded Systems}} (\bibinfo{year}{2018}), \bibinfo{pages}{238--268}.
\newblock


\bibitem[D’Anvers et~al\mbox{.}(2018)]%
        {d2018saber}
\bibfield{author}{\bibinfo{person}{Jan-Pieter D’Anvers},
  \bibinfo{person}{Angshuman Karmakar}, \bibinfo{person}{Sujoy Sinha~Roy},
  {and} \bibinfo{person}{Frederik Vercauteren}.}
  \bibinfo{year}{2018}\natexlab{}.
\newblock \showarticletitle{Saber: Module-LWR based key exchange, CPA-secure
  encryption and CCA-secure KEM}. In \bibinfo{booktitle}{\emph{Progress in
  Cryptology--AFRICACRYPT 2018}}. Springer, \bibinfo{pages}{282--305}.
\newblock


\bibitem[Espitau et~al\mbox{.}(2022)]%
        {espitau2022mitaka}
\bibfield{author}{\bibinfo{person}{Thomas Espitau},
  \bibinfo{person}{Pierre-Alain Fouque}, \bibinfo{person}{Fran{\c{c}}ois
  G{\'e}rard}, \bibinfo{person}{M{\'e}lissa Rossi}, \bibinfo{person}{Akira
  Takahashi}, {and} \bibinfo{person}{Tibouchi}.}
  \bibinfo{year}{2022}\natexlab{}.
\newblock \showarticletitle{Mitaka: a simpler, parallelizable, maskable variant
  of falcon}. In \bibinfo{booktitle}{\emph{Annual International Conf. on the
  Theory and Applications of Cryptographic Techniques}}. Springer.
\newblock


\bibitem[Fernández-Caramès and Fraga-Lamas(2020)]%
        {8967098}
\bibfield{author}{\bibinfo{person}{Tiago~M. Fernández-Caramès} {and}
  \bibinfo{person}{Paula Fraga-Lamas}.} \bibinfo{year}{2020}\natexlab{}.
\newblock \showarticletitle{Towards Post-Quantum Blockchain: A Review on
  Blockchain Cryptography Resistant to Quantum Computing Attacks}.
\newblock \bibinfo{journal}{\emph{IEEE Access}}  \bibinfo{volume}{8}
  (\bibinfo{year}{2020}), \bibinfo{pages}{21091--21116}.
\newblock
\urldef\tempurl%
\url{https://doi.org/10.1109/ACCESS.2020.2968985}
\showDOI{\tempurl}


\bibitem[Fernández-Caramés(2020)]%
        {8932459}
\bibfield{author}{\bibinfo{person}{Tiago~M. Fernández-Caramés}.}
  \bibinfo{year}{2020}\natexlab{}.
\newblock \showarticletitle{From Pre-Quantum to Post-Quantum IoT Security: A
  Survey on Quantum-Resistant Cryptosystems for the Internet of Things}.
\newblock \bibinfo{journal}{\emph{IEEE Internet of Things Journal}}
  \bibinfo{volume}{7}, \bibinfo{number}{7} (\bibinfo{year}{2020}),
  \bibinfo{pages}{6457--6480}.
\newblock
\urldef\tempurl%
\url{https://doi.org/10.1109/JIOT.2019.2958788}
\showDOI{\tempurl}


\bibitem[Fouque et~al\mbox{.}(2019)]%
        {fouque2019fast}
\bibfield{author}{\bibinfo{person}{PA Fouque}, \bibinfo{person}{J Hoffstein},
  \bibinfo{person}{P Kirchner}, \bibinfo{person}{V Lyubashevsky},
  \bibinfo{person}{T Pornin}, {and} \bibinfo{person}{T Prest}.}
  \bibinfo{year}{2019}\natexlab{}.
\newblock \bibinfo{title}{Fast-fourier lattice-based compact signatures over
  NTRU}.
\newblock
\newblock


\bibitem[Fouque et~al\mbox{.}(2022)]%
        {fouque2022bat}
\bibfield{author}{\bibinfo{person}{Pierre-Alain Fouque}, \bibinfo{person}{Paul
  Kirchner}, \bibinfo{person}{Thomas Pornin}, {and} \bibinfo{person}{Yang Yu}.}
  \bibinfo{year}{2022}\natexlab{}.
\newblock \showarticletitle{Bat: Small and fast kem over ntru lattices}.
\newblock \bibinfo{journal}{\emph{IACR Transactions on Cryptographic Hardware
  and Embedded Systems}} (\bibinfo{year}{2022}), \bibinfo{pages}{240--265}.
\newblock


\bibitem[Galbraith and Vercauteren(2018)]%
        {galbraith2018computational}
\bibfield{author}{\bibinfo{person}{Steven~D Galbraith} {and}
  \bibinfo{person}{Frederik Vercauteren}.} \bibinfo{year}{2018}\natexlab{}.
\newblock \showarticletitle{Computational problems in supersingular elliptic
  curve isogenies}.
\newblock \bibinfo{journal}{\emph{Quantum Information Processing}}
  \bibinfo{volume}{17} (\bibinfo{year}{2018}), \bibinfo{pages}{1--22}.
\newblock


\bibitem[Ganguly and Saxena(2023)]%
        {gangulynew}
\bibfield{author}{\bibinfo{person}{Anindya Ganguly} {and}
  \bibinfo{person}{Nitin Saxena}.} \bibinfo{year}{2023}\natexlab{}.
\newblock \showarticletitle{A New Multivariate Digital-Signature Scheme by
  Mixing Oil-Vinegar with Triangles}.
\newblock  (\bibinfo{year}{2023}).
\newblock


\bibitem[G{\"a}rtner(2023)]%
        {gartner2023ntwe}
\bibfield{author}{\bibinfo{person}{Joel G{\"a}rtner}.}
  \bibinfo{year}{2023}\natexlab{}.
\newblock \showarticletitle{NTWE: A Natural Combination of NTRU and LWE}.
\newblock \bibinfo{journal}{\emph{Cryptology ePrint Archive}}
  (\bibinfo{year}{2023}).
\newblock


\bibitem[Giroti and Malhotra(2022)]%
        {giroti2022quantum}
\bibfield{author}{\bibinfo{person}{Ishika Giroti} {and}
  \bibinfo{person}{Meenakshi Malhotra}.} \bibinfo{year}{2022}\natexlab{}.
\newblock \showarticletitle{Quantum Cryptography: A Pathway to Secure
  Communication}. In \bibinfo{booktitle}{\emph{2022 6th International
  Conference on Computation System and Information Technology for Sustainable
  Solutions (CSITSS)}}. IEEE, \bibinfo{pages}{1--6}.
\newblock


\bibitem[G{\"o}ttert et~al\mbox{.}(2012)]%
        {gottert2012design}
\bibfield{author}{\bibinfo{person}{Norman G{\"o}ttert}, \bibinfo{person}{Thomas
  Feller}, \bibinfo{person}{Michael Schneider}, \bibinfo{person}{Johannes
  Buchmann}, {and} \bibinfo{person}{Sorin Huss}.}
  \bibinfo{year}{2012}\natexlab{}.
\newblock \showarticletitle{On the design of hardware building blocks for
  modern lattice-based encryption schemes}. In
  \bibinfo{booktitle}{\emph{Cryptographic Hardware and Embedded Systems--CHES
  2012}}. Springer.
\newblock


\bibitem[Goy et~al\mbox{.}(2022)]%
        {HQC3}
\bibfield{author}{\bibinfo{person}{Guillaume Goy}, \bibinfo{person}{Antoine
  Loiseau}, {and} \bibinfo{person}{Philippe Gaborit}.}
  \bibinfo{year}{2022}\natexlab{}.
\newblock \showarticletitle{A New Key Recovery Side-Channel Attack On HQC With
  Chosen Ciphertext}. In \bibinfo{booktitle}{\emph{Post-Quantum Cryptography:
  13th International Workshop, PQCrypto 2022}}.
  \bibinfo{publisher}{Springer-Verlag}, \bibinfo{address}{Berlin, Heidelberg},
  \bibinfo{pages}{353–371}.
\newblock
\showISBNx{978-3-031-17233-5}


\bibitem[Grover(1996)]%
        {grover1996fast}
\bibfield{author}{\bibinfo{person}{Lov~K Grover}.}
  \bibinfo{year}{1996}\natexlab{}.
\newblock \showarticletitle{A fast quantum mechanical algorithm for database
  search}. In \bibinfo{booktitle}{\emph{Proceedings of the 28th ACM symp. on
  Theory of comp.}}
\newblock


\bibitem[Guo et~al\mbox{.}(2022a)]%
        {BIKE4}
\bibfield{author}{\bibinfo{person}{Qian Guo}, \bibinfo{person}{Clemens
  Hlauschek}, \bibinfo{person}{Thomas Johansson}, \bibinfo{person}{Norman
  Lahr}, \bibinfo{person}{Alexander Nilsson}, {and} \bibinfo{person}{{Robin
  Leander} Schr{\"o}der}.} \bibinfo{year}{2022}\natexlab{a}.
\newblock \showarticletitle{Don{\textquoteright}t Reject This: Key-Recovery
  Timing Attacks Due to Rejection-Sampling in HQC and BIKE}.
\newblock \bibinfo{journal}{\emph{IACR Trans. on Crypto. HW and Embed. Sys
  (TCHES)}}  \bibinfo{volume}{2022} (\bibinfo{year}{2022}).
\newblock


\bibitem[Guo et~al\mbox{.}(2022b)]%
        {Mc6}
\bibfield{author}{\bibinfo{person}{Qian Guo}, \bibinfo{person}{Andreas
  Johansson}, {and} \bibinfo{person}{Thomas Johansson}.}
  \bibinfo{year}{2022}\natexlab{b}.
\newblock \bibinfo{title}{A Key-Recovery Side-Channel Attack on Classic
  McEliece}.
\newblock \bibinfo{howpublished}{Cryptology ePrint Archive, Paper 2022/514}.
\newblock
\urldef\tempurl%
\url{https://eprint.iacr.org/2022/514}
\showURL{%
\tempurl}
\newblock
\shownote{\url{https://eprint.iacr.org/2022/514}}.


\bibitem[Guo et~al\mbox{.}(2016)]%
        {BIKE1}
\bibfield{author}{\bibinfo{person}{Qian Guo}, \bibinfo{person}{Thomas
  Johansson}, {and} \bibinfo{person}{Paul Stankovski}.}
  \bibinfo{year}{2016}\natexlab{}.
\newblock \showarticletitle{A key recovery attack on MDPC with CCA security
  using decoding errors}. In \bibinfo{booktitle}{\emph{Advances in Cryptology -
  ASIACRYPT 2016}}. \bibinfo{publisher}{Springer}.
\newblock


\bibitem[Hashimoto(2021)]%
        {hashimoto2021recent}
\bibfield{author}{\bibinfo{person}{Yasufumi Hashimoto}.}
  \bibinfo{year}{2021}\natexlab{}.
\newblock \showarticletitle{Recent developments in multivariate public key
  cryptosystems}. In \bibinfo{booktitle}{\emph{International Symposium on
  Mathematics, Quantum Theory, and Cryptography: Proceedings of MQC 2019}}.
  Springer Singapore, \bibinfo{pages}{209--229}.
\newblock


\bibitem[Hasija et~al\mbox{.}(2022)]%
        {9835864}
\bibfield{author}{\bibinfo{person}{Taniya Hasija}, \bibinfo{person}{K.~R.
  Ramkumar}, \bibinfo{person}{Amanpreet Kaur}, \bibinfo{person}{Sudesh Mittal},
  {and} \bibinfo{person}{Bhupendra Singh}.} \bibinfo{year}{2022}\natexlab{}.
\newblock \showarticletitle{A Survey on NIST Selected Third Round Candidates
  for Post Quantum Cryptography}. In \bibinfo{booktitle}{\emph{2022 7th
  International Conference on Communication and Electronics Systems (ICCES)}}.
  \bibinfo{pages}{737--743}.
\newblock


\bibitem[Heinz and P{\"o}ppelmann(2022)]%
        {Kyber15}
\bibfield{author}{\bibinfo{person}{Daniel Heinz} {and} \bibinfo{person}{Thomas
  P{\"o}ppelmann}.} \bibinfo{year}{2022}\natexlab{}.
\newblock \showarticletitle{Combined fault and DPA protection for lattice-based
  cryptography}.
\newblock \bibinfo{journal}{\emph{IEEE Trans. Comput.}} \bibinfo{volume}{72},
  \bibinfo{number}{4} (\bibinfo{year}{2022}).
\newblock


\bibitem[Hekkala et~al\mbox{.}(2023)]%
        {Hekkala2023}
\bibfield{author}{\bibinfo{person}{Julius Hekkala}, \bibinfo{person}{Mari
  Muurman}, \bibinfo{person}{Kimmo Halunen}, {and} \bibinfo{person}{Visa
  Vallivaara}.} \bibinfo{year}{2023}\natexlab{}.
\newblock \showarticletitle{Implementing Post-quantum Cryptography for
  Developers}.
\newblock \bibinfo{journal}{\emph{SN Computer Science}}  \bibinfo{volume}{4}
  (\bibinfo{year}{2023}).
\newblock
Issue 4.
\urldef\tempurl%
\url{https://doi.org/10.1007/s42979-023-01724-1}
\showDOI{\tempurl}


\bibitem[Hoffmann et~al\mbox{.}(2023)]%
        {hoffmann2023polka}
\bibfield{author}{\bibinfo{person}{Cl{\'e}ment Hoffmann},
  \bibinfo{person}{Beno{\^\i}t Libert}, \bibinfo{person}{Charles Momin},
  \bibinfo{person}{Thomas Peters}, {and} \bibinfo{person}{Fran{\c{c}}ois-Xavier
  Standaert}.} \bibinfo{year}{2023}\natexlab{}.
\newblock \showarticletitle{POLKA: Towards Leakage-Resistant Post-quantum
  CCA-Secure Public Key Encryption}. In \bibinfo{booktitle}{\emph{IACR
  International Conference on Public-Key Cryptography}}. Springer,
  \bibinfo{pages}{114--144}.
\newblock


\bibitem[Hoffstein et~al\mbox{.}(1998)]%
        {hoffstein1998ntru}
\bibfield{author}{\bibinfo{person}{Jeffrey Hoffstein}, \bibinfo{person}{Jill
  Pipher}, {and} \bibinfo{person}{Joseph~H Silverman}.}
  \bibinfo{year}{1998}\natexlab{}.
\newblock \showarticletitle{NTRU: A ring-based public key cryptosystem}. In
  \bibinfo{booktitle}{\emph{International algorithmic number theory
  symposium}}. Springer, \bibinfo{pages}{267--288}.
\newblock


\bibitem[Hounkpevi and Djimnaibeye(2023)]%
        {hounkpevieaglesign}
\bibfield{author}{\bibinfo{person}{Abiodoun~Clement Hounkpevi} {and}
  \bibinfo{person}{Sidoine Djimnaibeye}.} \bibinfo{year}{2023}\natexlab{}.
\newblock \showarticletitle{EagleSign: A new post-quantum ElGamal-like
  signature over lattices}.
\newblock  (\bibinfo{year}{2023}).
\newblock


\bibitem[Huang et~al\mbox{.}(2023)]%
        {HQC1}
\bibfield{author}{\bibinfo{person}{Senyang Huang}, \bibinfo{person}{Rui~Qi
  Sim}, \bibinfo{person}{Chitchanok Chuengsatiansup}, {and}
  \bibinfo{person}{Qian Guo}.} \bibinfo{year}{2023}\natexlab{}.
\newblock \bibinfo{title}{Cache-timing attack against HQC}.
\newblock \bibinfo{howpublished}{Cryptol Archive}.
\newblock


\bibitem[H{\"u}lsing et~al\mbox{.}(2013)]%
        {hulsing2013optimal}
\bibfield{author}{\bibinfo{person}{Andreas H{\"u}lsing}, \bibinfo{person}{Lea
  Rausch}, {and} \bibinfo{person}{Johannes Buchmann}.}
  \bibinfo{year}{2013}\natexlab{}.
\newblock \showarticletitle{Optimal parameters for XMSS MT}. In
  \bibinfo{booktitle}{\emph{Security Engineering and Intelligence Informatics:
  CD-ARES 2013 Workshops: MoCrySEn and SeCIHD, Regensburg, Germany, September
  2-6, 2013. Proceedings 8}}. Springer, \bibinfo{pages}{194--208}.
\newblock


\bibitem[Ishai et~al\mbox{.}(2007)]%
        {ishai2007zero}
\bibfield{author}{\bibinfo{person}{Yuval Ishai}, \bibinfo{person}{Eyal
  Kushilevitz}, \bibinfo{person}{Rafail Ostrovsky}, {and} \bibinfo{person}{Amit
  Sahai}.} \bibinfo{year}{2007}\natexlab{}.
\newblock \showarticletitle{Zero-knowledge from secure multiparty computation}.
  In \bibinfo{booktitle}{\emph{Proceedings of the thirty-ninth annual ACM
  symposium on Theory of computing}}. \bibinfo{pages}{21--30}.
\newblock


\bibitem[Jao et~al\mbox{.}(2017)]%
        {jao2017sike}
\bibfield{author}{\bibinfo{person}{David Jao}, \bibinfo{person}{Reza
  Azarderakhsh}, \bibinfo{person}{Matt Campagna}, {et~al\mbox{.}}}
  \bibinfo{year}{2017}\natexlab{}.
\newblock \showarticletitle{SIKE: Supersingular isogeny key encapsulation}.
\newblock  (\bibinfo{year}{2017}).
\newblock


\bibitem[Jao and De~Feo(2011)]%
        {jao2011towards}
\bibfield{author}{\bibinfo{person}{David Jao} {and} \bibinfo{person}{Luca
  De~Feo}.} \bibinfo{year}{2011}\natexlab{}.
\newblock \showarticletitle{Towards quantum-resistant cryptosystems from
  supersingular elliptic curve isogenies}. In
  \bibinfo{booktitle}{\emph{Post-Quantum Cryptography: 4th International
  Workshop, PQCrypto 2011, Taipei, Taiwan, November 29--December 2, 2011.
  Proceedings 4}}. Springer, \bibinfo{pages}{19--34}.
\newblock


\bibitem[Ji et~al\mbox{.}(2023)]%
        {ji2023hi}
\bibfield{author}{\bibinfo{person}{Xinyi Ji}, \bibinfo{person}{Jiankuo Dong},
  \bibinfo{person}{Pinchang Zhang}, \bibinfo{person}{Deng Tonggui},
  \bibinfo{person}{Hua Jiafeng}, {and} \bibinfo{person}{Fu Xiao}.}
  \bibinfo{year}{2023}\natexlab{}.
\newblock \showarticletitle{HI-Kyber: A novel high-performance implementation
  scheme of Kyber based on GPU}.
\newblock \bibinfo{journal}{\emph{Cryptology ePrint Archive}}
  (\bibinfo{year}{2023}).
\newblock


\bibitem[Kales and Zaverucha(2020)]%
        {kales2020improving}
\bibfield{author}{\bibinfo{person}{Daniel Kales} {and} \bibinfo{person}{Greg
  Zaverucha}.} \bibinfo{year}{2020}\natexlab{}.
\newblock \showarticletitle{Improving the performance of the picnic signature
  scheme}.
\newblock \bibinfo{journal}{\emph{Trans on Crypto. HW and Embed. Sys}}
  (\bibinfo{year}{2020}).
\newblock


\bibitem[Karabulut et~al\mbox{.}(2021)]%
        {dilit4}
\bibfield{author}{\bibinfo{person}{Emre Karabulut}, \bibinfo{person}{Erdem
  Alkim}, {and} \bibinfo{person}{Aydin Aysu}.} \bibinfo{year}{2021}\natexlab{}.
\newblock \showarticletitle{Single-trace side-channel attacks on $\omega$-small
  polynomial sampling: With applications to ntru, NTRU prime, and
  CRYSTALS-DILITHIUM}. In \bibinfo{booktitle}{\emph{2021 IEEE International
  Symposium on Hardware Oriented Security and Trust (HOST)}}. IEEE.
\newblock


\bibitem[Karabulut and Aysu(2021)]%
        {Falcon1}
\bibfield{author}{\bibinfo{person}{Emre Karabulut} {and} \bibinfo{person}{Aydin
  Aysu}.} \bibinfo{year}{2021}\natexlab{}.
\newblock \showarticletitle{Falcon down: Breaking falcon post-quantum signature
  scheme through side-channel attacks}. In \bibinfo{booktitle}{\emph{2021 58th
  ACM/IEEE Design Automation Conference (DAC)}}. IEEE,
  \bibinfo{pages}{691--696}.
\newblock


\bibitem[Katz and Lindell(2020)]%
        {katz2020introduction}
\bibfield{author}{\bibinfo{person}{Jonathan Katz} {and} \bibinfo{person}{Yehuda
  Lindell}.} \bibinfo{year}{2020}\natexlab{}.
\newblock \bibinfo{booktitle}{\emph{Introduction to modern cryptography}}.
\newblock \bibinfo{publisher}{CRC press}.
\newblock


\bibitem[Kim and Quisquater(2007)]%
        {SCA4}
\bibfield{author}{\bibinfo{person}{Chong~Hee Kim} {and}
  \bibinfo{person}{Jean-Jacques Quisquater}.} \bibinfo{year}{2007}\natexlab{}.
\newblock \showarticletitle{Faults, injection methods, and fault attacks}.
\newblock \bibinfo{journal}{\emph{IEEE Design \& Test of Computers}}
  \bibinfo{volume}{24}, \bibinfo{number}{6} (\bibinfo{year}{2007}).
\newblock


\bibitem[Kim and Park(2022)]%
        {kim2022ntru+}
\bibfield{author}{\bibinfo{person}{Jonghyun Kim} {and}
  \bibinfo{person}{Jong~Hwan Park}.} \bibinfo{year}{2022}\natexlab{}.
\newblock \showarticletitle{NTRU+: Compact Construction of NTRU Using Simple
  Encoding Method}.
\newblock \bibinfo{journal}{\emph{Cryptol Archive}} (\bibinfo{year}{2022}).
\newblock


\bibitem[Knuth(1976)]%
        {knuth1976complexity}
\bibfield{author}{\bibinfo{person}{Donald Knuth}.}
  \bibinfo{year}{1976}\natexlab{}.
\newblock \showarticletitle{The complexity of nonuniform random number
  generation}.
\newblock \bibinfo{journal}{\emph{Algorithm and Complexity, New Directions and
  Results}} (\bibinfo{year}{1976}).
\newblock


\bibitem[Kudinov et~al\mbox{.}(2022)]%
        {kudinov2022sphincs+}
\bibfield{author}{\bibinfo{person}{Mikhail Kudinov}, \bibinfo{person}{Andreas
  H{\"u}lsing}, {and} \bibinfo{person}{Ronen}.}
  \bibinfo{year}{2022}\natexlab{}.
\newblock \showarticletitle{SPHINCS+ C: compressing Sphincs+ with (almost) no
  cost}.
\newblock \bibinfo{journal}{\emph{Cryptol Arch.}} (\bibinfo{year}{2022}).
\newblock


\bibitem[Kumar et~al\mbox{.}(2022)]%
        {doi.org/10.1002/spy2.200}
\bibfield{author}{\bibinfo{person}{Adarsh Kumar}, \bibinfo{person}{Carlo
  Ottaviani}, \bibinfo{person}{Sukhpal~Singh Gill}, {and}
  \bibinfo{person}{Rajkumar Buyya}.} \bibinfo{year}{2022}\natexlab{}.
\newblock \showarticletitle{Securing the future internet of things with
  post-quantum cryptography}.
\newblock \bibinfo{journal}{\emph{SECURITY AND PRIVACY}} \bibinfo{volume}{5},
  \bibinfo{number}{2} (\bibinfo{year}{2022}), \bibinfo{pages}{e200}.
\newblock


\bibitem[Kumar(2022)]%
        {KUMAR2022100242}
\bibfield{author}{\bibinfo{person}{Manish Kumar}.}
  \bibinfo{year}{2022}\natexlab{}.
\newblock \showarticletitle{Post-quantum cryptography Algorithm's
  standardization and performance analysis}.
\newblock \bibinfo{journal}{\emph{Array}}  \bibinfo{volume}{15}
  (\bibinfo{year}{2022}), \bibinfo{pages}{100242}.
\newblock


\bibitem[Kumar and Pattnaik(2020)]%
        {9286147}
\bibfield{author}{\bibinfo{person}{Manoj Kumar} {and} \bibinfo{person}{Pratap
  Pattnaik}.} \bibinfo{year}{2020}\natexlab{}.
\newblock \showarticletitle{Post Quantum Cryptography(PQC) - An overview:
  (Invited Paper)}. In \bibinfo{booktitle}{\emph{2020 IEEE High Performance
  Extreme Computing Conference (HPEC)}}. \bibinfo{pages}{1--9}.
\newblock
\urldef\tempurl%
\url{https://doi.org/10.1109/HPEC43674.2020.9286147}
\showDOI{\tempurl}


\bibitem[Kumari et~al\mbox{.}(2022)]%
        {doi.org/10.1002/spe.3121}
\bibfield{author}{\bibinfo{person}{Swati Kumari}, \bibinfo{person}{Maninder
  Singh}, \bibinfo{person}{Raman Singh}, {and} \bibinfo{person}{Hitesh
  Tewari}.} \bibinfo{year}{2022}\natexlab{}.
\newblock \showarticletitle{Post-quantum cryptography techniques for secure
  communication in resource-constrained Internet of Things devices: A
  comprehensive survey}.
\newblock \bibinfo{journal}{\emph{Software: Practice and Experience}}
  \bibinfo{volume}{52}, \bibinfo{number}{10} (\bibinfo{year}{2022}),
  \bibinfo{pages}{2047--2076}.
\newblock


\bibitem[Kundu et~al\mbox{.}(2020)]%
        {kundu2020post}
\bibfield{author}{\bibinfo{person}{Nibedita Kundu},
  \bibinfo{person}{Sumit~Kumar Debnath}, \bibinfo{person}{Dheerendra Mishra},
  {and} \bibinfo{person}{Tanmay Choudhury}.} \bibinfo{year}{2020}\natexlab{}.
\newblock \showarticletitle{Post-quantum digital signature scheme based on
  multivariate cubic problem}.
\newblock \bibinfo{journal}{\emph{Journal of Information Security and
  Applications}}  \bibinfo{volume}{53} (\bibinfo{year}{2020}),
  \bibinfo{pages}{102512}.
\newblock


\bibitem[Li et~al\mbox{.}(2022)]%
        {li2022hash}
\bibfield{author}{\bibinfo{person}{Lingyun Li}, \bibinfo{person}{Xianhui Lu},
  {and} \bibinfo{person}{Kunpeng Wang}.} \bibinfo{year}{2022}\natexlab{}.
\newblock \showarticletitle{Hash-based signature revisited}.
\newblock \bibinfo{journal}{\emph{Cybersecurity}} \bibinfo{volume}{5},
  \bibinfo{number}{1} (\bibinfo{year}{2022}), \bibinfo{pages}{1--26}.
\newblock


\bibitem[Li et~al\mbox{.}(2023)]%
        {li2023high}
\bibfield{author}{\bibinfo{person}{Xiang Li}, \bibinfo{person}{Jiahao Lu},
  \bibinfo{person}{Dongsheng Liu}, \bibinfo{person}{Aobo Li},
  \bibinfo{person}{Shuo Yang}, {and} \bibinfo{person}{Tianze Huang}.}
  \bibinfo{year}{2023}\natexlab{}.
\newblock \showarticletitle{A High Speed Post-Quantum Crypto-Processor for
  Crystals-Dilithium}.
\newblock \bibinfo{journal}{\emph{IEEE Transactions on Circuits and Systems II:
  Express Briefs}} (\bibinfo{year}{2023}).
\newblock


\bibitem[Lindner and Peikert(2011)]%
        {lindner2011better}
\bibfield{author}{\bibinfo{person}{Richard Lindner} {and}
  \bibinfo{person}{Chris Peikert}.} \bibinfo{year}{2011}\natexlab{}.
\newblock \showarticletitle{Better key sizes (and attacks) for LWE-based
  encryption}. In \bibinfo{booktitle}{\emph{Topics in Cryptology--CT-RSA 2011:
  The Cryptographers’ Track at the RSA Conference 2011, San Francisco, CA,
  USA, February 14-18, 2011. Proceedings}}. Springer,
  \bibinfo{pages}{319--339}.
\newblock


\bibitem[Liu and Zhao(2023)]%
        {liu2023emle}
\bibfield{author}{\bibinfo{person}{D Liu} {and} \bibinfo{person}{R~K Zhao}.}
  \bibinfo{year}{2023}\natexlab{}.
\newblock \showarticletitle{eMLE-Sig 2.0: A Signature Scheme based on Embedded
  Multilayer Equations with Heavy Layer Randomization}.
\newblock  (\bibinfo{year}{2023}).
\newblock


\bibitem[Liu et~al\mbox{.}(2022)]%
        {liu2022new}
\bibfield{author}{\bibinfo{person}{Fukang Liu}, \bibinfo{person}{Willi Meier},
  \bibinfo{person}{Santanu Sarkar}, {and} \bibinfo{person}{Takanori Isobe}.}
  \bibinfo{year}{2022}\natexlab{}.
\newblock \showarticletitle{New low-memory algebraic attacks on LowMC in the
  picnic setting}.
\newblock \bibinfo{journal}{\emph{IACR Transactions on Symmetric Cryptology}}
  (\bibinfo{year}{2022}), \bibinfo{pages}{102--122}.
\newblock


\bibitem[Lyubashevsky and Seiler(2019)]%
        {lyubashevsky2019nttru}
\bibfield{author}{\bibinfo{person}{Vadim Lyubashevsky} {and}
  \bibinfo{person}{Gregor Seiler}.} \bibinfo{year}{2019}\natexlab{}.
\newblock \showarticletitle{NTTRU: truly fast NTRU using NTT}.
\newblock \bibinfo{journal}{\emph{Cryptology ePrint Archive}}
  (\bibinfo{year}{2019}).
\newblock


\bibitem[Malina et~al\mbox{.}(2021)]%
        {9363165}
\bibfield{author}{\bibinfo{person}{Lukas Malina}, \bibinfo{person}{Petr
  Dzurenda}, {et~al\mbox{.}}} \bibinfo{year}{2021}\natexlab{}.
\newblock \showarticletitle{Post-Quantum Era Privacy Protection for Intelligent
  Infrastructures}.
\newblock \bibinfo{journal}{\emph{IEEE Access}} (\bibinfo{year}{2021}).
\newblock


\bibitem[Malina et~al\mbox{.}(2018)]%
        {malina2018feasibility}
\bibfield{author}{\bibinfo{person}{Lukas Malina}, \bibinfo{person}{Lucie
  Popelova}, \bibinfo{person}{Petr Dzurenda}, \bibinfo{person}{Jan Hajny},
  {and} \bibinfo{person}{Zdenek Martinasek}.} \bibinfo{year}{2018}\natexlab{}.
\newblock \showarticletitle{On feasibility of post-quantum cryptography on
  small devices}.
\newblock \bibinfo{journal}{\emph{IFAC-PapersOnLine}} \bibinfo{volume}{51},
  \bibinfo{number}{6} (\bibinfo{year}{2018}), \bibinfo{pages}{462--467}.
\newblock


\bibitem[Marsaglia and Tsang(2000)]%
        {marsaglia2000ziggurat}
\bibfield{author}{\bibinfo{person}{George Marsaglia} {and}
  \bibinfo{person}{Wai~Wan Tsang}.} \bibinfo{year}{2000}\natexlab{}.
\newblock \showarticletitle{The ziggurat method for generating random
  variables}.
\newblock \bibinfo{journal}{\emph{Journal of statistical software}}
  \bibinfo{volume}{5} (\bibinfo{year}{2000}), \bibinfo{pages}{1--7}.
\newblock


\bibitem[Marzougui et~al\mbox{.}(2022)]%
        {dilit6}
\bibfield{author}{\bibinfo{person}{Soundes Marzougui}, \bibinfo{person}{Vincent
  Ulitzsch}, \bibinfo{person}{Mehdi Tibouchi}, {and}
  \bibinfo{person}{Jean-Pierre Seifert}.} \bibinfo{year}{2022}\natexlab{}.
\newblock \showarticletitle{Profiling side-channel attacks on Dilithium: A
  small bit-fiddling leak breaks it all}.
\newblock \bibinfo{journal}{\emph{Cryptology ePrint Archive}}
  (\bibinfo{year}{2022}).
\newblock


\bibitem[Matsumoto and Imai(1988)]%
        {matsumoto1988public}
\bibfield{author}{\bibinfo{person}{Tsutomu Matsumoto} {and}
  \bibinfo{person}{Hideki Imai}.} \bibinfo{year}{1988}\natexlab{}.
\newblock \showarticletitle{Public quadratic polynomial-tuples for efficient
  signature-verification and message-encryption}. In
  \bibinfo{booktitle}{\emph{Advances in Cryptology—EUROCRYPT’88}}.
  Springer.
\newblock


\bibitem[McCarthy et~al\mbox{.}(2019)]%
        {Falcon4}
\bibfield{author}{\bibinfo{person}{Sarah McCarthy}, \bibinfo{person}{James
  Howe}, \bibinfo{person}{Neil Smyth}, \bibinfo{person}{S{\'e}amus Brannigan},
  {and} \bibinfo{person}{M{\'a}ire O’Neill}.}
  \bibinfo{year}{2019}\natexlab{}.
\newblock \showarticletitle{BEARZ attack FALCON: implementation attacks with
  countermeasures on the FALCON signature scheme}.
\newblock \bibinfo{journal}{\emph{Cryptology ePrint Archive}}
  (\bibinfo{year}{2019}).
\newblock


\bibitem[McEliece(1978)]%
        {mceliece1978public}
\bibfield{author}{\bibinfo{person}{Robert~J McEliece}.}
  \bibinfo{year}{1978}\natexlab{}.
\newblock \showarticletitle{A public-key cryptosystem based on algebraic}.
\newblock \bibinfo{journal}{\emph{Coding Thv}}  \bibinfo{volume}{4244}
  (\bibinfo{year}{1978}), \bibinfo{pages}{114--116}.
\newblock


\bibitem[McGrew et~al\mbox{.}(2019)]%
        {mcgrew2019rfc}
\bibfield{author}{\bibinfo{person}{David McGrew}, \bibinfo{person}{Michael
  Curcio}, {and} \bibinfo{person}{Scott Fluhrer}.}
  \bibinfo{year}{2019}\natexlab{}.
\newblock \bibinfo{title}{RFC 8554: Leighton-Micali hash-based signatures}.
\newblock
\newblock


\bibitem[Mehic et~al\mbox{.}(2020)]%
        {mehic2020quantum}
\bibfield{author}{\bibinfo{person}{Miralem Mehic}, \bibinfo{person}{Marcin
  Niemiec}, \bibinfo{person}{Stefan Rass}, \bibinfo{person}{Jiajun Ma},
  \bibinfo{person}{Momtchil Peev}, \bibinfo{person}{Alejandro Aguado},
  \bibinfo{person}{Vicente Martin}, \bibinfo{person}{Stefan Schauer},
  \bibinfo{person}{Andreas Poppe}, \bibinfo{person}{Christoph Pacher},
  {et~al\mbox{.}}} \bibinfo{year}{2020}\natexlab{}.
\newblock \showarticletitle{Quantum key distribution: a networking
  perspective}.
\newblock \bibinfo{journal}{\emph{ACM Computing Surveys (CSUR)}}
  \bibinfo{volume}{53}, \bibinfo{number}{5} (\bibinfo{year}{2020}),
  \bibinfo{pages}{1--41}.
\newblock


\bibitem[Melchor et~al\mbox{.}(2018)]%
        {melchor2018hamming}
\bibfield{author}{\bibinfo{person}{Carlos~Aguilar Melchor},
  \bibinfo{person}{Nicolas Aragon}, \bibinfo{person}{Slim Bettaieb},
  \bibinfo{person}{Lo{\i}c Bidoux}, \bibinfo{person}{Olivier Blazy},
  \bibinfo{person}{Jean-Christophe Deneuville}, \bibinfo{person}{Philippe
  Gaborit}, \bibinfo{person}{Edoardo Persichetti}, \bibinfo{person}{Gilles
  Z{\'e}mor}, {and} \bibinfo{person}{IC Bourges}.}
  \bibinfo{year}{2018}\natexlab{}.
\newblock \showarticletitle{Hamming quasi-cyclic (HQC)}.
\newblock \bibinfo{journal}{\emph{NIST PQC Round}} \bibinfo{volume}{2},
  \bibinfo{number}{4} (\bibinfo{year}{2018}), \bibinfo{pages}{13}.
\newblock


\bibitem[Micciancio and Goldwasser(2002)]%
        {micciancio2002complexity}
\bibfield{author}{\bibinfo{person}{Daniele Micciancio} {and}
  \bibinfo{person}{Shafi Goldwasser}.} \bibinfo{year}{2002}\natexlab{}.
\newblock \bibinfo{booktitle}{\emph{Complexity of lattice problems: a
  cryptographic perspective}}.
\newblock \bibinfo{publisher}{Springer Science \& Business Media}.
\newblock


\bibitem[Moody et~al\mbox{.}(2020)]%
        {moody2020status}
\bibfield{author}{\bibinfo{person}{Dustin Moody}, \bibinfo{person}{Gorjan
  Alagic}, \bibinfo{person}{Daniel~C Apon}, \bibinfo{person}{David~A Cooper},
  \bibinfo{person}{Quynh~H Dang}, \bibinfo{person}{John~M Kelsey},
  \bibinfo{person}{Yi-Kai Liu}, \bibinfo{person}{Carl~A Miller},
  \bibinfo{person}{Rene~C Peralta}, \bibinfo{person}{Ray~A Perlner},
  {et~al\mbox{.}}} \bibinfo{year}{2020}\natexlab{}.
\newblock \showarticletitle{Status report on the second round of the NIST
  post-quantum cryptography standardization process}.
\newblock  (\bibinfo{year}{2020}).
\newblock


\bibitem[Moraitis et~al\mbox{.}(2023)]%
        {moraitis2023securing}
\bibfield{author}{\bibinfo{person}{Michail Moraitis}, \bibinfo{person}{Yanning
  Ji}, \bibinfo{person}{Martin Brisfors}, \bibinfo{person}{Elena Dubrova},
  \bibinfo{person}{Niklas Lindskog}, {et~al\mbox{.}}}
  \bibinfo{year}{2023}\natexlab{}.
\newblock \showarticletitle{Securing CRYSTALS-Kyber in FPGA Using Duplication
  and Clock Randomization}.
\newblock \bibinfo{journal}{\emph{IEEE Design \& Test}} (\bibinfo{year}{2023}).
\newblock


\bibitem[Mosca and Piani(2021)]%
        {mosca2021quantum}
\bibfield{author}{\bibinfo{person}{Michele Mosca} {and} \bibinfo{person}{Marco
  Piani}.} \bibinfo{year}{2021}\natexlab{}.
\newblock \showarticletitle{Quantum threat timeline report 2020}.
\newblock \bibinfo{journal}{\emph{Global Risk Insitute.}}
  (\bibinfo{year}{2021}).
\newblock


\bibitem[Nejatollahi et~al\mbox{.}(2019)]%
        {nejatollahi2019post}
\bibfield{author}{\bibinfo{person}{Hamid Nejatollahi}, \bibinfo{person}{Nikil
  Dutt}, \bibinfo{person}{Sandip Ray}, \bibinfo{person}{Francesco Regazzoni},
  \bibinfo{person}{Indranil Banerjee}, {and} \bibinfo{person}{Rosario
  Cammarota}.} \bibinfo{year}{2019}\natexlab{}.
\newblock \showarticletitle{Post-quantum lattice-based cryptography
  implementations: A survey}.
\newblock \bibinfo{journal}{\emph{ACM Computing Surveys (CSUR)}}
  \bibinfo{volume}{51}, \bibinfo{number}{6} (\bibinfo{year}{2019}),
  \bibinfo{pages}{1--41}.
\newblock


\bibitem[Ni et~al\mbox{.}(2023)]%
        {ni2023hpka}
\bibfield{author}{\bibinfo{person}{Ziying Ni} {et~al\mbox{.}}}
  \bibinfo{year}{2023}\natexlab{}.
\newblock \showarticletitle{HPKA: A High-Performance CRYSTALS-Kyber Accelerator
  Exploring Efficient Pipelining}.
\newblock \bibinfo{journal}{\emph{IEEE Trans. Comput.}} (\bibinfo{year}{2023}).
\newblock


\bibitem[P~C et~al\mbox{.}(2022)]%
        {9787987}
\bibfield{author}{\bibinfo{person}{Sajimon P~C}, \bibinfo{person}{Kurunandan
  Jain}, {and} \bibinfo{person}{Prabhakar Krishnan}.}
  \bibinfo{year}{2022}\natexlab{}.
\newblock \showarticletitle{Analysis of Post-Quantum Cryptography for Internet
  of Things}. In \bibinfo{booktitle}{\emph{2022 6th International Conference on
  Intelligent Computing and Control Systems (ICICCS)}}.
  \bibinfo{pages}{387--394}.
\newblock
\urldef\tempurl%
\url{https://doi.org/10.1109/ICICCS53718.2022.9787987}
\showDOI{\tempurl}


\bibitem[Patarin(1996)]%
        {patarin1996hidden}
\bibfield{author}{\bibinfo{person}{Jacques Patarin}.}
  \bibinfo{year}{1996}\natexlab{}.
\newblock \showarticletitle{Hidden fields equations (HFE) and isomorphisms of
  polynomials (IP): Two new families of asymmetric algorithms}. In
  \bibinfo{booktitle}{\emph{International Conference on the Theory and
  Applications of Cryptographic Techniques}}. Springer,
  \bibinfo{pages}{33--48}.
\newblock


\bibitem[Patarin(1997)]%
        {patarin1997oil}
\bibfield{author}{\bibinfo{person}{Jacques Patarin}.}
  \bibinfo{year}{1997}\natexlab{}.
\newblock \showarticletitle{The oil and vinegar signature scheme}. In
  \bibinfo{booktitle}{\emph{Presented at the Dagstuhl Workshop on Cryptography
  September 1997}}.
\newblock


\bibitem[Peikert et~al\mbox{.}(2016)]%
        {peikert2016decade}
\bibfield{author}{\bibinfo{person}{Chris Peikert} {et~al\mbox{.}}}
  \bibinfo{year}{2016}\natexlab{}.
\newblock \showarticletitle{A decade of lattice cryptography}.
\newblock \bibinfo{journal}{\emph{Foundations and trends{\textregistered} in
  theoretical computer science}} \bibinfo{volume}{10}, \bibinfo{number}{4}
  (\bibinfo{year}{2016}), \bibinfo{pages}{283--424}.
\newblock


\bibitem[Qassim et~al\mbox{.}(2017)]%
        {qassim2017post}
\bibfield{author}{\bibinfo{person}{Yousef Qassim},
  \bibinfo{person}{Mario~Edgardo Maga{\~n}a}, {and} \bibinfo{person}{Attila
  Yavuz}.} \bibinfo{year}{2017}\natexlab{}.
\newblock \showarticletitle{Post-quantum hybrid security mechanism for MIMO
  systems}. In \bibinfo{booktitle}{\emph{2017 International Conference on
  Computing, Networking and Communications (ICNC)}}. IEEE,
  \bibinfo{pages}{684--689}.
\newblock


\bibitem[Raavi et~al\mbox{.}(2021)]%
        {10.1007/978-3-030-78375-4_17}
\bibfield{author}{\bibinfo{person}{Manohar Raavi}, \bibinfo{person}{Simeon
  Wuthier}, {and} \bibinfo{person}{Pranav Chandramouli}.}
  \bibinfo{year}{2021}\natexlab{}.
\newblock \showarticletitle{Security Comparisons and Performance Analyses of
  Post-quantum Signature Algorithms}. In \bibinfo{booktitle}{\emph{Applied
  Cryptography and Network Security}}. \bibinfo{publisher}{Springer
  International Publishing}.
\newblock
\showISBNx{978-3-030-78375-4}


\bibitem[Randolph and Diehl(2020)]%
        {SCA2}
\bibfield{author}{\bibinfo{person}{Mark Randolph} {and}
  \bibinfo{person}{William Diehl}.} \bibinfo{year}{2020}\natexlab{}.
\newblock \showarticletitle{Power side-channel attack analysis: A review of 20
  years of study for the layman}.
\newblock \bibinfo{journal}{\emph{Crypto.}} (\bibinfo{year}{2020}).
\newblock


\bibitem[Ravi et~al\mbox{.}(2022)]%
        {Kyber10}
\bibfield{author}{\bibinfo{person}{Prasanna Ravi}, \bibinfo{person}{Anupam
  Chattopadhyay}, \bibinfo{person}{Jan~Pieter D’Anvers}, {and}
  \bibinfo{person}{Anubhab Baksi}.} \bibinfo{year}{2022}\natexlab{}.
\newblock \showarticletitle{Side-channel and fault-injection attacks over
  lattice-based post-quantum schemes (Kyber, Dilithium): Survey and new
  results}.
\newblock \bibinfo{journal}{\emph{ACM Transactions on Embedded Computing
  Systems}} (\bibinfo{year}{2022}).
\newblock


\bibitem[Ravi and Howe(2021)]%
        {ravi2021lattice}
\bibfield{author}{\bibinfo{person}{Prasanna Ravi} {and}
  \bibinfo{person}{Howe}.} \bibinfo{year}{2021}\natexlab{}.
\newblock \showarticletitle{Lattice-based key-sharing schemes: A survey}.
\newblock \bibinfo{journal}{\emph{ACM Computing Surveys (CSUR)}}
  \bibinfo{volume}{54}, \bibinfo{number}{1} (\bibinfo{year}{2021}),
  \bibinfo{pages}{1--39}.
\newblock


\bibitem[Ravi et~al\mbox{.}(2018)]%
        {dilit1}
\bibfield{author}{\bibinfo{person}{Prasanna Ravi},
  \bibinfo{person}{Mahabir~Prasad Jhanwar}, \bibinfo{person}{James Howe},
  \bibinfo{person}{Anupam Chattopadhyay}, {and} \bibinfo{person}{Shivam
  Bhasin}.} \bibinfo{year}{2018}\natexlab{}.
\newblock \showarticletitle{Side-channel assisted existential forgery attack on
  Dilithium-a NIST PQC candidate}.
\newblock \bibinfo{journal}{\emph{Cryptology ePrint Archive}}
  (\bibinfo{year}{2018}).
\newblock


\bibitem[Richter-Brockmann et~al\mbox{.}(2021)]%
        {richter2021racing}
\bibfield{author}{\bibinfo{person}{Jan Richter-Brockmann},
  \bibinfo{person}{Ming-Shing Chen}, \bibinfo{person}{Santosh Ghosh}, {and}
  \bibinfo{person}{Tim G{\"u}neysu}.} \bibinfo{year}{2021}\natexlab{}.
\newblock \showarticletitle{Racing BIKE: Improved polynomial multiplication and
  inversion in hardware}.
\newblock \bibinfo{journal}{\emph{Cryptology ePrint Archive}}
  (\bibinfo{year}{2021}).
\newblock


\bibitem[Rodriguez(2023)]%
        {rodriguez2023hppc}
\bibfield{author}{\bibinfo{person}{Borja~Gomez Rodriguez}.}
  \bibinfo{year}{2023}\natexlab{}.
\newblock \showarticletitle{HPPC: Hidden Product of Polynomial Composition}.
\newblock \bibinfo{journal}{\emph{Cryptology ePrint Archive}}
  (\bibinfo{year}{2023}).
\newblock


\bibitem[Rodriguez et~al\mbox{.}(2023)]%
        {Kyber8}
\bibfield{author}{\bibinfo{person}{Rafael~Carrera Rodriguez},
  \bibinfo{person}{Florent Bruguier}, \bibinfo{person}{Emanuele Valea}, {and}
  \bibinfo{person}{Pascal Benoit}.} \bibinfo{year}{2023}\natexlab{}.
\newblock \showarticletitle{Correlation electromagnetic analysis on an FPGA
  implementation of CRYSTALS-Kyber}. In \bibinfo{booktitle}{\emph{2023 18th
  Conference on Ph. D Research in Microelectronics and Electronics (PRIME)}}.
  IEEE, \bibinfo{pages}{217--220}.
\newblock


\bibitem[Roy and Kalita(2019)]%
        {survey24}
\bibfield{author}{\bibinfo{person}{Kumar~Sekhar Roy} {and}
  \bibinfo{person}{Hemanta~Kumar Kalita}.} \bibinfo{year}{2019}\natexlab{}.
\newblock \showarticletitle{A Survey on Post-Quantum Cryptography for
  Constrained Devices}. In \bibinfo{booktitle}{\emph{International Journal of
  Applied Engineering Research}}, Vol.~\bibinfo{volume}{14}.
  \bibinfo{pages}{2608--2615}.
\newblock


\bibitem[{\c{S}}ahin and Akleylek(2023)]%
        {csahin2023survey}
\bibfield{author}{\bibinfo{person}{Meryem~Soysald{\i} {\c{S}}ahin} {and}
  \bibinfo{person}{Sedat Akleylek}.} \bibinfo{year}{2023}\natexlab{}.
\newblock \showarticletitle{A survey of quantum secure group signature schemes:
  Lattice-based approach}.
\newblock \bibinfo{journal}{\emph{Journal of Information Security and
  Applications}}  \bibinfo{volume}{73} (\bibinfo{year}{2023}),
  \bibinfo{pages}{103432}.
\newblock


\bibitem[Sarker et~al\mbox{.}(2022a)]%
        {Kyber14}
\bibfield{author}{\bibinfo{person}{Ausmita Sarker},
  \bibinfo{person}{Alvaro~Cintas Canto}, {and}
  \bibinfo{person}{Mehran~Mozaffari Kermani}.}
  \bibinfo{year}{2022}\natexlab{a}.
\newblock \showarticletitle{Error Detection Architectures for Hardware/Software
  Co-design Approaches of Number-Theoretic Transform}.
\newblock \bibinfo{journal}{\emph{IEEE Transactions on Computer-Aided Design of
  Integrated Circuits and Systems}} (\bibinfo{year}{2022}).
\newblock


\bibitem[Sarker et~al\mbox{.}(2022b)]%
        {Falcon5}
\bibfield{author}{\bibinfo{person}{Ausmita Sarker},
  \bibinfo{person}{Mehran~Mozaffari Kermani}, {and} \bibinfo{person}{Reza
  Azarderakhsh}.} \bibinfo{year}{2022}\natexlab{b}.
\newblock \showarticletitle{Efficient Error Detection Architectures for
  Postquantum Signature Falcon’s Sampler and KEM SABER}.
\newblock \bibinfo{journal}{\emph{IEEE Transactions on Very Large Scale
  Integration (VLSI) Systems}} \bibinfo{volume}{30}, \bibinfo{number}{6}
  (\bibinfo{year}{2022}), \bibinfo{pages}{794--802}.
\newblock


\bibitem[Sayakkara et~al\mbox{.}(2019)]%
        {SCA3}
\bibfield{author}{\bibinfo{person}{Asanka Sayakkara}, \bibinfo{person}{Nhien-An
  Le-Khac}, {and} \bibinfo{person}{Mark Scanlon}.}
  \bibinfo{year}{2019}\natexlab{}.
\newblock \showarticletitle{A survey of electromagnetic side-channel attacks
  and discussion on their case-progressing potential for digital forensics}.
\newblock \bibinfo{journal}{\emph{Digital Investigation}}  \bibinfo{volume}{29}
  (\bibinfo{year}{2019}), \bibinfo{pages}{43--54}.
\newblock


\bibitem[Schamberger et~al\mbox{.}(2020)]%
        {HQC2}
\bibfield{author}{\bibinfo{person}{Thomas Schamberger}, \bibinfo{person}{Julian
  Renner}, \bibinfo{person}{Georg Sigl}, {and} \bibinfo{person}{Antonia
  Wachter-Zeh}.} \bibinfo{year}{2020}\natexlab{}.
\newblock \showarticletitle{A Power Side-Channel Attack on the CCA2-Secure HQC
  KEM}. In \bibinfo{booktitle}{\emph{Smart Card Research and Advanced
  Applications}}. \bibinfo{publisher}{Springer-Verlag},
  \bibinfo{pages}{119–134}.
\newblock
\showISBNx{978-3-030-68486-0}


\bibitem[Semaev and Submitter(2023)]%
        {semaevdigital}
\bibfield{author}{\bibinfo{person}{Igor Semaev} {and}
  \bibinfo{person}{Auxiliary Submitter}.} \bibinfo{year}{2023}\natexlab{}.
\newblock \showarticletitle{DIGITAL SIGNATURE ALGORITHMS EHTV3 AND EHTV4
  SUBMISSION TO NIST PQC}.
\newblock  (\bibinfo{year}{2023}).
\newblock


\bibitem[Shim(2022)]%
        {9646494}
\bibfield{author}{\bibinfo{person}{Kyung-Ah Shim}.}
  \bibinfo{year}{2022}\natexlab{}.
\newblock \showarticletitle{A Survey on Post-Quantum Public-Key Signature
  Schemes for Secure Vehicular Communications}.
\newblock \bibinfo{journal}{\emph{IEEE Transactions on Intelligent
  Transportation Systems}} \bibinfo{volume}{23}, \bibinfo{number}{9}
  (\bibinfo{year}{2022}), \bibinfo{pages}{14025--14042}.
\newblock
\urldef\tempurl%
\url{https://doi.org/10.1109/TITS.2021.3131668}
\showDOI{\tempurl}


\bibitem[Shor(1994)]%
        {shor1994algorithms}
\bibfield{author}{\bibinfo{person}{Peter~W Shor}.}
  \bibinfo{year}{1994}\natexlab{}.
\newblock \showarticletitle{Algorithms for quantum computation: discrete
  logarithms and factoring}. In \bibinfo{booktitle}{\emph{35th annual symp. on
  found. of CS}}. Ieee.
\newblock


\bibitem[Shoufan et~al\mbox{.}(2009)]%
        {Mc2}
\bibfield{author}{\bibinfo{person}{Abdulhadi Shoufan}, \bibinfo{person}{Falko
  Strenzke}, \bibinfo{person}{H.~Gregor Molter}, {and} \bibinfo{person}{Marc
  Stöttinger}.} \bibinfo{year}{2009}\natexlab{}.
\newblock \showarticletitle{A Timing Attack against Patterson Algorithm in the
  McEliece PKC}, Vol.~\bibinfo{volume}{5984}. \bibinfo{pages}{161--175}.
\newblock
\showISBNx{978-3-642-14422-6}
\urldef\tempurl%
\url{https://doi.org/10.1007/978-3-642-14423-3_12}
\showDOI{\tempurl}


\bibitem[Sim et~al\mbox{.}(2019)]%
        {BIKE5}
\bibfield{author}{\bibinfo{person}{Bo-Yeon Sim}, \bibinfo{person}{Jihoon Kwon},
  \bibinfo{person}{Kyu Choi}, \bibinfo{person}{Jihoon Cho},
  \bibinfo{person}{Aesun Park}, {and} \bibinfo{person}{Dong-Guk Han}.}
  \bibinfo{year}{2019}\natexlab{}.
\newblock \showarticletitle{Novel Side-Channel Attacks on Quasi-Cyclic
  Code-Based Cryptography}.
\newblock \bibinfo{journal}{\emph{IACR Transactions on Cryptographic Hardware
  and Embedded Systems}} (\bibinfo{date}{08} \bibinfo{year}{2019}),
  \bibinfo{pages}{180--212}.
\newblock


\bibitem[Sivasubramanian(2020)]%
        {sivasubramanian2020comparative}
\bibfield{author}{\bibinfo{person}{Kimsukha~Selvi Sivasubramanian}.}
  \bibinfo{year}{2020}\natexlab{}.
\newblock \bibinfo{title}{A comparative analysis of Post-Quantum Hash-based
  Signature Algorithm}.
\newblock
\newblock


\bibitem[Smith-Tone(2021)]%
        {smith2021new}
\bibfield{author}{\bibinfo{person}{Daniel Smith-Tone}.}
  \bibinfo{year}{2021}\natexlab{}.
\newblock \showarticletitle{New practical multivariate signatures from a
  nonlinear modifier}. In \bibinfo{booktitle}{\emph{Post-Quantum Cryptography:
  12th International Workshop, PQCrypto 2021, Daejeon, South Korea, July
  20--22, 2021, Proceedings 12}}. Springer, \bibinfo{pages}{79--97}.
\newblock


\bibitem[Spreitzer et~al\mbox{.}(2017)]%
        {SCA1}
\bibfield{author}{\bibinfo{person}{Raphael Spreitzer},
  \bibinfo{person}{Veelasha Moonsamy}, \bibinfo{person}{Thomas Korak}, {and}
  \bibinfo{person}{Stefan Mangard}.} \bibinfo{year}{2017}\natexlab{}.
\newblock \showarticletitle{Systematic classification of side-channel attacks:
  A case study for mobile devices}.
\newblock \bibinfo{journal}{\emph{IEEE communications surveys \& tutorials}}
  \bibinfo{volume}{20}, \bibinfo{number}{1} (\bibinfo{year}{2017}),
  \bibinfo{pages}{465--488}.
\newblock


\bibitem[Srivastava et~al\mbox{.}(2023)]%
        {srivastava2023overview}
\bibfield{author}{\bibinfo{person}{Vikas Srivastava}, \bibinfo{person}{Anubhab
  Baksi}, {and} \bibinfo{person}{Sumit~Kumar Debnath}.}
  \bibinfo{year}{2023}\natexlab{}.
\newblock \showarticletitle{An Overview of Hash Based Signatures}.
\newblock \bibinfo{journal}{\emph{Cryptology ePrint Archive}}
  (\bibinfo{year}{2023}).
\newblock


\bibitem[Srivastava and Gupta(2023)]%
        {srivastava2023ascon}
\bibfield{author}{\bibinfo{person}{Vikas Srivastava} {and}
  \bibinfo{person}{Gupta}.} \bibinfo{year}{2023}\natexlab{}.
\newblock \showarticletitle{Ascon-Sign}.
\newblock \bibinfo{journal}{\emph{NIST PQC Additional Round}}
  \bibinfo{volume}{1} (\bibinfo{year}{2023}).
\newblock


\bibitem[Ueno et~al\mbox{.}(2021)]%
        {BIKE3}
\bibfield{author}{\bibinfo{person}{Rei Ueno}, \bibinfo{person}{Keita Xagawa},
  \bibinfo{person}{Yutaro Tanaka}, \bibinfo{person}{Akira Ito},
  \bibinfo{person}{Junko Takahashi}, {and} \bibinfo{person}{Naofumi Homma}.}
  \bibinfo{year}{2021}\natexlab{}.
\newblock \bibinfo{title}{Curse of Re-encryption: A Generic Power/EM Analysis
  on Post-Quantum KEMs}.
\newblock \bibinfo{howpublished}{Cryptology ePrint Archive, Paper 2021/849}.
\newblock
\urldef\tempurl%
\url{https://eprint.iacr.org/2021/849}
\showURL{%
\tempurl}
\newblock
\shownote{\url{https://eprint.iacr.org/2021/849}}.


\bibitem[Weger et~al\mbox{.}(2022)]%
        {weger2022survey}
\bibfield{author}{\bibinfo{person}{Violetta Weger}, \bibinfo{person}{Niklas
  Gassner}, {and} \bibinfo{person}{Joachim Rosenthal}.}
  \bibinfo{year}{2022}\natexlab{}.
\newblock \showarticletitle{A survey on code-based cryptography}.
\newblock \bibinfo{journal}{\emph{arXiv preprint arXiv:2201.07119}}
  (\bibinfo{year}{2022}).
\newblock


\bibitem[Xing and Li(2021)]%
        {xing2021compact}
\bibfield{author}{\bibinfo{person}{Yufei Xing} {and} \bibinfo{person}{Shuguo
  Li}.} \bibinfo{year}{2021}\natexlab{}.
\newblock \showarticletitle{A compact hardware implementation of CCA-secure key
  exchange mechanism CRYSTALS-KYBER on FPGA}.
\newblock \bibinfo{journal}{\emph{IACR Transactions on Cryptographic Hardware
  and Embedded Systems}} (\bibinfo{year}{2021}), \bibinfo{pages}{328--356}.
\newblock


\bibitem[Xu et~al\mbox{.}(2023)]%
        {xu2023overview}
\bibfield{author}{\bibinfo{person}{Guobin Xu}, \bibinfo{person}{Jianzhou Mao},
  \bibinfo{person}{Eric Sakk}, {and} \bibinfo{person}{Shuangbao~Paul Wang}.}
  \bibinfo{year}{2023}\natexlab{}.
\newblock \showarticletitle{An Overview of Quantum-Safe Approaches: Quantum Key
  Distribution and Post-Quantum Cryptography}. In
  \bibinfo{booktitle}{\emph{2023 57th Annual Conference on Information Sciences
  and Systems (CISS)}}. IEEE, \bibinfo{pages}{1--6}.
\newblock


\bibitem[Yang et~al\mbox{.}(2023)]%
        {yang2023stamp}
\bibfield{author}{\bibinfo{person}{Bolin Yang}, \bibinfo{person}{Prasanna
  Ravi}, {and} \bibinfo{person}{Zhang}.} \bibinfo{year}{2023}\natexlab{}.
\newblock \showarticletitle{STAMP-Single Trace Attack on M-LWE Pointwise
  Multiplication in Kyber}.
\newblock \bibinfo{journal}{\emph{Cryptology Archive}} (\bibinfo{year}{2023}).
\newblock


\bibitem[Yang(2023)]%
        {yang2023survey}
\bibfield{author}{\bibinfo{person}{Zebo Yang}.}
  \bibinfo{year}{2023}\natexlab{}.
\newblock \showarticletitle{A Survey of Important Issues in Quantum Computing
  and Communications}.
\newblock \bibinfo{journal}{\emph{IEEE Comm. Surveys \& Tutorials}}
  (\bibinfo{year}{2023}).
\newblock


\bibitem[Yasuda(2018)]%
        {yasuda2018multivariate}
\bibfield{author}{\bibinfo{person}{Takanori Yasuda}.}
  \bibinfo{year}{2018}\natexlab{}.
\newblock \showarticletitle{Multivariate encryption schemes based on the
  constrained MQ problem}. In \bibinfo{booktitle}{\emph{Provable Security: 12th
  International Conference, ProvSec 2018, Jeju, South Korea, October 25-28,
  2018, Proceedings 12}}. Springer, \bibinfo{pages}{129--146}.
\newblock


\bibitem[Yasuda et~al\mbox{.}(2020)]%
        {yasuda2020multivariate}
\bibfield{author}{\bibinfo{person}{Takanori Yasuda}, \bibinfo{person}{Yacheng
  Wang}, {and} \bibinfo{person}{Tsuyoshi Takagi}.}
  \bibinfo{year}{2020}\natexlab{}.
\newblock \showarticletitle{Multivariate encryption schemes based on polynomial
  equations over real numbers}. In \bibinfo{booktitle}{\emph{Post-Quantum
  Cryptography: 11th International Conference, PQCrypto 2020, Paris, France,
  April 15--17, 2020, Proceedings 11}}. Springer, \bibinfo{pages}{402--421}.
\newblock


\bibitem[Yavuz and Behnia(2022)]%
        {yavuz2022frog}
\bibfield{author}{\bibinfo{person}{Attila~A Yavuz} {and}
  \bibinfo{person}{Rouzbeh Behnia}.} \bibinfo{year}{2022}\natexlab{}.
\newblock \showarticletitle{FROG: Forward-Secure Post-Quantum Signature}.
\newblock \bibinfo{journal}{\emph{arXiv preprint arXiv:2205.07112}}
  (\bibinfo{year}{2022}).
\newblock


\bibitem[Zhang et~al\mbox{.}(2023)]%
        {Falcon3}
\bibfield{author}{\bibinfo{person}{Shiduo Zhang}, \bibinfo{person}{Xiuhan Lin},
  \bibinfo{person}{Yang Yu}, {and} \bibinfo{person}{Weijia Wang}.}
  \bibinfo{year}{2023}\natexlab{}.
\newblock \showarticletitle{Improved Power Analysis Attacks on Falcon}. In
  \bibinfo{booktitle}{\emph{Annual International Conference on the Theory and
  Applications of Cryptographic Techniques}}. Springer,
  \bibinfo{pages}{565--595}.
\newblock


\end{thebibliography}


\end{document}